 \documentclass[longauth]{aa}

\usepackage{graphicx}
\usepackage{orcidlink}
\usepackage{subcaption}
\usepackage{txfonts}
\usepackage{amsmath}
\hypersetup{
    colorlinks=true,     % Enlace coloreado
    linkcolor=blue,     % Color para enlaces internos
    citecolor=blue,     % Color para citas
    urlcolor=blue       % Color para URLs
}
\definecolor{green_new}{rgb}{0.0, 0.5, 0.0}

\begin{document}

   \title{The \textit{AGORA} High-Resolution Galaxy Simulations Comparison Project. VII.}

   \subtitle{Satellite quenching  in zoom-in simulation of a Milky Way-mass halo}
   \authorrunning{R. Rodríguez-Cardoso et al.}

   \author{R.~Rodríguez-Cardoso\inst{1, 2, 3}$^\star$\orcidlink{0000-0002-9158-195X},
    S.~Roca-Fàbrega\inst{4}$^\star$\orcidlink{0000-0002-6299-152X},
    Minyong~Jung\inst{5}$^\star$\orcidlink{0000-0002-9144-1383},
    Th\d{i}nh~H.~Nguy$\tilde{\hat{\rm{e}}}$n\inst{6, 7}$^\star$\orcidlink{0009-0002-2290-8039},
    Ji-hoon~Kim\inst{5, 8, 9}\orcidlink{0000-0003-4464-1160},
    Joel~Primack\inst{10}\orcidlink{0000-0001-5091-5098} ,
    Oscar~Agertz\inst{4}\orcidlink{0000-0002-4287-1088} , 
    Kirk~S.~S.~Barrow\inst{6}\orcidlink{0000-0002-8638-1697},
    Jesus~Gallego\inst{1, 3}\orcidlink{0000-0003-1439-7697},
    Kentaro~Nagamine\inst{11, 12, 13, 14, 15}\orcidlink{0000-0001-7457-8487},
    Johnny~W.~Powell\inst{16}\orcidlink{0000-0002-3764-2395},
    Yves~Revaz\inst{17}\orcidlink{0000-0002-6227-0108},
    Hector~Velázquez\inst{18},
    Anna~Genina\inst{19}\orcidlink{0000-0003-0073-3012},
    Hyeonyong~Kim\inst{5}\orcidlink{0000-0002-7820-2281},
    Alessandro~Lupi\inst{20, 21}\orcidlink{0000-0001-6106-7821},
    Tom~Abel\inst{22, 23, 24}\orcidlink{0000-0002-5969-1251},
    Renyue~Cen \inst{25, 26} \orcidlink{0000-0001-8531-9536},
    Daniel~Ceverino\inst{27, 28}\orcidlink{0000-0002-8680-248X}, 
    Avishai~Dekel\inst{29}\orcidlink{0000-0003-4174-0374},
    Boon~Kiat~Oh\inst{5, 30}\orcidlink{0000-0003-4597-6739},
    Thomas~R.~Quinn\inst{31}\orcidlink{0000-0001-5510-2803},
    \\
          \and
           The AGORA Collaboration \inst{32}
          }

   \institute{Departamento de Física de la Tierra y Astrofísica, Fac. de C.C. Físicas, Universidad
Complutense de Madrid, E-28040 Madrid, Spain\\
              \email{ramorodr@ucm.es}
         \and
             GMV, Space and Avionics Equipment, Isaac Newton, 11 Tres Cantos, E-28760 Madrid, Spain
        \and Instituto de Física de Partículas y del Cosmos, IPARCOS, Fac. C.C. Físicas, Universidad
Complutense de Madrid, E-28040 Madrid, Spain
        \and Lund Observatory, Division of Astrophysics, Department of Physics, Lund University, SE-221 00 Lund, Sweden\\
                    \email{santi.roca\_fabrega@fysik.lu.se }
        \and Center for Theoretical Physics, Department of Physics and Astronomy, Seoul National University, Seoul 08826, Republic of Korea\\
                        \email{wispedia@snu.ac.kr}
        \and Department of Astronomy, University of Illinois at Urbana-Champaign, Urbana, IL 61801, USA\\
                    \email{thinhhn2@illinois.edu}
        \and Center for AstroPhysical Surveys, National Center for Supercomputing Applications, Urbana, IL, 61801, USA
        \and Institute for Data Innovation in Science, Seoul National University, Seoul 08826, Korea
        \and Seoul National University Astronomy Research Center, Seoul 08826, Republic of Korea
        \and  Department of Physics, University of California at Santa Cruz, Santa Cruz, CA 95064, USA
        \and Theoretical Astrophysics, Department of Earth and Space Science, Graduate School of Science, Osaka University, Toyonaka, Osaka 560-0043, Japan
        \and Theoretical Joint Research, Forefront Research Center, Graduate School of Science, Osaka University, Toyonaka, Osaka 560-0043, Japan
        \and  Kavli IPMU (WPI), The University of Tokyo, 5-1-5 Kashiwanoha, Kashiwa, Chiba, 277-8583, Japan    
        \and Department of Physics \& Astronomy, University of Nevada, Las Vegas, 4505 S. Maryland Pkwy, Las Vegas, NV 89154-4002, USA
        \and Nevada Center for Astrophysics, University of Nevada, Las Vegas, 4505 S. Maryland Pkwy, Las Vegas, NV 89154-4002, USA
        \and  Department of Physics, Reed College, Portland, OR 97202, USA
        \and Institute of Physics, Laboratoire d’Astrophysique, École Polytechnique Fédérale de Lausanne (EPFL), CH-1015 Lausanne, Switzerland
        \and  Instituto de Astronomía, Universidad Nacional Autónoma de México, A.P. 70-264, 04510, Mexico, D.F., Mexico
        \and Max-Planck-Institut für Astrophysik, Karl-Schwarzschild-Str. 1, D-85748, Garching, Germany
        \and Como Lake Center for Astrophysics, DiSAT, Università degli Studi dell’Insubria, via Valleggio 11, IT-22100, Como, Italy
        \and  INAF - Osservatorio di Astrofisica e Scienza dello Spazio di Bologna, Via Gobetti 93/3, 40129 Bologna, ItalyThe 
        \and  Kavli Institute for Particle Astrophysics and Cosmology, Stanford University, Stanford, CA 94305, USA
        \and Department of Physics, Stanford University, Stanford, CA 94305, USA
        \and SLAC National Accelerator Laboratory, Menlo Park, CA 94025, USA
        \and Center for Cosmology and Computational Astrophysics, Institute for Advanced Study in Physics, Zhejiang University, Hangzhou 310027, China
        \and Institute of Astronomy, School of Physics, Zhejiang University, Hangzhou 310027, China
        \and  Departamento de Física Teórica, Facultad de Ciencias, Universidad Autónoma de Madrid, Cantoblanco, E-28049 Madrid, Spain
        \and CIAFF, Facultad de Ciencias, Universidad Autónoma de Madrid, E-28049 Madrid, Spain
        \and Center for Astrophysics and Planetary Science, Racah Institute of Physics, The Hebrew University, Jerusalem 91904, Israel
        \and  Department of Physics, University of Connecticut, U-3046, Storrs, CT 06269, USA
        \and Department of Astronomy, University of Washington, Seattle, WA 98195, USA
        }

   \date{Received 31 December, 2024; accepted 16 April, 2025}

% \abstract{}{}{}{}{} 
% 5 {} token are mandatory
  \abstract
   {Satellite galaxies experience multiple physical processes when interacting with their host halos, often leading to the quenching of star formation. In the Local Group, satellite quenching has been shown to be highly efficient, affecting nearly all satellites except the most massive ones. While recent surveys study Milky Way-analogs to assess how representative our Local Group is, the dominant physical mechanisms behind satellite quenching in Milky Way-mass halos remain under debate.}
   {We analyze satellite quenching within the same Milky Way-mass halo simulated using various widely used astrophysical codes, each using different hydrodynamic methods and implementing different supernovae feedback recipes. The goal is to determine whether quenched fractions, quenching timescales, and the dominant quenching mechanisms are consistent across codes or if they show sensitivity to the specific hydrodynamic method and supernovae feedback physics employed.}
       {We used a subset of high-resolution cosmological zoom-in simulations of a Milky Way-mass halo from the multiple-code AGORA \texttt{CosmoRun} suite. Our analysis focuses on comparing satellite quenching across the different models and against observational data. We also analyzed the dominant mechanisms driving satellite quenching in each model. }
   { 
   We find that the quenched fraction is consistent with the latest SAGA Survey results within its $1\sigma$ host-to-host scatter across all the models. 
   Regarding quenching timescales, all the models reproduce the trend observed in the ELVES survey, Local Group observations, and previous simulations: The less massive the satellite, the shorter its quenching timescale. All of our models converge on the dominant quenching mechanisms: Strangulation halts cold gas accretion in all satellites, while ram pressure stripping is the predominant mechanism for gas removal, and it is particularly effective in satellites with $\rm{M}_* < 10^8\, \rm{M}_\odot$. Nevertheless, the efficiency of the stripping mechanisms differs among the codes, showing a strong sensitivity to the different supernovae feedback implementations and/or hydrodynamic methods employed.}
   {}


   \maketitle
%
%-------------------------------------------------------------------
\defcitealias{Kim_2014_paper1}{Paper I}
\defcitealias{Kim_2016_paper2}{Paper II}
\defcitealias{santi_paper3_agora}{Paper III}
\defcitealias{santi_paper_4_agora}{Paper IV}
\defcitealias{minyong_paper5}{Paper V}
\defcitealias{Strawn_2024}{Paper VI}
\section{Introduction} \label{sec:intro}
\renewcommand{\thefootnote}{}
\footnotetext{$^{\star}$ Corresponding authors.}
\footnotetext{$^{32}$ \href{https://sites.google.com/site/santacruzcomparisonproject/} {https://www.agorasimulations.org/}}
Understanding how, when, and why star formation ceases in both satellite and field galaxies remains a topic of great interest in astronomy \citep{Peng_2010_quenching, Schaye_2010_physics_quench, Wetzel_2013, Nelso_2018_bimodality_TNG, Donnari_2021_TNG}. Ultimately, the mechanisms that quench galaxies must reduce the amount of cool gas available for star formation. This reduction can be achieved by either directly removing gas from galaxies or stifling the rate at which the gas is replenished via accretion. In the current galaxy formation scenario, gas surrounding halos falls into their potential well, fueling star formation of the galaxies embedded in them. However, if the mass of the halo is large enough such that its gravitational dynamical time is much shorter than the cooling time of the gas, the kinetic energy of the infalling gas is converted into thermal energy, heating the gas to the halo's virial temperature \citep{White_frenk_91}. These accretion shocks are expected to form close to the virial radius of the halo and are therefore commonly referred to as virial shocks. Although virial shocks may be unstable and thus do not survive around low-mass halos, they are expected to be an inevitable consequence of structure formation for halos above a few times $10^{11}\, \rm{M}_\odot$ \citep{Birnboim_dekel_2003, keres_2005}. This points to a mass quenching due to halting of gas accretion of halos more massive than the so-called "golden mass" around $\rm{M}_{\rm{halo}} \sim 10^{12}\, \rm{M}_\odot$ \citep{dekel2019_goldenmass}. Thus, this scenario naturally leads to the observed bimodality between central galaxies above and below $\rm{M}_* \sim 10^{10.5}\, \rm{M}_\odot$ \citep{Kauffman_2003b, DB06}.

\renewcommand{\thefootnote}{\arabic{footnote}}
\setcounter{footnote}{0}

In the case of satellite galaxies, the picture is further complicated by the fact that both internal and external processes can act simultaneously. The interplay between these different processes and how they cause the eventual quenching of satellite galaxies remains poorly understood in the literature \citep{Fillingham_2016, MERLUZZI_2016, Wright_2022, Samuel_2023, Cramer_2024, Wang_saga_5_2024}. Satellite galaxies are expected to undergo a variety of physical processes during their infall to their host halos. 
The strong tidal forces experienced by the satellite due to the presence of a central galaxy can remove dark matter (DM), gas, and stars from the galaxy through a process called "tidal stripping" \citep{Read_2006_tidal}. Tidal forces may also perturb the structure of galaxies and change a satellite's morphology \citep{Mayer_2001_tidal_morph}. Moreover, as the satellite travels through the dense host circumgalactic medium (CGM), the gas bound to the satellite experiences a drag force due to the relative motion of the two fluids. If the drag force exceeds the restoring force due to the satellite's own gravity, its gas will be stripped in a process called "ram pressure stripping" \citep{Gunn_Gott_1972, Abadi_1999}. In addition to ram pressure stripping, the interaction between the satellite's gas and the hot or warm host CGM can lead to a slower but continuous gas loss, usually referred to as "continuous" or "viscous or turbulent stripping," caused by the Kelvin–Helmholtz (KH) instability induced by the host gas \citep{Nulsen_1982_viscous, Quilis_2000_viscous, Schulz_2001_viscous}. There are other mechanisms that are usually invoked as contributing to the loss of gas from the satellite galaxy, such as "galaxy harassment," where satellite galaxies are progressively “heated" by high-speed encounters with other satellite galaxies and become more prone to disruption by the potential well of the halo \citep{Moore_96_harassment}. Whereas the aforementioned mechanisms are in charge of gas removal, satellites living inside virial shocked halos also experience a cut off of their cold gas supply \citep{Gabor_2015_strang} in a process often called "strangulation" in the literature \citep{Peng_2015_strang} that prevents the satellite from replenishing its gas reservoirs.

Within the Local Group (LG), satellite galaxies are almost all gas poor and quiescent, except for some of the most massive satellites such as the Magellanic Clouds, LGS3, and IC10 (e.g., \citealt{Grcevich_Putman_HI_2009,kirby_2013, Mcconnachie_2012, Spekkens_2014, Wetzel_2015b}), suggesting that their gas is efficiently removed by some stripping mechanism. Due to the unmatched depth and completeness of observations of its satellite population, the Milky Way (MW) has become the primary reference for simulations exploring the fundamental physics of satellite quenching and their timescales  \citep{Mayer_2006, Simpson_2018_ram_pressure_MW, Akins_2021, Samuel_fire_2022, Samuel_2023}. Consequently, it is essential to determine if our LG is representative. This goal has prompted significant efforts to observe MW analogs and analyze their satellite populations. Examples include the pioneering work by \citeauthor{zaritsky_93} \citeyearpar{zaritsky_93, Zaritsky_1997} to the recent Satellites Around Galactic Analogs (SAGA) Survey \citep{Geha_2017, Mao_2021, mao2024}, in which the authors study the properties of 378 satellite galaxies around 101 MW analogs. 

Comparisons between the satellite properties and quenched fractions from the SAGA Survey, the LG, and other MW-analog surveys, such as the Exploration of Local VolumE Satellites (ELVES) survey \citep{Carlsten_2020, Carlsten_2022, Greene_2023}, have revealed some tension. These comparisons suggest that our LG has a higher fraction of quenched satellites than their analogs, which has usually been attributed to potential differences in the assembly history of the host halo \citep{Haussaman_2019}. However, this tension has been alleviated somewhat with the latest SAGA data release \citep{Geha_2024} after increasing their host statistics and applying a correction method to their quenched fraction for spectroscopic incompleteness. Recent efforts from the simulation side have begun to understand this discrepancy (e.g., \citealt{karunakaran_2021} or \citealt{font_2022}), but the representativeness of the LG is still under question.

The unresolved questions surrounding quenching mechanisms affecting satellites in a MW-mass halo, along with the sensitivity of these mechanisms to factors such as stellar feedback, code architecture, and virial shock formation, make this a compelling area of study for the Assembling Galaxies of Resolved Anatomy (AGORA) code comparison project, whose earlier simulations are shown in \cite{Kim_2014_paper1, Kim_2016_paper2} (hereafter \citetalias{Kim_2014_paper1} and \citetalias{Kim_2016_paper2}, respectively). The AGORA project aims to enhance the predictive capabilities of numerical galaxy formation simulations by comparing high-resolution galaxy-scale calculations across multiple code platforms. In this large international collaboration, leading simulation code researchers are engaged in examining how different simulation codes converge or diverge when applied to the same initial conditions \textcolor{black}{(ICs)} while keeping physical implementations as consistent as possible. In this paper, we analyze the quenching of the satellite population around a MW-mass target halo using \texttt{CosmoRun} simulations described in \cite{santi_paper3_agora, santi_paper_4_agora} (hereafter \citetalias{santi_paper3_agora} and \citetalias{santi_paper_4_agora}, respectively).  The satellite population in these simulations has been carefully studied in \cite{minyong_paper5} (hereafter, \citetalias{minyong_paper5}), whereas the differences in the CGM across the different models has been presented in \cite{Strawn_2024} (hereafter, \citetalias{Strawn_2024}). Historically, code comparisons between grid-based and smoothed particle hydrodynamics (SPH) methods revealed significant discrepancies in the treatment of fluid instabilities, such as the suppression of viscous stripping in SPH codes, since they were unable to reproduce dynamical instabilities \citep{Agertz_2007}. Although substantial efforts have been made to improve the treatment of instabilities in SPH codes \citep{Price_2008_SPH, Wadsley_2008_SPH, Read_2010_sph}, comparing how satellite quenching occurs across different codes employing distinct hydrodynamic techniques remains an essential task in numerical astrophysics in order to assess the reproducibility of results. 
In this paper, we compare five\footnote{We note that in contrast with the eight codes presented in Papers IV-VI, we use only five codes here, as they are the ones that reach $\rm{z} < 1$ and thus cover host masses close to the critical threshold where virial shocks are expected to form. The lowest redshift achieved by each code group does not reflect the performance of the code but instead the availability of manpower and CPU
time at the computational facilities each group had access to.} hydrodynamic \texttt{CosmoRun} simulations, all of which are performed with the state-of-the-art galaxy simulation codes widely used in the numerical galaxy formation community, and we study the evolution of the properties of their satellites during the interaction with the host halo. This approach allows us to determine if our results are consistent regardless of the code architecture and stellar feedback physics employed. Conversely, any divergence in the results will help us understand the impact of varying feedback and code models, providing us with a more informed interpretation of our observables.

This paper is organized as follows: Section \ref{sec:methodology} provides an overview of the AGORA \texttt{CosmoRun} simulation, including details on subhalo identification, stellar particle assignment, and our definition of quiescence. Section \ref{sec:results} presents the quenched fraction of the satellite population and its evolution with cosmic time across all models as well as the satellite quenching timescales for each model. We then assess the contribution of different quenching mechanisms for each model and identify the dominant mechanism. Section \ref{sec:conclusion} summarizes the main convergences and divergences between models and highlights the significant contributions of comparison projects such as this one to the field of galaxy simulations. Finally, in Section \ref{sec:caveats} we outline key caveats of our methodology and simulations.
%-------------------------------------------
\section{Methodology} \label{sec:methodology}
\subsection{The AGORA \texttt{CosmoRun} simulation suite}

Throughout this paper we use a subset of simulations from the \texttt{CosmoRun} simulation suite described in \citetalias{santi_paper3_agora} and \citetalias{santi_paper_4_agora}. \texttt{CosmoRun} is a suite of high-resolution cosmological zoom-in simulations of a MW-mass halo ($\sim 10^{12} \rm{M_{\odot}}$ at $\rm{z} = 0$) across multiple code platforms. The simulations analyzed herein started from the same cosmological initial conditions, created using the software \textsc{MUSIC}, which generates a realistic distribution of DM and primordial gas at a starting redshift $\rm{z} = 100$. The adopted cosmological parameters are $\Omega_{\Lambda} = 0.728$, $\Omega_{\text{matter}} = 0.272$, $\Omega_{\text{DM}} = 0.227$, $\sigma_8 = 0.807$, $\rm{n_s} = 0.961$, and $\rm{h} = 0.702$. The subset of simulations used in this paper consists of 5 out of the original 8 codes in \citetalias{santi_paper_4_agora}, for which we already have snapshots at $\rm{z}\leq 0.3$. This subset includes: adaptive mesh refinement (AMR) codes ART-I \citep{Kravtsov_97_art} and ENZO \citep{Bryan_2014_enzo, Brummel_smith_2019_enzo}, SPH codes GADGET-3 (an updated version of GADGET-2; \citealt{Springel_2005_gadget2}) and GEAR \citep{Revaz_Jablonka_2012_gear}; and the moving-mesh (MM) code AREPO-T \citep{Springel_2010}. We refer to the version used of AREPO as AREPO-T, which represents the AREPO code with thermal feedback  (the details about AREPO-T model are illustrated in footnote 55 of \citetalias{santi_paper_4_agora}). How galaxies formed and evolved has been studied using all these different approaches, each with its own advantages and disadvantages. In \citetalias{santi_paper3_agora} and \citetalias{santi_paper_4_agora} was shown that all the codes reached an overall agreement in the stellar properties of the target halo and in its mass assembly history, after a series of calibration steps. At $\rm{z=0}$, all the codes converge to roughly $\rm{M}_{halo} \sim 10^{12}\rm{M}_\odot$ and $\rm{M}_* \sim 10^{11}\rm{M}_\odot$, a more detailed analysis can be found in \citetalias{santi_paper_4_agora}.

Codes participating in AGORA share much of the physics that governs their operation: gas heating and cooling parameters, implemented by the common package GRACKLE \citep{smith_2017}. Redshift dependent cosmic ultraviolet background \citep{Haardt_2012}, also provided by GRACKLE. Star formation criteria are also identical for all the codes, with the exception of choosing the stochastic or deterministic nature of this process. Regarding to the code-dependent physics, \textcolor{black}{each code group is given the freedom to choose its own feedback scheme for energy and metals}. Details of both the common and code-dependent physics of each code are described with great detail in \citetalias{santi_paper3_agora} and \citetalias{santi_paper_4_agora}. 

Particle-based (i.e., SPH and MM) codes in \texttt{CosmoRun} simulations have a gravitational force softening length in the highest-resolution region of 800 comoving pc until $\rm{z} = 9$ and 80 proper pc afterward. In the case of grid-based codes, the finest cell size is set to 163 comoving pc, or 12 additional refinement levels for a $128^3$ root resolution in a $(60$ comoving $\rm{h}^{-1} \text{Mpc}^3)$ box. A cell is adaptively refined into 8 child cells on particle over-densities of 4. For details on runtime parameters, we refer the readers to \citetalias{santi_paper3_agora}.

In \citetalias{minyong_paper5}, it was shown that the population of satellites in all \texttt{CosmoRun} simulations is comparable to that of MW or M31 in their stellar masses and stellar velocity dispersions, probing that by implementing the common baryonic physics adopted in AGORA and the stellar feedback prescription commonly used in each code, the so-called ``missing satellite problem” is fully resolved across all participating codes. Some systematic differences in the stellar to halo mass relation were reported, with ART-I and GEAR \textcolor{black}{showing larger} $\rm{M}_*/\rm{M_{halo}}$ than ENZO or AREPO-T. 

The differences in the CGM properties of the target halo, primarily driven by the varying feedback prescriptions of each code - such as their ability to expel metals - are analyzed in detail in \citetalias{Strawn_2024}.

\subsection{Star assignment}

To study the satellite galaxies in our simulations, we need to identify the subhalos that formed stars. \textcolor{black}{We assign stellar particles to a halo} following the process outlined in \citetalias{minyong_paper5}, summarized below. Firstly, all the stellar particles located within $0.8R_{\text{vir}}$ from the halo are identified. Next, we narrow down the selection to those with velocities relative to the halo that are less than twice the halo’s maximum circular velocity. We then calculate the radius that contains 90\% of the stellar particles ($\rm{R}_{90}$) and the stellar velocity dispersion ($\sigma_{\text{vel}}$). To refine our selection, we filter the stellar particles by applying two additional criteria: (1) they must be located within $1.5\rm{R}_{90}$ of the center of mass of the halo and stellar particles, and (2) their velocities relative to the halo must be less than $2\sigma_{\text{vel}}$. This process is iterated, recalculating $\rm{R_{90}}$ and $\rm{\sigma_{\text{vel}}}$ for the selected particles until the values converge to within 99\% of the previous iteration. We proceed from the most massive halos down to smaller ones, ensuring that no stellar particles are reassigned once they have been already associated with a more massive halo. Finally, we define “galaxies” as those with stellar masses at least six times the approximate stellar particle mass resolution (i.e., $\rm{M}_{\text{star}} > 6m_{\text{gas,IC}} = 3.39\times 10^5 \,\rm{M}_{\odot}$; see Section 3.1 of \citetalias{santi_paper3_agora}).
\label{Star_assigment}

\subsection{ Definition of quiescence}
\label{Quiescence_definition}

In observations, the most accepted criteria to define when a galaxy is quiescent relies on the star formation rate (SFR) determined using tracers such as H$\alpha$, UV/FUV, HI or CO emission (e.g., \citealt{Leroy_2008,Grcevich_2009_hi, Spekkens_2014, Putman_2021}). The former two provide insights into recent star formation over the past few 10 Myr (H$\alpha$) or few 100 Myr (UV/FUV), while the latter two reflect the abundance of cold gas, serving as an indicator of instantaneous star formation. Our simulations enable the computation of both the complete time-resolved star formation history and the gas content. Hence, we intend to utilize information from both metrics to define quiescence in our galaxies, following a methodology akin to that presented in \cite{Samuel_fire_2022}. Unless otherwise specified, throughout this paper, we categorize a galaxy as quenched if it fulfils these three conditions: \textit{(i)} it has not undergone star formation activity within the last 200 Myrs, \textit{(ii)}  its star-forming gas mass content ($\rm{T}<10^{4}\,\rm{K}$ and $\rm{n_H}>1\, \rm{cm}^{-3}$) is lower than six times the approximate mass resolution of gas particles (i.e., $\rm{M}_{\text{SF, gas}} < \rm{m}_{\text{gas, IC}} = 3.39\times 10^5 \, \rm{M}_{\odot}$), and \textit{(iii)} the galaxy fulfills the previous two conditions for the rest of its lifetime.
We verified that our results were consistent with alternative definitions, such as applying a condition of $\rm{sSFR} < 10^{-11} \,\rm{yr}^{-1}$ alone, without altering the conclusions throughout the paper.
\subsection{Gas assigment}
\label{gas_assigment}

To understand the quenching process of our satellites and correctly determine their timescale, it is crucial to accurately compute their gas content, ensuring no contamination from unbound gas from the host CGM. In particle-based simulations, this step is straightforward as we assign gas particles to each halo based on gravitational binding. Conversely, in grid-based simulations, direct gas particle assignment is not feasible. Therefore, throughout this paper, we employ a different approach to avoid mixing the gas from the host halo with that of the subhalos:

\begin{enumerate}
    \item We compute an average radial gas density profile of the gas contained in the host halo, $\rm{\rho_{CGM}(r)}$, extending out to a radius $2\rm{R_{vir}^{host}}$ in each snapshot. The average
    radial gas density profile is computed using all gas cells that are not within any subhalo.
    \item For each subhalo at a distance $r < 2\rm{R_{vir}^{host}}$, we compute the gas mass within its virial radius, $\rm{M_{gas}^{sub}(r)}$, in each snapshot during its infall. We then estimate the host gas mass at that distance $\rm{r}$ for a sphere of $\rm{R_{vir}^{sub}}$, considering snapshots before and after the subhalo was located at that position, $\rm{r}$:
    \begin{equation}
        \rm{M_{CGM}(r) = \rho_{CGM}(r) \frac{4\pi}{3}({R_{vir}^{sub}})^3},
    \end{equation}
    
    and we average the values from the previous ($\rm{M_{CGM}^{prior}}$) and subsequent ($\rm{M_{CGM}^{post}}$) snapshots of the subhalo passage. We can accurately capture the CGM gas mass just before and after the passage of the subhalo thanks to the availability of many snapshots with small temporal separation (see \citetalias{santi_paper3_agora} and \citetalias{santi_paper_4_agora} where the short timesteps are described).
    \item Finally, the gas content associated with the subhalo is computed as
    \begin{equation}
       \rm{M_{bound\,gas}^{sub} (r) = M_{gas}^{sub}(r) - M_{CGM}(r)}, 
    \end{equation}
    
    where $\rm{M_{CGM}(r) = \left(M_{CGM}^{prior}(r)+ M_{CGM}^{post}(r)\right)/2}$.
    This process is repeated for the star-forming, cold, cool, and hot gas components. 
\end{enumerate}

\begin{figure}
    \centering
    \includegraphics[width=0.9\linewidth]{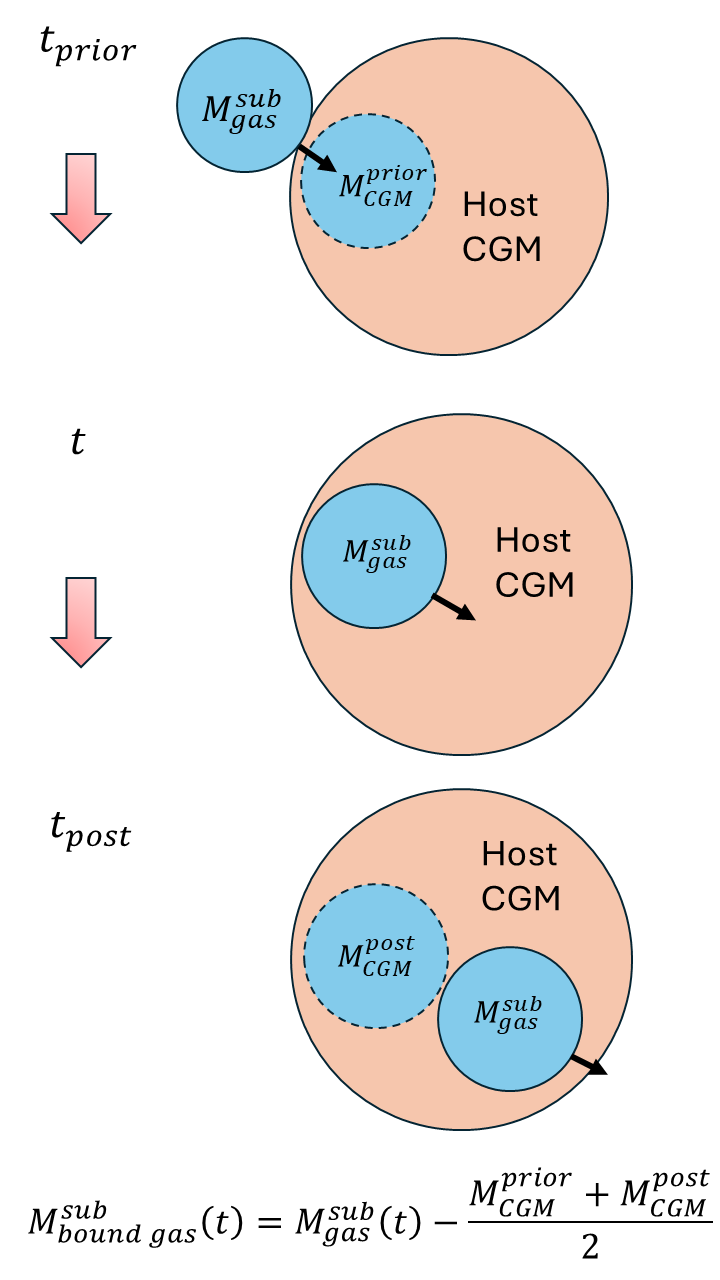}
    \caption{Schematic illustrating the process used to determine the subhalo's gas mass in grid-based codes. This involves correcting the total gas mass within the subhalo's virial radius by subtracting the host's CGM  that is momentarily  inside the subhalo’s virial radius but not truly associated with it, as described in Section \ref{gas_assigment}. The host's CGM mass at the snapshots before and after the subhalo's passage is computed using the average gas density profile, considering a sphere of the subhalo's virial radius during the passage. A comparison between the gas mass obtained using this method and that using bound gas particles in particle-based codes is presented in Figure \ref{fig:appendix} in Appendix \ref{appendix:inflows}.}
    \label{fig:schema_gas}
\end{figure}

A visual schematic of this method is shown in Figure \ref{fig:schema_gas}. Using this approach, we correct the gas content of our subhalo by excluding the CGM that is momentarily inside the subhalo's virial radius but not truly associated with it. Our strategy for estimating the mass of the host gas at some radial distance, using the average gas density profile, robustly captures the order of magnitude of the mass contribution related to the CGM for each component. However, satellites in our simulations may encounter many local perturbations in the CGM along their orbital path. These variations arise from feedback-driven winds from the host galaxy, clumpy gas accretion from the IGM, and
stripped gas from other satellites. Our approach neglects these local variations, \textcolor{black}{that} can lead to artificial spikes and drops of gas mass. To mitigate this, we smooth the evolution of the gas mass, by applying a median filter with a kernel \textcolor{black}{window} of 200 Myr, to focus on consistent physical drops or spikes. This approach enables us to pinpoint when the gas in our satellite is stripped or consumed and not replenished, thereby determining the quenching time of our satellite. To test the reliability of this method, we determine the gas mass in particle-based codes using both this method and by identifying the gas particles actually bound to the subhalo, finding good convergence between the two approaches. Results comparing both approaches are shown in Appendix \ref{appendix:inflows}.

\subsection{Subhalo finding}
\label{sec:halofinder}
The reliable identification of subhalos during their interaction with the host remains a nuanced challenge in numerical simulations. Although reasonable agreement exists among halo finders regarding the positions and attributes of isolated halos \citep{Knebe_2012_galaxycomparison}, the identification of subhalos is notably more challenging, owing to their tendency to blend into the variable background density of their larger host (see \citealt{Symfind_Mansfield_2023} for an overview on subhalo finding). Even widely used algorithms \textcolor{black}{such} as ROCKSTAR \citep{Rockstar_behroozi_2012}, which employ 6D phase-space friends-of-friends algorithms to group particles, and \textsc{SubFind} \citep{Subfind_springel_2001} struggle to discern subhalos when exposed to substantial mass loss due to strong tidal stripping \citep{Onions_2012_subhalofindercomparison,Diemer_sparta_2023}. 

The initial approach followed on this paper consists of using ROCKSTAR halofinder first, which identifies all the (sub)halos for a single snapshot by looking for overdensities on matter distribution using the 6D phase-space. Then, we use the merger tree code Consistent-Trees \citep{Consistent_behroozi_2012} to establish connections between halos and subhalos across temporal instances. Hereafter, we refer to this combination of ROCKSTAR + Consistent-Trees as "RCT". 

The reliability of RCT for tracking a subhalo lies in its consistent detection across snapshots. Challenges arise when the subhalo is close to the host material, hindering the identification of associated density contrast \citep{Muldrew_2010_subhalodetection, Knebe_2011, Han_HBT_2012}, when unbound streams stripped from the subhalo have a substantial mass compared to the subhalo itself \citep{Han_2018}. This can cause RCT to lose or misidentify a fraction of subhalos \citep{Symfind_Mansfield_2023, Diemer_sparta_2023}, especially during and after the pericenter passage, after experiencing a significant loss of mass. In such cases, the subhalo may be mistakenly identified as already merged with the host when it is, in fact, still a separate bound substructure. Alternatively, it could be wrongly identified as a different subhalo, which is not truly associated with the subhalo's particles.

To prevent premature loss of subhalos, we employed a method similar to the approaches used by the SYMFIND \citep{Symfind_Mansfield_2023} and SPARTA \citep{Diemer_sparta_2023} algorithms. Both of them have proved how RCT combined with "particle-tracking" of subhalos reliably extends their lifetime during their infall to the host, as long as they remain a distinct sub-structure. As in these algorithms, we use RCT output as the input for our method and track the particles originally belonging to a subhalo prior to its first infall, relying  only on these particles to identify the subhalo at later times. In the subsections below, we outline the general structure of our particle-tracking approach, highlighting the different steps and decisions involved.

\begin{figure*}
    \centering
    \includegraphics[width=\linewidth]{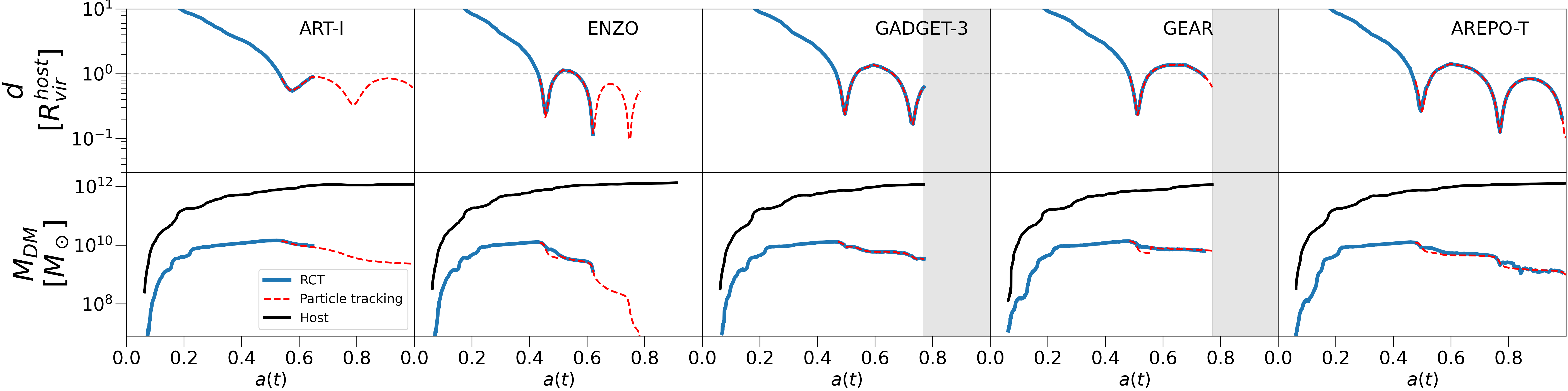}
    \caption{Evolution of the same subhalo over time for all the models, as measured by both RCT (ROCKSTAR + Consistent-Trees; blue solid line) and our algorithm (red dashed line), until the last snapshot available for each run. The time domain where snapshots are still not available for each specific code are indicated as a gray shaded region. \textbf{Top:} Subhalo's trajectory during its infall to the host halo. The host virial radius is indicated by the horizontal gray dashed line. \textbf{Bottom}: Evolution of the subhalo mass compared to the host halo (black dashed line).}
    \label{tracking}
\end{figure*}

\begin{figure*}
    \centering
    \includegraphics[width=\linewidth]{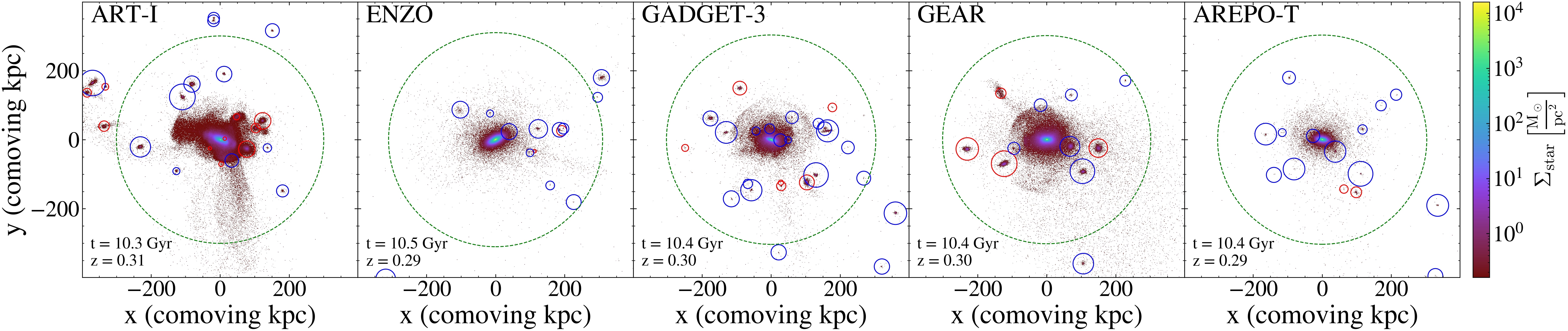}
    \caption{\textcolor{black}{Stellar surface density} at $\rm{z}\sim0.3$ for all the models. The host virial radius is indicated by the green dashed circle. "Reliable" satellite galaxies (see Section \ref{sec:halofinder} for reliability definition) and field galaxies are marked with colored solid circles representing $0.5 \rm{R_{vir}^{sub}}$. Blue circles denote galaxies identified by both RCT and our particle-tracking method, while red circles indicate those identified as "reliable" only by our particle-tracking method, showing that RCT has lost them at any snapshot, thus preventing the study of the evolution of the subhalo properties.}
    \label{stellar_surface_density}
\end{figure*}

\subsubsection{Merger tree post-processing}

We post-process the RCT merger trees by removing spurious branches as follows; we discard any branch of the merger tree that originated as a subhalo within a more massive halo, which in simulations with fine snapshots indicates a numerical artifact \citep{Symfind_Mansfield_2023}. Such spurious branches are often generated due to subhalos that are incorrectly identified as merged, even though they remain independent substructures of the host halo. Consequently, when ROCKSTAR re-detects them in later snapshots, Consistent-Trees mistakenly considers them as new branches originating within the host halo. Occasionally, Consistent-Trees does not consider them as a new branch but assigns them as the evolution of a smaller subhalo branch nearby in 6D phase space. To clean our merger tree in this case, we remove all branches that exhibit an unphysical abrupt mass increase (by a factor of 10) between two consecutive snapshots.

\subsubsection{Subhalo member particles}
We identified the particles belonging to each subhalo at the snapshot prior to infall into the host to avoid contaminating the subhalo member particles with host particles. To determine subhalo member particles, we adopted a similar definition to that used in \cite{Diemer_sparta_2023}. We designated all DM particles within $2\rm{R_{vir}^{sub}}$ as candidate members, and then we required the fulfillment of at least one of two conditions: (i) the particle entered $\rm{R_{vir}^{sub}}$ for the first time at least $2\rm{R_{vir}^{host}}$ from the host center, and/or (ii) the particle's total energy is negative.
The former condition excludes host particles that happen to currently be co-located with the subhalo, while the latter takes into account particles that become physically bound to the subhalo as it travels through the host halo’s outskirts.

\subsubsection{Subhalo tracking and properties}
Once our subhalo crosses host virial radius, we no longer rely on RCT output. Instead, for the subsequent snapshots we find our subhalo by tracking its \textcolor{black}{DM} member particles prior to accretion. We impose the condition that no additional particles are accreted by the subhalo during its infall. While this assumption may not hold true in all instances, particularly during major mergers, the number of accreted particles should typically be small in comparison to the initial mass of the subhalo \citep{Behroozi_2015_majormergerscomparison, Diemer_sparta_2023}. Even when a host particle's existing trajectory leads it to come close enough to be gravitationally bound to a subhalo, it is rapidly lost again and has little impact on the subhalo's mass long-term evolution \citep{Han_HBT_2012}. This simplification significantly reduces computational time. As an exception, if a subhalo contains its own (sub)subhalo population before crossing the host's virial radius, we allow the particles of these (sub)subhalos to be accreted by the subhalo during its infall.

At each snapshot, we track the member particles and we perform an unbinding, removing  particles that are not gravitational bound. Each subhalo's position is determined by the position of the 32 most gravitationally \textcolor{black}{DM} bound particles (or the $10\%$ most bound if the number of member particles is lower than 320), since the mean particle position is not a good estimator when the distribution is anisotropic. 

Subhalo properties are computed using particles inside $\rm{R_{vir}^{sub}}$ which is computed using bound subhalo particles and following \cite{Byan_Norman_1998} definition. We determine the maximum rotational velocity as $\rm{v_{max}=max\left(\rm{\sqrt{\frac{\rm{Gm(<r)}}{\rm{r}}}}\right)}$. 

\subsubsection{Ending a subhalo}

The last step in our algorithm involves checking whether our subhalo remains a distinct substructure or if it has been disrupted or merged with its host. We define a subhalo as disrupted if the number of bound particles ($\rm{N_{bound}}$) is lower than 20. However, certain subhalos can sink to the host center and remain there indefinitely \citep{Han_2018}. As all of their particles are on low-radius orbits, they can sustain $\rm{N_{bound}} > 20$ even though their particles are no longer significantly distinguishable from those of their host \citep{Diemer_sparta_2023}. Consequently, we terminate a subhalo if its distance to the center of the host halo is less than the subhalo's half mass radius ($\rm{r_{half}}$) for five consecutive snapshots. Upon subhalo's ending, we consider the subhalo merged with its host. If the subhalo's host merges with a new halo before the subhalo's ending, we designate this new halo as the subhalo's new host.

\subsubsection{Comparing with RCT}

Results for applying this particle-tracking approach to a representative subhalo are shown in Figure \ref{tracking} for all the ComsoRun models analyzed in this paper. We note that the method used to match subhalos between codes, ensuring we are tracking the same subhalo, is described in Section \ref{sec:pairing}. Moreover, the evolution of the properties of this subhalo across the different models is studied in detail in Section \ref{sec:case_study}. On the top panel in Figure \ref{tracking} we plot the trajectory in host virial radius units, whereas on the bottom one the mass evolution with respect to the host. \textcolor{black}{Before comparing both methods, it is worth mentioning that although the subhalo originates from the same IC in all models, differences arise in the number of pericenters and the depth of each pericenter depending on the model. One possible factor behind these differences is the intercode timing discrepancies in accretion times detected in \citetalias{santi_paper_4_agora} (see Appendix C in that paper), where it was found that small differences at high redshift can lead to changes in the impact parameters of infalling subhalos. Additionally, these timing discrepancies affect the mass of the main halo, causing slight variations in host halo mass. While these variations are small, due to the stochastic nature of gravitational interactions, even minor differences in mass can influence the orbits of our satellites.} The comparison between the output for RCT and our particle-tracking output are plotted as colored lines. For the snapshots when RCT is still detecting the subhalo, there is complete agreement in the subhalo position. The mass evolution is almost identical for both methods, with the particle-tracking predicting slightly lower masses during pericenters. This could come from a small amount of host matter being incidentally associated with the subhalo by ROCKSTAR when the subhalos is crossing the densest regions of the halo, as they detect in \cite{Symfind_Mansfield_2023}. Figure \ref{tracking} shows how the subhalo analyzed was wrongly identified as merged by RCT for several \texttt{CosmoRun} models. This issue is particularly evident for ART-I and ENZO,  where our particle-tracking method enables the subhalo to be followed for nearly two additional orbits. In addition, for GEAR and AREPO-T models, RCT is also losing our subhalos during the final snapshots. \textcolor{black}{Differences in RCT's ability to track subhalos across models are mainly due to variations in their orbits, which pose different challenges for RCT and lead to the satellite being considered merged at different times depending on the model.} Our approach is capable of tracking the subhalo until the actual merger or the end of the simulation (indicated by the gray shaded region), whereas RCT prematurely loses subhalos even when they remain relatively massive (several times $10^9\text{M}_\odot$) and still easily detectable by eye.

In Figure \ref{stellar_surface_density}, we present the \textcolor{black}{stellar surface density} for all the models at $\rm{z} \sim 0.3$. We compare `reliable' satellite galaxies identified using our particle-tracking approach with those identified using only RCT. We define a `reliable' satellite galaxy as a subhalo that was not originally formed within the host's virial radius and contained stellar particles before becoming a subhalo. Figure \ref{stellar_surface_density} illustrates that RCT often fails to track a significant number of satellites. Although ROCKSTAR may identify these subhalos due to their now detectable density contrast, Consistent-Trees may have lost them previously and now erroneously associates them with branches of other subhalos. Consequently, these subhalos are not considered reliable by our criteria, as their evolution cannot be investigated. In contrast, our method substantially increases the number of reliably detected satellites by tracking subhalos even after substantial mass loss and associating them with their appropriate branch.

A more extensive analysis integrating this particle-tracking approach in a different halo finder and comparing with various codes will be covered in a future paper. Here, we emphasize that our method ensures accurate tracking of subhalos, which is essential for studying their evolutionary properties as they interact with the host halo.

\subsection{Intercode satellite pairing}
\label{sec:pairing}
By leveraging the \texttt{CosmoRun} suite, where all the codes start from the same initial condition at $\rm{z} = 100$, we match the same subhalos between the codes. We conduct a galaxy-by-galaxy inspection, utilizing the method and pipeline described in \citetalias{minyong_paper5} and  \cite{schaller2015}; \cite{lovell2021}. \textcolor{black}{This approach allowed us to find a pair of matched halos between two simulations that share initial conditions but not particle IDs, as only SPH codes share particle IDs.}

The idea consists of identifying a pair of halos originating from the same DM patch in a nearly homogeneous early universe. Initially, we select the 40 particles closest to a target halo’s center, for instance, in the ENZO \texttt{CosmoRun} at $\rm{z} \sim 0.3$. Each DM particle's trajectory is traced backward in time to determine its position at $\rm{z} = 100$, corresponding to the initial conditions. Subsequently, for each of the 40 particles in the ENZO run, a corresponding particle in the GADGET-3 run is located as the nearest particle in position in the initial condition of GADGET-3. The ``matched'' subhalos are selected if a subhalo at the same step ($\rm{z}\sim 0.3$) possesses more than half of the corresponding particles. Conversely, by performing the same procedure in reverse, another link is established — that is, the 40 most bound particles in the GADGET-3 run are initially identified, followed by locating their counterpart particles in the ENZO run. A pair of two halos that are bijectively mapped (bidirectionally connected) between the two simulations are considered as a ``matched" pair. We expand the matching process to identify matching subhalos across all participating codes.

In addition, given the manageable number of subhalos compared between codes along this paper, we visually verify the consistency of their matching by checking their \textcolor{black}{position in the 6D phase space} during infall and comparing the host's evolutionary stage when the subhalos cross the host’s virial radius. This ensures that we are comparing the same subhalos across different codes.

\section{Results} \label{sec:results}
\subsection{Evolution of satellite population}
\label{sec:sat_evolution}
\begin{figure}
    \centering
    \includegraphics[width=0.99\linewidth]{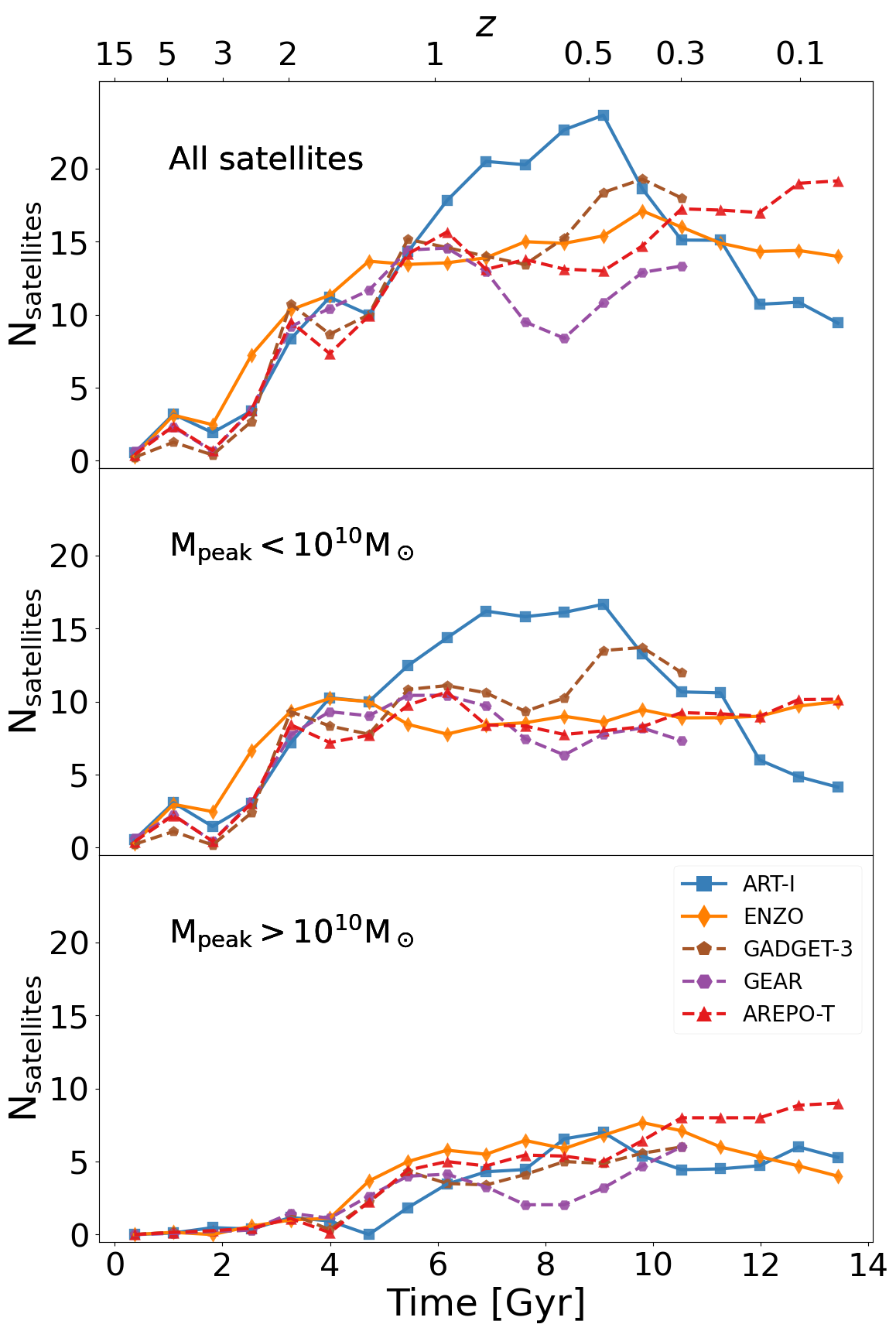}
    \caption{Evolution of the satellites' number count across cosmic time for all \texttt{CosmoRun} models analyzed in this paper. We count all satellite galaxies located within  the host virial radius and the `splashback' galaxies. Markers and lines represent the average number of satellites for each time bin. \textbf{Top panel:} Evolution of the total number of satellite galaxies for all models. \textbf{Central and bottom panels:} Evolution of the number of satellite galaxies with $\rm{M_{peak}<10^{10}\,\rm{M_\odot}}$ and $\rm{M_{peak}>10^{10}\,\rm{M_\odot}}$, respectively.}
    \label{sat_evolution}
\end{figure}
The different choices of stellar feedback 
significantly influence whether the subhalos population will host star-forming gas and how much of this will be available during its evolution, determining the amount of stellar mass that our subhalos can produce. Thus, the number of satellites and their evolution over cosmic time can provide insights into the effects of the various feedback recipes used. This evolution for each \texttt{CosmoRun} model is shown in Figure \ref{sat_evolution}. We consider as satellite galaxies both galaxies contained inside the host's virial radius, and the ones that are temporally outside due to their orbit but that were previously inside - so called “backsplash” galaxies \citep{Diemer_2021_splashback}. For the convenience of the reader, we continue to refer to the different \texttt{CosmoRun} models by the code name rather than specifying each time that it is the model with the specific feedback implementation from the \texttt{CosmoRun} simulations. This implies that other simulation groups using an AGORA code but with a different feedback implementation should be cautious when comparing their results.

In the top panel of Figure \ref{sat_evolution}, we present the evolution for the total number of satellite galaxies. The population of satellites increases until $\rm{z}\sim1$ across all models, after which it remains roughly constant, \textcolor{black}{as the host halo was  selected for its quiet merger history}. Overall, there is good agreement between the models on the evolution of the number of satellites, with the exception of ART-I (blue line). ART-I shows a slightly higher number count of satellites at $\rm{z}\sim1 - 0.5$ followed by a decline, ultimately converging with the other models \textcolor{black}{and eventually even exhibiting a lower number of subhalos. This sharp decline in ART-I is primarily due to a significant fraction of subhalos being disrupted during the last major merger around $\rm{z}\sim0.5$}.  In the central and bottom panels we show the result of splitting the satellites population in two different mass ranges (low versus high mass), using the peak halo mass\footnote{We define peak halo mass as the maximum DM halo mass achieved by a halo during its evolution.}. The number of satellites with $\rm{M_{peak}}>10^{10}\,\rm{M}_\odot$ is roughly consistent across all models. This is in line with our expectations, as in this mass range \textcolor{black}{all halos host galaxies} regardless of the feedback recipe.  \textcolor{black}{Consequently, the number of satellites reflects the number of subhalos in this mass range, which is quite similar between codes}. In \citetalias{Kim_2014_paper1}, we showed that all codes generate similar DM structures no matter the code. This is still true for the \texttt{CosmoRun} as shown in \citetalias{santi_paper3_agora} and \citetalias{minyong_paper5}. Nonetheless, the small differences observed in this mass range arise from variations in their accretion times (\textcolor{black}{for a detailed discussion about intercode timing discrepancies, see Appendix C} in \citetalias{santi_paper_4_agora}), and on their disruption times which may change due to orbital variations and/or slightly different density profiles. Regarding the satellite number counts with $\rm{M_{peak}}<10^{10}\,\rm{M}_\odot$, there is good agreement between models, with some scatter since $\rm{z} \sim 1$, when ART-I and GADGET-3 show higher counts than GEAR,  AREPO-T and ENZO  \textcolor{black}{until the sharp decline of ART-I at $\rm{z} \sim 0.5$. Interestingly, we found that ART-I and GADGET, due to their SN feedback implementations, exhibit the highest efficiency in star-formation in low-mass halos, which may explain the higher number of subhalos in these models between $\rm{z} \sim 1$ and $\rm{z} \sim 0.3$.}

\begin{figure}
    \centering
    \includegraphics[width=0.99\linewidth]{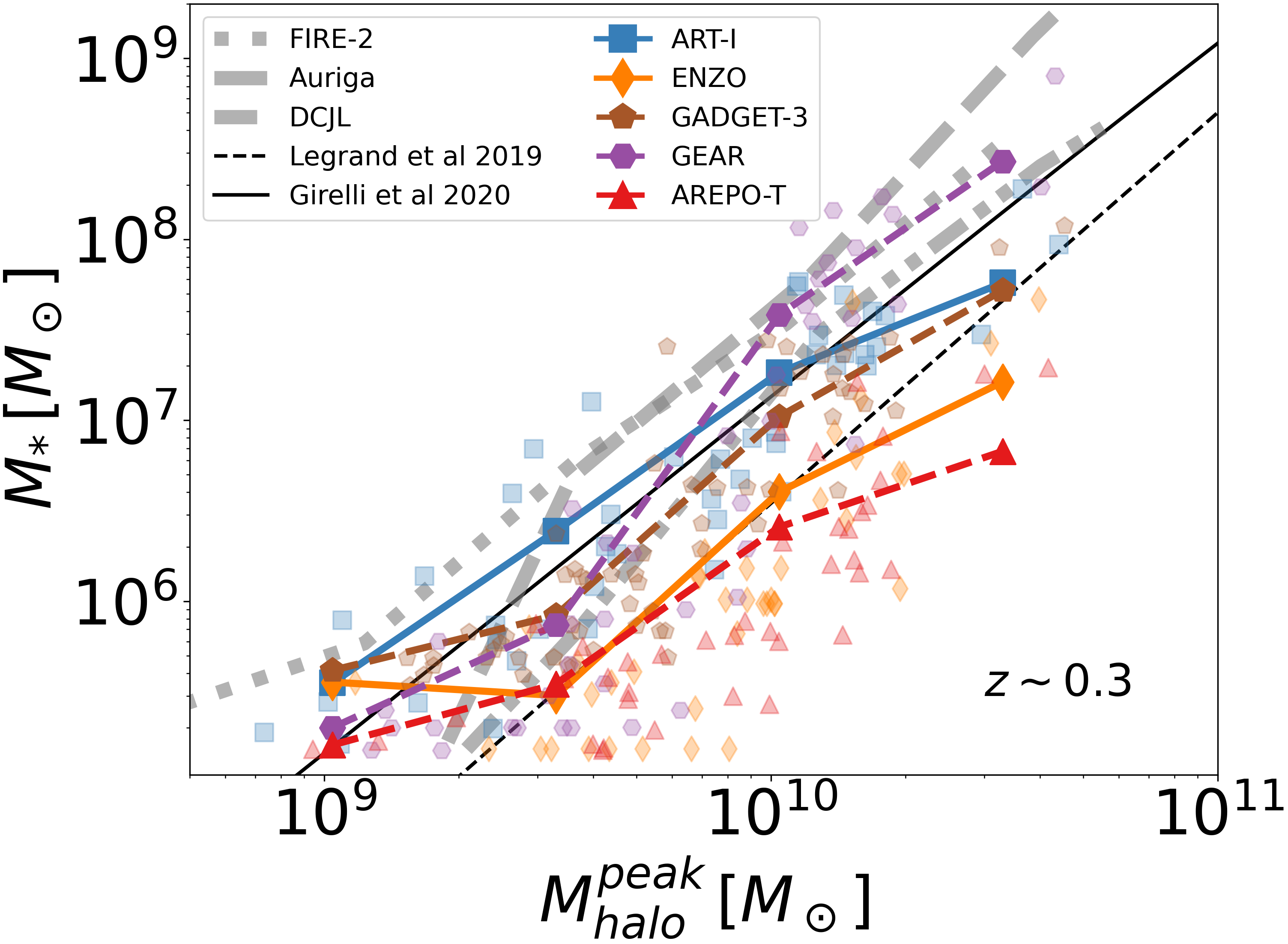}
    \caption{Stellar-to-halo mass relation at $\rm{z}\sim 0.3$ for the five \texttt{CosmoRun} models reaching $\rm{z}<1$. Each galaxy is plotted as a single marker. Solid and dashed colored lines represent the mean value of stellar masses in each peak halo mass bin for each model. The gray dotted, dashed and dot-dashed lines are for dwarf galaxies in other cosmological zoom-in simulations at $\rm{z}=0$ (FIRE-2, Auriga and DC Justice League, respectively; \citep{Hopkins_2018_overquenching, Grand_2021, Munshi_2021}). The solid and dashed black lines without markers are semi-empirical models for $0.2 < \rm{z} < 0.5$ with extrapolation to low-mass galaxies \citep{Legrand_2019, Girelli_2020}.}
    \label{fig:stellar-halo}
\end{figure}

Figure \ref{fig:stellar-halo} shows the stellar-to-halo mass relation (SHMR) at $\rm{z}\sim 0.3$ for the five \texttt{CosmoRun} models analyzed through this paper. One may notice that halos in some models produce considerably higher stellar mass than in others, highlighting the impact of the different feedback models and codes in the stellar mass produced. These differences are particularly pronounced for halos with $\rm{M}_\mathrm{peak} > 10^{10} \, \rm{M_\odot}$, where the stellar mass in GEAR halos exceeds that of AREPO-T halos by around 2 dex.

While the \texttt{CosmoRun} simulations are calibrated to yield the same stellar mass at $\rm{z} = 4$ for the main galaxy, and this convergence persists down to lower redshifts (see Figure 4 of \citetalias{santi_paper_4_agora}), the scatter among different models is greater for less massive galaxies. This can be understood as a consequence of the different supernova (SN) feedback recipes, which have a more significant impact on low-mass halos due to their shallower gravitational potential wells. In these low-mass halos, supernova winds can expel gas more effectively, whereas for halos with \textcolor{black}{total} masses above a few times $10^{11}\,\rm{M}_\odot$, SN feedback becomes not efficient \citep{Dekel_1986_SN,dekel2019_goldenmass}, so differences due to different SN feedback implementations are expected to be lower.

By comparing our SHMR at $\rm{z}=0.3$, \textcolor{black}{shown in Figure \ref{fig:stellar-halo}}, with the one at $\rm{z} = 2$ presented in Figure 9 of \citetalias{minyong_paper5}, we can observe that the differences between models are greater at $\rm{z}=0.3$. Additionally, in general, we find lower stellar masses at $\rm{z}=0.3$ for the same halo mass at $\rm{z}=2$. This can be understood as a reflection of the peak in stellar efficiency occurring at $\rm{z}=2$. In contrast, at lower redshifts, star formation becomes less efficient while the halo continues to grow in mass, leading to lower stellar masses for the same halo masses. Overall, the trends identified in \citetalias{minyong_paper5} remain consistent. Attending to halos with peak halo masses greater than $10^{10}\,\rm{M_\odot}$, GEAR is the model forming the highest stellar masses, followed by ART-I and GADGET-3, while ENZO and AREPO-T form significantly fewer stars for the same halos. These differences on the stellar mass of our satellites \textcolor{black}{are} visually evident in Figure \ref{stellar_surface_density}, where we presented the \textcolor{black}{stellar surface density} of the satellite population (along with some nearby field galaxies) at $\rm{z}\sim 0.3$.

In Figure \ref{fig:stellar-halo}, we also compare the SHMR of our models with previous results of other cosmological zoom-in simulations. The thick gray dotted, dashed and dot-dashed represent the SHMR for dwarf galaxies at $\rm{z}=0$ in the FIRE-2, Auriga and DC Justice League simulations, respectively; \citep{Hopkins_2018_overquenching, Grand_2021, Munshi_2021}. The dashed and solid black lines without markers represent the predictions extracted from \textcolor{black}{semi-analytical} models at $0.2 < \rm{z} < 0.5$ with extrapolation to dwarf galaxies \citep{Legrand_2019, Girelli_2020}. Overall, our models fit, within their scatter, the predictions of the \textcolor{black}{semi-analytical} models. On the other hand, in general, the SHMR of our models is slightly lower than that obtained in other simulations at $\rm{z}=0$. However, it is worth noting that the other simulations use the halo mass at $\rm{z}=0$ instead of the peak halo mass. For field galaxies, the peak halo mass usually coincides with the halo mass at $\rm{z}=0$, but for satellite galaxies, tidal stripping removes DM from the halo while the stellar mass often remains intact. Therefore, by including satellites in their sample and using the halo mass at $\rm{z}=0$, they tend to have higher stellar masses for the same halo mass.

\subsection{Satellite quenched fraction across host halo evolution}
\label{sec: fq_evolution}

The evolution of the host halo’s CGM can have a strong impact on the properties of satellites when they enter the CGM region. In particular, it is expected that the quenched satellites fraction (hereafter $\rm{f_{q,\,sat}}$) increases when a warm/hot coronna of gas is present around the host halo.

\begin{figure*}
    \centering
    \includegraphics[width=\linewidth]{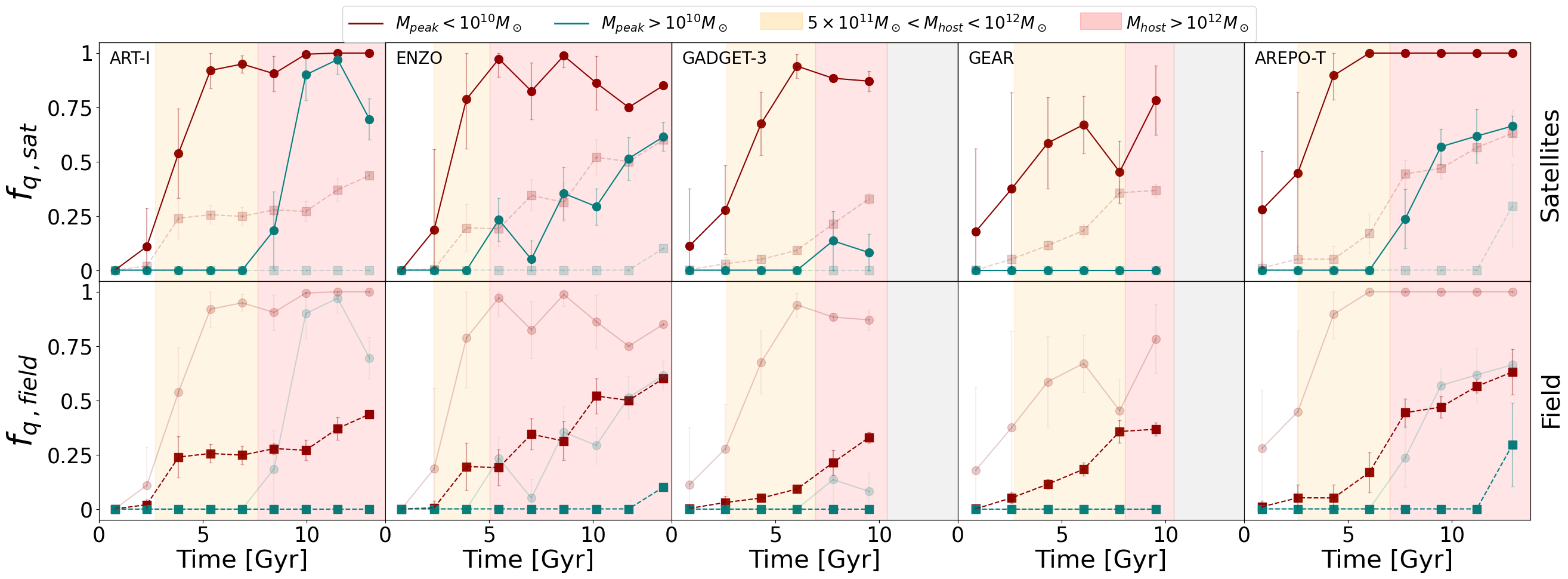}
    \caption{Evolution of the fraction of quenched satellite galaxies (\textbf{top} and circle markers) across cosmic time compared with the quenched fraction of field galaxies (\textbf{bottom} and square markers) for each model. \textcolor{black}{Markers indicate the mean $\rm{f_q}$ for each time bin, and the error bars represent the standard deviation.} In the row where we show the quenched fraction of the satellites, we also display that of the field galaxies with higher transparency and vice versa to facilitate comparison. We follow the quenching definition described in Section \ref{Quiescence_definition}. The orange and red shaded regions represent the epochs when the host halo is more massive than $5\times10^{11}\,\rm{M}_\odot$ and $10^{12}\, \rm{M}_\odot$, respectively. The time domain where snapshots are still not available for each specific code are indicated as a gray shaded region. Quenched fractions for satellite galaxies and field galaxies above and below $\rm{M}_{peak} = 10^{10}\rm{M}_\odot$ are shown in different colors. }
    \label{fq_evolution}
\end{figure*}

In Figure \ref{fq_evolution}, top row, we show the evolution of $\rm{f_{q,\,sat}}$ for all the models; whereas in the bottom row, we show the evolution for field\footnote{
Similar to the approach followed in \citetalias{minyong_paper5}, we define field halos using the following criteria: \textit{(i)} A field halo must reside beyond $2\rm{R_{vir}^{host}}$ of our target host halo,\textit{(ii)}  It must be more massive than $10^7 \rm{h^{-1} M_{\odot}}$ in DM, and \textit{(iii)}  It must not be a subhalo of another halo (i.e., satellites of other halos are excluded). After assigning stellar particles to these halos using the method described in Section \ref{Star_assigment}, we consider only the field galaxies whose stellar masses are heavier than $6\rm{m}_{\text{gas},\text{IC}} = 3.39 \times 10^5 \, \rm{M}_{\odot}$
}
galaxies well outside the host influence zone ($\rm{f_{q,\,field}}$). By comparing the different quenched fractions in both satellite and field galaxies,  we identify how the evolution of the host CGM influences satellite properties in contrast to dwarf galaxies population not affected by the central galaxy. To better visualize the effects of the different CGM states on our satellite quenching, red and orange shaded regions in both panels are highlighting the epochs when the host halo is more massive than $5\times10^{11}\,\rm{M}_\odot$ and $10^{12}\, \rm{M}_\odot$, close to the mass when it is expected from theory that the central galaxy \textcolor{black}{develops} a warm-hot CGM \citep{keres_2005} (see Figures 6 and 9 in \citetalias{Strawn_2024}). The mass of the satellite galaxies is also a variable that needs to be accounted for when studying satellites quenching \citep{Fillingham_2015}. The gravitational pull in more massive satellites can retain gas mass against the stripping processes in a more effective way than less massive ones. To further analyze this effect, in this figure, we divide our satellite population sample into two mass bins, one with subhalos with a peak halo mass above  $10^{10}\, \rm{M}_\odot$ and another below. We use the peak halo mass to ensure that we are comparing the quenching of the same subhalos across different models, as stellar mass would differ depending on the model, as shown in Section \ref{sec:sat_evolution}. From Figure \ref{fq_evolution}, we can conclude the following:
\begin{enumerate}
    \item Low-mass subhalos undergo quenching earlier than the high-mass subhalos across all models. The $\rm{f_{q,\,sat}}$ for them is higher than that for high-mass subhalos throughout all the evolution. This suggests that the quenching mechanism is both more effective and rapid for low-mass subhalos.

    \item All models agree that $\rm{f_{q,\,sat}}$ for low-mass subhalos is consistently higher than $\rm{f_{q,\,field}}$ for low-mass satellite galaxies, which is \textcolor{black}{especially remarkable} when the host halo mass surpasses $5\times10^{11}\,\rm{M}_\odot$. 
    
    \item In all models, the high-mass satellite galaxies only quench when the host halo exceeds $10^{12}\,\rm{M}_\odot$ (increasing $\rm{f_{q,\,sat}}$). Meanwhile, the field galaxies with the same mass remain unquenched ($\rm{f_{q,\,field} = 0}$), except for the last time bin in AREPO-T.

    \item The value of $\rm{f_{q,\,sat}}$ differs among models, particularly for high-mass subhalos. While ART-I and AREPO-T achieve $\rm{f_{q,\,sat}}$ above 80\% and 60\%, respectively, for high-mass subhalos at $\rm{z} \sim 0.3$, GADGET-3  barely exceeds 15\% and GEAR is not able to quench any subhalo above $\rm{M}_{peak} = 10^{10}\, \rm{M}_\odot$. Although the models show better agreement for $\rm{f_{q,\,sat}}$ in low-mass subhalos, GEAR also displays slightly lower $\rm{f_{q,\,sat}}$ compared to the other models. \textcolor{black}{The causes behind these intercode differences, in relation to the different quenching mechanisms, are analyzed in Sec \ref{sec:dominance_mechanisms}}
\end{enumerate}

As previously discussed, the suppression of cold inflows is a predicted outcome for halos with masses exceeding a few times $10^{11}\,\rm{M}_\odot$. In this context, the more massive the host halo is, the hotter its CGM becomes. The infalling satellites will encounter this hostile medium, so they will experience stronger ram pressure when entering halos with larger masses. The tidal stripping will also be naturally larger in halos with larger masses (i.e., stronger gravitational potential gradients). Due to the deepening of the gravitational potential, \textcolor{black}{statistically}, the number of subhalos and their velocities also grow, increasing the possibility of high speed satellite-satellite \textcolor{black}{encounters}. Our results regarding the increase of $\rm{f_{q,\,sat}}$ when approaching $10^{12}\,\rm{M}_\odot$ are thus in good agreement with these expectations, as all quenching mechanisms are more efficient when the host halo is more massive.

\begin{figure}
    \centering
    \begin{subfigure}[b]{0.99\linewidth}
        \includegraphics[width=\linewidth]{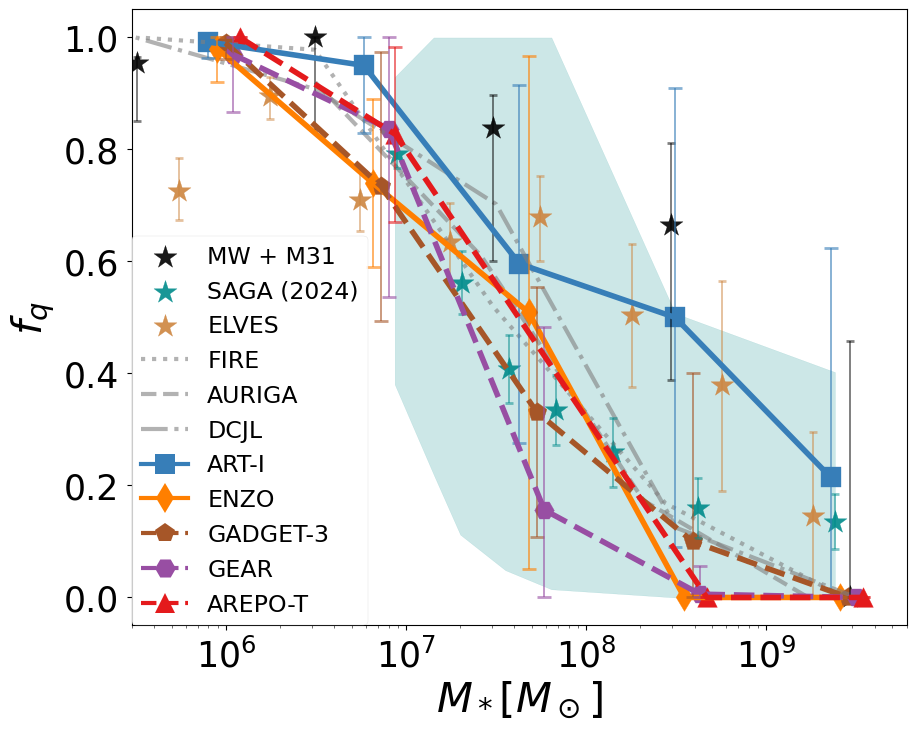}
    \end{subfigure}
    \caption{Quenched fractions of satellite galaxies of the \texttt{CosmoRun} target galaxy as a function of stellar mass for each \texttt{CosmoRun} model. Each model is represented by a colored line. In order to match with observations, here a satellite galaxy is defined as quenched if $\rm{sSFR}  < 10^{-11} \rm{yr}^{-1}$. Due to a lack of host statistics, we used \textcolor{black}{a common number of 50} snapshots from $\rm{z}<1$, when our host halo is above the critical mass for virial shock formation in all models, up to $\rm{z}\sim 0.3$, the last available snapshot for GADGET-3 and GEAR, to ensure a consistent comparison between models. \textcolor{black}{Each marker represents the mean value across different snapshots for each mass bin, with error bars indicating the standard deviation}. Markers have been slightly displaced for clarity, but they represent the same stellar mass bin. We compare our quenched fractions with observations of the LG (from \citealt{Wetzel_2015b} updated with \citealt{Putman_2021} as in \citealt{Samuel_fire_2022}), as well as data from the \textcolor{black}{"Gold" and "Silver" sample of the} SAGA Survey \citep{Geha_2024} and findings from the ELVES survey \citep{Carlsten_2022}. The cyan shaded region represents the SAGA host-to-host 1$\sigma$ scatter. Additionally, quenched fractions from other cosmological simulations of Milky Way analogs are shown with gray dotted, dashed, and dash-dotted lines: FIRE \citep{Samuel_fire_2022}, AURIGA \citep{Simpson_2018_ram_pressure_MW}, and the ChaNGa DC Justice League (DCJL) \citep{Akins_2021}, respectively. We note that both the observational data and the simulations from other groups we compare against are at $\rm{z}=0$. Since the range used in our case to calculate the mean quenched fraction for each mass bin is between $\rm{z}=1$ and $\rm{z}\sim0.3$, our $\rm{f_q}$ can be interpreted as a lower limit for the expected quenched fraction at $\rm{z}=0$.}
    \label{fq_LG}
\end{figure}

Figure \ref{fq_LG} displays a comparison of $\rm{f_{q,\,sat}}$ for each model with observations of LG (\citealt{Wetzel_2015b} and \citealt{Putman_2021}) and MW-analogs from the SAGA \citep{Geha_2024} and ELVES \citep{Carlsten_2022} surveys, as well as with other cosmological simulations of MW-mass halos \citep{Simpson_2018_ram_pressure_MW,Akins_2021,Samuel_fire_2022}. \textcolor{black}{To facilitate comparison with Figure \ref{fq_evolution}, we note that $\rm{f_{q,\,sat}}$ was computed in stellar mass bins instead of halo mass bins. Since the SHMR varies between models, halos with $\rm{M_{halo}} \sim 10^{10}\,\rm{M_\odot}$ correspond now to different stellar mass ranges depending on the model. For instance, in AREPO-T, these halos typically have stellar masses of a few times $10^{6}\,\rm{M_\odot}$, whereas in GEAR, they characteristically reach around $10^{8}\,\rm{M_\odot}$.}
Consistent with the findings in Figure \ref{fq_evolution}, there is generally good agreement among models regarding the quenched fraction of low stellar mass satellites with $\rm{M}_* < 10^7\,\rm{M}_\odot$, that is close to unity. However, the quenched fraction between models shows larger variation in the stellar mass range of $\sim 10^7-10^8\,\rm{M}_\odot$, where the quenched fraction for GEAR is below 20\%, whereas for ART-I is close to 60\%. These results suggest a higher efficiency on the satellite quenching in this mass range in ART-I compared to GEAR, similar to the trends observed in Figure \ref{fq_evolution}. For more massive satellites, with stellar masses above $10^8\, \rm{M}_\odot$, nearly all remain star-forming across all models, except for ART-I, which exhibits a higher quenched fraction.

Before comparing with observations and other cosmological simulations from different groups, it is important to clarify how we calculate the quenched fractions. Due to the lack of host statistics, we use \textcolor{black}{a common number of 50} snapshots starting from $\rm{z} < 1$, when our host halo exceeds the critical mass for virial shock formation in all models, up to $\rm{z} \sim 0.3$, the last available snapshot for GADGET-3 and GEAR, to ensure a consistent comparison between models. However, both the observations and the simulations we compare against are at $\rm{z}=0$. As shown in Figure \ref{fq_evolution}, $\rm{f_q}$ is expected to increase as redshift decreases and the host halo evolves, as satellites interact with the warm corona gas of the host's CGM for a longer period. Therefore, the $\rm{f_q}$ should be interpreted as a lower limit compared to the $\rm{f_q}$ that would be computed at $\rm{z}=0$, if host statistics were available, where a higher quenched fraction is expected. Moreover, in relation to the observational data from the SAGA and ELVES surveys, it is important to note that their satellite samples may be biased by interlopers, that is, field galaxies incorrectly identified as satellites due to projection effects. This could lead to an artificially lower quenched fraction, as these field galaxies would likely be star-forming.

Despite the differences in $\rm{f_q}$ among our models, when compared to those $\rm{f_q}$ observed in Figure \ref{fq_LG}, all our models are consistent with the latest SAGA data within its 1$\sigma$ host-to-host scatter (cyan diamonds and cyan shaded region). For $\rm{M}_*< 10^{7}\rm{M}_\odot$ quenched fractions for all the models are close to 100\%, consistent with the ones observed in the LG and in the ELVES survey. For satellites with stellar masses in the range $10^7 - 10^8 \,\rm{M}_\odot$, our models generally show quenched fractions more in line with the SAGA Survey than with the LG and ELVES, which observe slightly higher $\rm{f_q}$. However, the $\rm{f_q}$ from our models in this mass range is also consistent with that observed in the LG and ELVES within the scatter, especially when considering that our $\rm{f_q}$ is expected to be higher at $\rm{z}=0$.  For satellites with stellar masses above $10^{8}\rm{M}_\odot$, the ART-I model is the only one that matches the quenched fractions observed in the LG and ELVES survey, whereas the other show almost no quenching of satellites above this stellar mass, which is consistent with SAGA data and previous findings \citep{Fillingham_2016, Akins_2021}. All our models are also in agreement with results from other cosmological simulations, considering the large uncertainties due to the lack of host statistics. In general, our models predict a lower $\rm{f_q}$, especially for the mass range between $10^7$ and $10^8,\rm{M}_\odot$. This may be primarily due to the fact that the quenched fractions in our models should be understood as lower limits compared to the expected quenched fractions at $\rm{z=0}$, as mentioned above.

These findings illustrate how varying feedback implementations and code architecture for the same host and satellites can result in different quenched fractions. Consequently, some models align with the quenched fractions observed in the LG, while others fall within the lower quenched fraction scatter of the SAGA Survey. See Section \ref{sec:dominance_mechanisms} for a more detailed analysis of the causes of the intercode differences.

\subsection{Satellite quenching timescales}
\label{sec:when}
\begin{figure*}
    \centering
    \begin{subfigure}[b]{0.99\linewidth}
        \includegraphics[width=\linewidth]{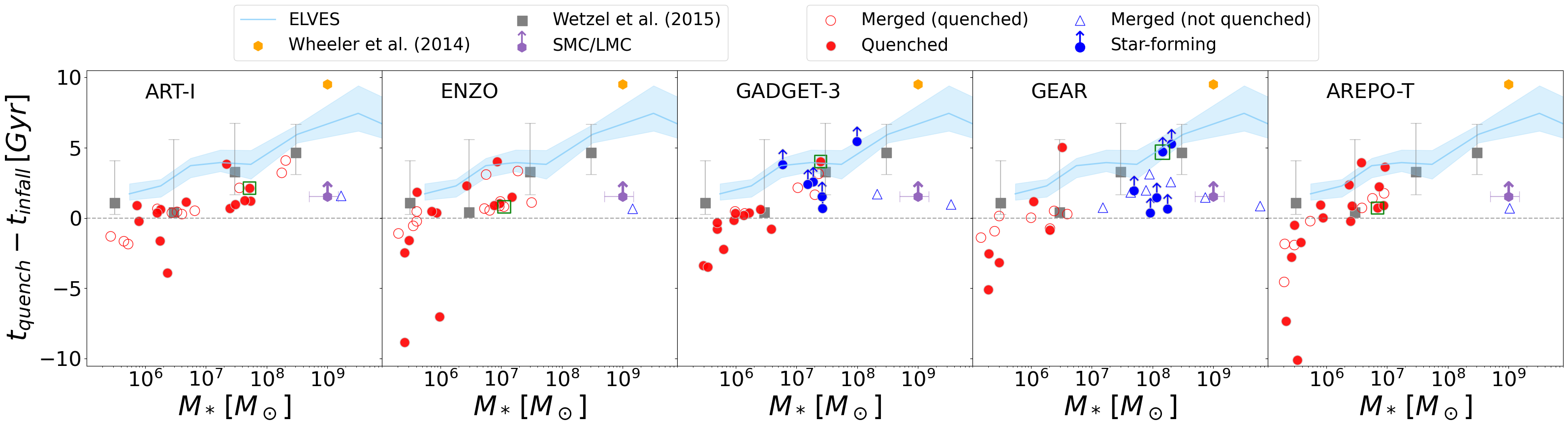}
    \end{subfigure}
    \caption{Quenching delay time of satellite galaxies as a function of their stellar mass. Quenched satellites are represented by red filled circles, while star-forming satellites are indicated by blue arrows, marking a lower limit on their quenching delay times. Satellites that merged when $\rm{M}_{host} > 5\times10^{11}\,\rm{M}_\odot$ are also shown: red open circles denote those quenched before merging, and blue open triangles denote those that were not (they still have either ongoing star formation or star-forming gas reservoirs at the time of merger). Observational estimates from \cite{Wetzel_2015b}, and \cite{Wheeler_2014} are shown in gray squares and orange hexagons, respectively. An estimate of the quenching timescale for the SMC/LMC system is shown in purple, using the infall time predicted by \cite{Kallivayalil_2013}. The quenching time for the ELVES survey \citep{Greene_2023} is also shown in blue. The open green square enclosing a single data point in each panel indicates the same satellite across the different models and we compare its evolution in Section \ref{sec:case_study} and in Figures \ref{fig:gas_density_comparison} and \ref{fig:mechanism_case_comparison}.}
    \label{quench_timescales}
\end{figure*}

To investigate the quenching timescales of satellites, we defined the quenching delay time as $\rm{t}_{\text{quench}} - \rm{t}_{\text{infall}}$, which measures the time it takes for a satellite to quench relative to infall into the host halo.
In Figure \ref{quench_timescales}, we show the quenching delay time versus satellite stellar mass for the different models. We include all surviving satellites in the lowest-z available snapshot and those that merged with the host when $\rm{M}_{host} > 5\times10^{11}\,\rm{M}_\odot$. We distinguish between quenched and star-forming satellites (red versus blue, respectively) and between those that quenched before merging and those that did not (red open circles versus blue open triangles, respectively). Star-forming satellites are represented by arrows as their lookback infall time can be interpreted as a lower
limit on the potential quenching timescale.

All the models follow the same trend identified in observations: the less massive the satellite galaxy is, the faster its quenching. Satellites with stellar masses above $10^7\,\rm{M}_\odot$ are resistant to rapid environmental quenching for all the models, whereas satellites below $10^7\,\rm{M}_\odot$ are compatible with fast and efficient quenching showing quenching delay times below $\sim 2\,\text{Gyr}$. These trends are generally consistent with the observational estimations for MW satellites \citep{Wetzel_2015b}, \textcolor{black}{for satellites in MW analogs \citep{Greene_2023_elves_quenchingtimes} and for satellites in more massive groups \citep{Wheeler_2014}}; as well as with results from cosmological simulations such as \cite{Akins_2021} and \cite{Samuel_fire_2022}. \textcolor{black}{Infall times in observations are estimated by using the most likely infall time from cosmological simulations for a surviving satellite of a specific stellar mass. A relevant caveat when comparing our results with observational estimates is that the latter assume all satellites were star-forming prior to their first infall. This assumption may not be true, for example, if the satellites were affected by reionization or pre-processing.} 

It is worth noting  the spread that exists in  the quenching delay time at a given stellar mass, even for the same model. For example, for AREPO-T the quenching time scales for satellites with $\rm{M}_* \sim 10^6 - 10^7\,\rm{M}_\odot$ spread from 0 to 4 Gyr. These quenching timescales are generally within the uncertainty range provided by \citep{Wetzel_2015b}. This scatter reflects, in part, that stellar mass by itself does not completely determine how long a satellite can retain its cold gas against quenching processes. Other factors, such as the eccentricity of the satellite's orbit, the initial gas and DM mass, and the different gas and mass concentration, also play significant roles. This is explored in more detail in Section \ref{sec:how?}, where we also study the different quenching timescales among models by focussing on the physical mechanisms behind quenching. 

Some discrepancies between models can be highlighted. The ART-I model is the only one that quenches satellites with masses above $10^8\,\rm{M}_\odot$ (as shown in Figure \ref{fq_LG}), while for the rest of the models all galaxies with masses above this limit remain star-forming, similar to the findings in \citealt{Akins_2021}. While satellites above $10^8\,\rm{M}_\odot$ in ART-I  quench in timescales around $\sim 3 - 4\,\text{Gyr}$, in  GEAR they remain star-forming even after $\sim 5\,$Gyr of evolution, indicating discrepancies in the efficiency of quenching high-mass satellites, as shown in Figure \ref{fq_evolution}. GEAR and GADGET-3 have a higher abundance of star-forming galaxies, pointing to higher quenching delay times than the other models. In addition, GEAR also has a higher number of satellites that were disrupted prior to quenching, indicating differences in satellite-host interactions relative to the other models. Another noteworthy detail is that the only satellite with $\rm{M}_*>10^9\,\rm{M}_\odot$ in all models merges within very short timescales ($1$ Gyr or less) while still star-forming, due to its more radial trajectory. In Section \ref{sec:dominance_mechanisms} we performed a more detailed analysis of the possible sources of these intercode differences.

Many of the lowest-mass satellites, below $\sim 10^6\,\rm{M}_\odot$, undergo quenching before infall. Here, we summarize the many potential mechanisms for the early quenching. Although we have studied some of them (e.g., the quenching time versus reionization), it is not within our scope to provide the reader with a final answer, as this would require a much deeper analysis than the one presented in this paper. Physical processes unrelated to the host halo can lead to the loss of cold gas, resulting in the quenching of star formation. Such low-mass halos are particularly vulnerable to these processes, \textcolor{black}{such} as cosmic reionization \citep{Brown_2014}. However, since not all of our low-mass satellites that are early quenched do so during the reionization epoch, alternative scenarios need to be considered. One possibility is that reionization suppresses gas accretion, and any remaining gas will either be expelled by stellar feedback \citep{Benitez-llambay_2015} or consumed in star formation through self-shielding \citep{katz_2020}. Moreover, heating from the UV background (initially mentioned by \citealt{Bullock_2000_Reionization}), which peaked around $\rm{z} \sim 2$ in our model \citep{Haardt_2012}, may also contribute to the quenching of low-mass galaxies. Environmental quenching outside the host halo is another potential mechanism. Ram pressure stripping due to the host halo's gas can be effective up to distances of approximately $4\rm{R_{vir}}$ \citep{Cen_2014}, while pre-processing within low-mass groups has been noted as a significant factor in the quenching of MW satellites \citep{Wetzel_2015a, Samuel_fire_2022}. \textcolor{black}{Nevertheless, in addition to these physical processes, numerical overquenching might also be a factor, particularly in galaxies close to the resolution limit \citep{Hopkins_2018_overquenching}, as satellites with stellar masses below $10^6,{\rm{M}_\odot}$ in \texttt{CosmoRun} simulations may experience artificial suppression of star formation due to limited resolution (see Section \ref{sec:caveats}).}

\subsection{Investigating how satellites quench}
\label{sec:how?}

Timescales shown in Figure \ref{quench_timescales} suggest, for all the \texttt{CosmoRun} models, the presence of an environmental quenching mechanism that quenches rapidly and effectively low-mass satellites  and quenches in a less efficient way, with timescales around their crossing time, intermediate-mass satellites. For the more massive satellites, this environmental quenching mechanism is not so effective leading to large quenching delay times. In this section, we analyze the contribution of several quenching mechanisms of satellite galaxies often proposed in literature. First, we examine in detail the quenching of the same individual satellite across all our models and present the methodology that will be used for determining the contribution and interaction of different quenching mechanisms. Then, we apply this methodology to all satellites within each model and analyze the statistical results. 

Before beginning the analysis, we describe the approach followed to compute the parameters used to characterize each quenching mechanism.
\subsubsection{Quenching mechanisms}
\paragraph{\textbullet\hspace{0.5em}Strangulation. }
To assess whether there was a cutoff of cold inflows upon entering the virial-shocked halo of the main galaxy, we computed the mass ratio of inflows penetrating the satellite galaxy, by determining the amount of cold gas accreted by the satellite galaxy at each snapshot. This step is straightforward for particle-based codes, as we have the IDs of the gas particles, so we can determine the number of particles accreted by comparing the IDs at different snapshots. However, the determination of inflows become more complex for grid-based codes, where we cannot track individual particles. 

Therefore, we calculated the cool/cold gas accreted by the satellite galaxy using gas cells. We considered as inflows all cells that fulfill all the following conditions: i) they reside between \(0.2 \, \rm{R_{vir}^{sub}}\) and \(0.3 \, \rm{R_{vir}^{sub}}\), ii) they move with a sufficiently negative radial velocity to cross the shell between \(0.2 \, \rm{R_{vir}^{sub}}\) and \(0.3 \, \rm{R_{vir}^{sub}}\) within 100 Myr, and iii) with lower temperature than \(10^{4.5}\) K and density $\rm{n_H > 10^{-2.5}\,cm^{-3}}$. However, we must be careful not to introduce CGM gas that the satellite encounters during its infall as inflows. To address this, we corrected for the satellite's motion through the CGM in order to determine the gas streams that are actually falling toward the satellite with negative radial velocities, and not merely due to the subhalo's movement.  To ensure that our method reliably captures the inflow mass, we tested it on particle-based codes, using the \texttt{yt} build-in grid. We compared our results using this method with those obtained by tracking particle IDs to identify which particles were actually accreted, and we achieved good convergence. The comparison between the two methods and more details about how the inflows are computed can be found in Appendix \ref{appendix:inflows}.

\paragraph{\textbullet\hspace{0.5em}Ram pressure stripping. }
In order to compute the ram pressure felt by the gas and the restoring force exerted by the subhalo, we follow a similar apporach to that presented in \cite{Simpson_2018_ram_pressure_MW}. First, we use the classical formula of \cite{Gunn_Gott_1972} to determine the ram pressure: $\rm{P}_{\text{ram}} = \rho_{\text{CGM}}v_{\text{sat}}^2$, where $\rho_{\text{CGM}}$ is the density of the medium through which the satellite galaxy is moving and $\rm{v_{sat}}$ is the relative velocity of the satellite with respect to the surrounding gas. For calculating the ram pressure, we adopt as $\rm{v_{sat}}$ the velocity of the satellite with respect to the host. For $\rho_{\rm{CGM}}$ we compute an average radial gas density profile extending out to a radius of $4\rm{R_{vir}}$ for the host halo in each snapshot, as in Section \ref{gas_assigment}. As we mentioned above, this radially averaged estimate neglects local perturbations. However, it robustly captures the effect of radial infall that drives the main change in ram pressure, which can vary by orders of magnitude \citep{Simpson_2018_ram_pressure_MW}.
 
The restoring force per area on the satellite’s gas can be expressed as $\rm{P}_{\text{rest}} = \left|\frac{\partial \Phi}{\partial z_{\text{h}}} \right|_{\rm{max}}\Sigma_{\text{gas}}$, where $\Sigma_{\text{gas}}$ is the satellite’s gas surface density, $\rm{z}_{\text{h}}$ is the direction of motion (and gas displacement), $\Phi$ is the gravitational potential, and $\left|\frac{\partial \Phi}{\partial z_{\text{h}}} \right|_{\rm{max}}$ represents the maximum of the derivative of $\Phi$ along $\rm{z}_{\text{h}}$ \citep{Roediger_2005}. For approximating the restoring force for the whole satellite, we adopt a simple estimate for $\left|\frac{\partial \Phi}{\partial \rm{z}_{\text{h}}} \right|_{\rm{max}}$ and $\Sigma_{\text{gas}}$. The gas surface density is estimated from the radius enclosing half the gas mass ($\rm{r}_{\text{gas}}^{1/2}$), such that $\Sigma_{\text{gas}} = \frac{M_{\text{gas}}}{2\pi(\rm{r}_{\text{gas}}^{1/2})^2}$ (where $\rm{M}_{\text{gas}}$ is the total mass in gas). We estimate $\left|\frac{\partial \Phi}{\partial \rm{z}_{\text{h}}} \right|_{\rm{max}}\sim \frac{\rm{v}_{\text{max}}^2}{\rm{r}_{\text{max}}}$, where $\rm{v_{\text{max}}}$ is the maximum velocity of the spherically averaged subhalo rotation curve, and $\rm{r}_{\text{max}}$ is the radius where this peak occurs.
Therefore, we estimate that ram pressure stripping is effective when
\begin{equation}
    \rm{\rho_{\text{CGM}}v_{\text{sat}}^2 > \frac{v_{\text{max}}^2}{r_{\text{max}}}\frac{M_{\text{gas}}}{2\pi(r_{\text{gas}}^{1/2})^2}}
    \label{ram_pressure_condition}.
\end{equation}

Finally, we can compute the ram pressure radius, $\rm{r_{ram}}$. This radius defines the boundary beyond which all the gas should be stripped out due to ram pressure exceeding the restoring force.  For that, we use the same expression as in \cite{zhu2024itsbreezecircumgalacticmedium}. Specifically, $\rm{r_{ram}}$ is identified as the radius at which these forces are balanced:
\begin{equation}
    \rm{\rho_{\text{CGM}}(r){v_{\text{sat}}(r)}^2  = \alpha\frac{GM_{tot}(r)\rho_{gas}(r)}{r}}
    \label{ram_radius_condition}
\end{equation}

where $\alpha$ is a geometric factor of order unity from integration along the projection \citep{mccarthy_2018}.

\paragraph{\textbullet\hspace{0.5em}Tidal stripping. }
We computed the tidal radius ($\rm{r_{tidal}}$) of each satellite using the same approach followed in \cite{henriques_tidal_2010}, where they use the isothermal sphere approximation for the mass distribution of the central halo and satellite galaxy, and they assume that the satellite follows a circular orbit: 
\begin{equation}
    \rm{r_{tidal} \sim \frac{1}{\sqrt{2}}\frac{\sigma_{sat}}{\sigma_{host}}r_{sat}}
    \label{tidal_radius},
\end{equation}

where $\sigma_{\rm{sat}}$ and $\sigma_{\rm{host}}$ are the velocity dispersions for the satellite and the host halo, respectively, and $\rm{r_{sat}}$ is the radial distance of the satellite to the host. While this approach does not take into account the disk potential, which may contribute to tidal stripping \citep{Green_2021}, it allows us to estimate up to what radius the gas will resist being stripped by tidal forces.

\paragraph{\textbullet\hspace{0.5em}Harassment. }
To estimate the amount of energy produced in a high-velocity satellite-satellite encounter, we followed the approach described in \cite{Marasco_2016}. The amount of heat $E_s$ that an extended satellite of total mass $\rm{M}_s$ gains during an encounter with a point-like system of total mass $\rm{M}_p$ can be computed via the impulsive approximation as

\begin{equation}
    \rm{E_s \sim \frac{4}{3}G^2M_s\left(\frac{M_p}{v}\right)^2\frac{\left<r^2\right>}{b^4}}
    \label{harassment},
\end{equation}
where $\rm{v}$ is the relative velocity between the two objects, $\rm{b}$ is the impact parameter, and \( \langle \rm{r^2} \rangle \) is the mass-weighted mean square radius \(\left(\frac{\sum \rm{m_i r_i^2}}{\sum \rm{m_i}}\right)\) of the extended system (e.g., \cite{BinneyTremaine2008}, p. 660). We use this expression only as a proxy to detect when satellites undergoing a high-speed encounter. For each satellite galaxy and each snapshot, we compute the maximum $\rm{E_s}$ by considering all subhalos as possible encounters and substituting $\rm{v}$ and their actual distance for the impact parameter. Then, we compare $\rm{E_s}$ with the value of the total (kinetic + potential) internal energy of the satellite $\rm{E_{int}}$.

\subsubsection{Case study}
\label{sec:case_study}
In this section, we analyze a single satellite identified across all the models (see green squares in Figure \ref{quench_timescales}). We choose to show only one case to present the methodology that we use in the following sections to study the impact of the different quenching mechanisms considered. In Section \ref{sec:dominance_mechanisms}, we show the result of applying this method to all satellites in the simulations.

\begin{figure*}
    \centering
    \includegraphics[width=0.99\linewidth]{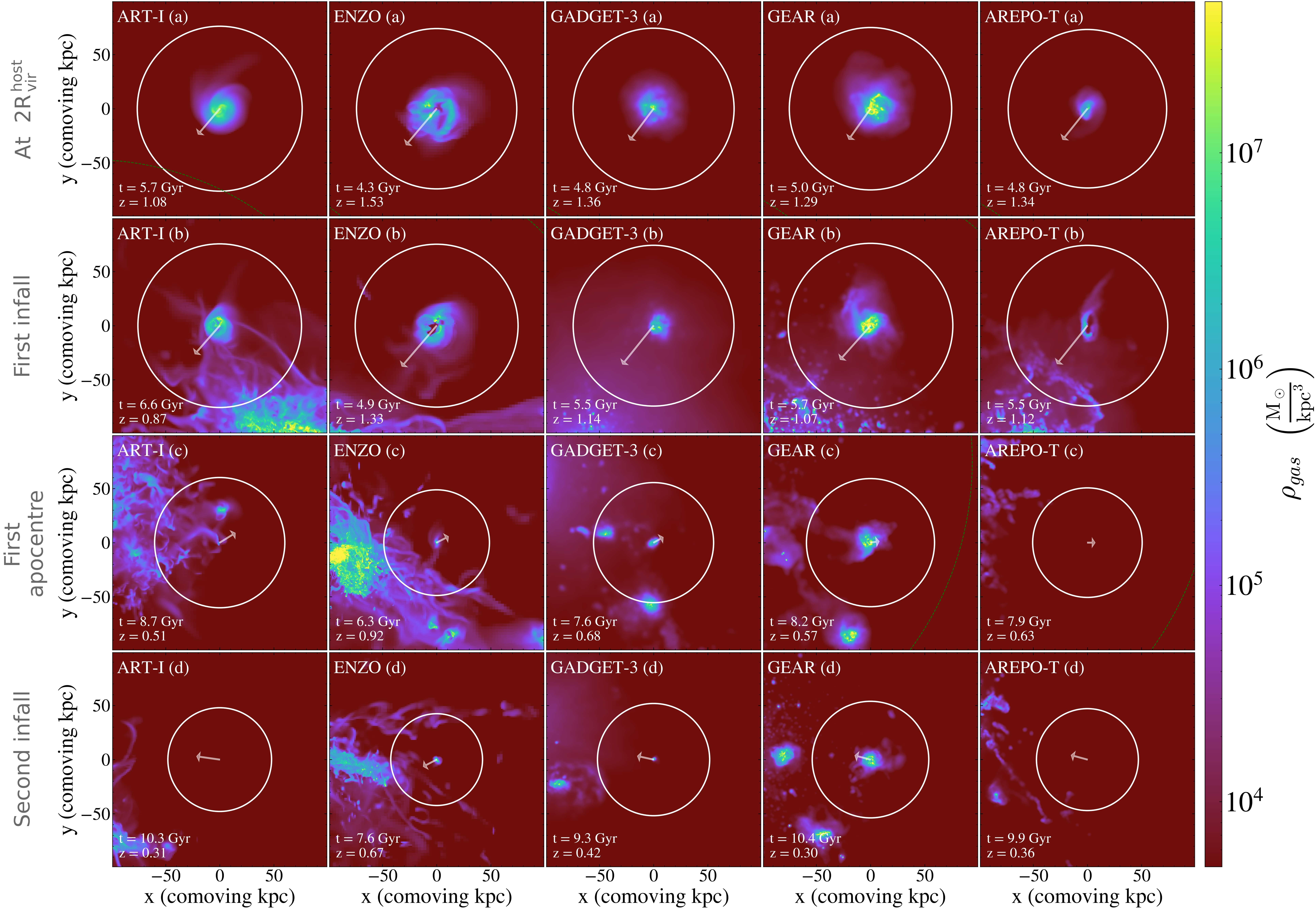}
    \caption{Gas density of the same subhalo across the different models during their infall to the host halo. \textcolor{black}{Gas density is weighted by itself to emphasize the dense regions.} Each column \textcolor{black}{represents} a different model, while each row indicates the same infall stage. These stages are highlighted in Figure \ref{fig:mechanism_case_comparison}: \textbf{(a)}  first snapshot at a distance to the host lower than $2\rm{R_{vir}^{host}}$, \textbf{(b)} at $\rm{t_{infall}}$, \textbf{(c)} during the first apocenter, and \textbf{(d)} during the second infall, slightly before the second pericenter. White solid circle show the subhalo virial radius, while the white arrow provides information about the subhalo's velocity, showing the predicted position of the center of the subhalo in $100\, \text{Myr}$ assuming the same velocity as in the current snapshot. An animation showcasing the complete temporal evolution of the subhalo's gas density and temperature during the infall is available at this \href{https://drive.google.com/drive/folders/1hlsPFaEK9HC8iFW9HDNFx9I58eZli3YS?dmr=1&ec=wgc-drive-globalnav-goto}{link} (see footnote \ref{footnote_movie}) .}
    \label{fig:gas_density_comparison}
\end{figure*}

\begin{figure*}
    \centering
    \includegraphics[width=0.99\linewidth]{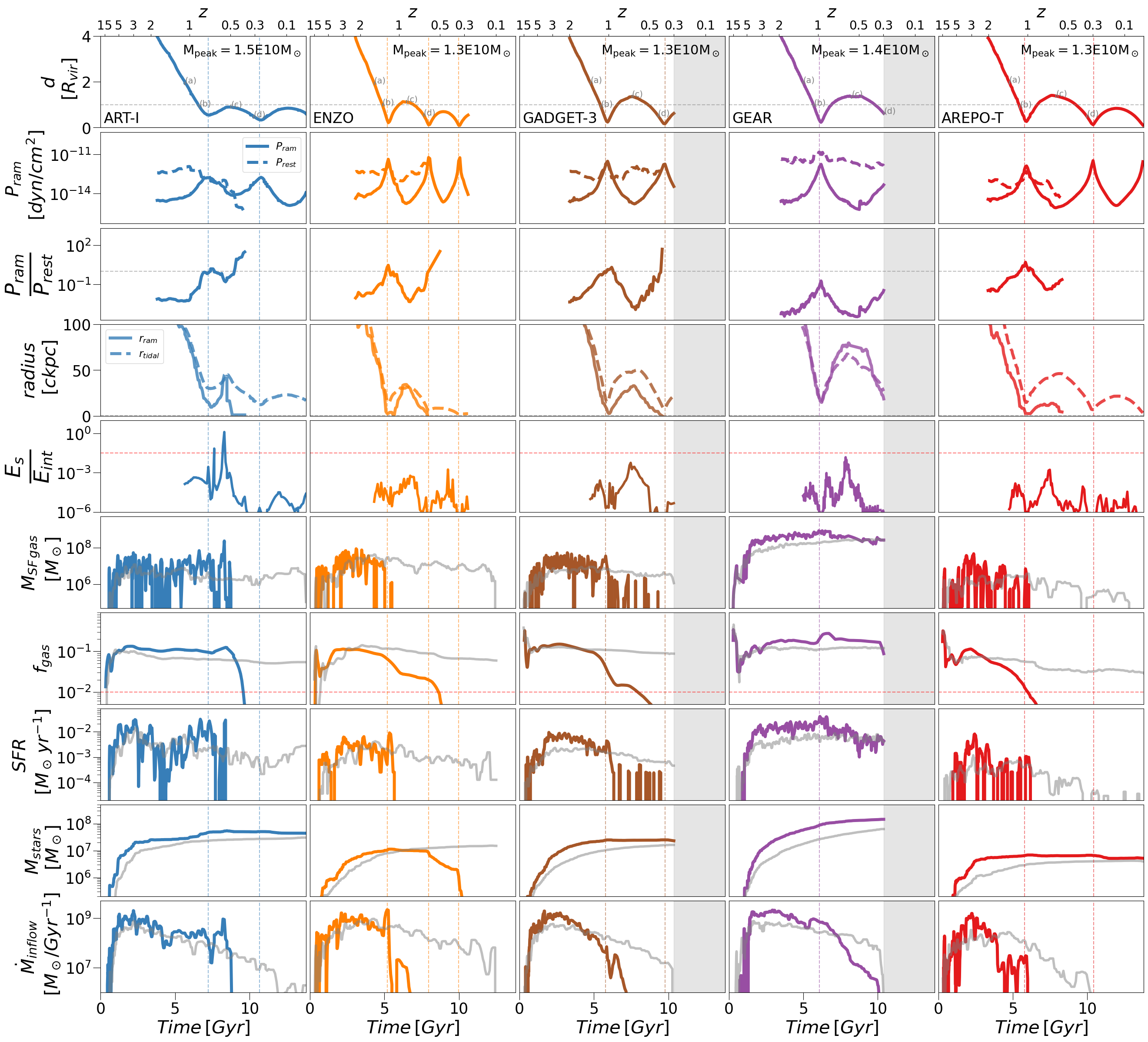}
    \caption{Evolution of the same satellite as in Figure \ref{fig:gas_density_comparison} across all the models. \textbf{First row:} Trajectory during their infall into the host halo. \textbf{Second row:} Evolution of the ram pressure (solid line) felt by the gas and the restoring force (dashed line) per unit area. Both lines begin when the satellite comes within $4\rm{R_{vir}^{host}}$  and $\rm{P_{rest}}$ is defined as long as the satellite contain gas. \textbf{Third row:}  Evolution of the ratio of ram pressure to the restoring force per unit area. Gray horizontal dashed line marks $\rm{P_{ram}/P_{rest}} = 1$ \textbf{Fourth row:} Comparison between the tidal and ram pressure radius, represented as dashed and solid lines, respectively. \textbf{Fifth row:} Ratio between $\rm{E_{s}}$ and $\rm{E_{int}}$ to identify high velocity satellite-satellite encounter. Red horizontal dashed line marks the threshold $\rm{log\left(E_s/E_{int}\right)}>-1.5$. \textbf{Sixth row:} Evolution of the star-forming gas mass, defined as gas with $\rm{n_H>1\,cm^{-3}}$ and a temperature below $10^{4}\,\text{K}$. \textbf{Seventh row:}  Evolution of the ratio $\rm{M_{gas}/M_{halo}}$, the 1\% is marked as a red dashed line for reference. \textbf{Eighth row:} SFR evolution, usually named Star Formation History (SFH). \textbf{Ninth row:} Evolution of the stellar mass. Stellar mass loss serves as an effective tracer for tidal stripping in the ISM. \textbf{Tenth row:} Mass inflow rate evolution for studying the effect of strangulation. Gray background lines in \textbf{rows six to ten} represent the evolution of the same parameter for field galaxies with $\rm{M_{peak}}>10^{10}\,\rm{M}_\odot$ for each model. The time domain where snapshots are still not available for each specific code are indicated as a gray shaded region. Vertical dashed lines mark the pericenter passages for each model. The same plot for a sample of (intercode matched) satellite galaxies can be found at this \href{https://drive.google.com/drive/folders/1T5GcqfGgDetak15O0dEDWGZwmNdhgncy?dmr=1&ec=wgc-drive-globalnav-goto}{link}, where examples of galaxies of different masses and following more radial or more tangential orbits are provided.}
    \label{fig:mechanism_case_comparison}
\end{figure*}

By simply observing the differences in the same satellite across the different models in the Figure \ref{quench_timescales}, the variations in stellar mass and quenching delay time are evident. Quenching timescales in ENZO and AREPO-T models are very short, close to $1$ Gyr. On the other hand, for GADGET-3 and GEAR we have higher quenching delay times, specially for GEAR for which our satellite remains star-forming after $\sim 5$ Gyr of evolution. Sitting between both behaviors is ART-I, which has a quenching timescale of approximately 2 Gyr, close to the crossing time for MW-mass halos. To better illustrate this and to analyze the quenching process in the different \texttt{CosmoRun} models, we plot the gas density for each model in different stages during the infall to the host\footnote{A movie showcasing the complete temporal evolution of the subhalo's gas density and temperature during the infall is available at this \href{https://drive.google.com/drive/folders/1hlsPFaEK9HC8iFW9HDNFx9I58eZli3YS?dmr=1&ec=wgc-drive-globalnav-goto}{link}. The movie is synchronized across codes so that each frame corresponds to the same orbital stage of the trajectory relative to the host and spans from when the satellites are at $6\rm{R_{vir}^{host}}$ to their second apocenter (or until the last available snapshot of the simulation if they have not reached it yet). \label{footnote_movie}} in Figure \ref{fig:gas_density_comparison}. Each row corresponds to a different evolutionary stage: The top row is the first snapshot at $r < 2\rm{R_{vir}^{host}}$ (a); the second row is at $\rm{t_{infall}}$, that is, first infall to the host (b); the third row is during the first apocenter (c); and the bottom row is during the second infall, which is slightly before the second pericenter (d). We highlight the time and position of these evolutionary stages on the top panel of Figure \ref{fig:mechanism_case_comparison}, using the (a), (b), (c), and (d) labels. 

In the top row (a) of Figure \ref{fig:gas_density_comparison}, we see the initial gas density distribution for all models,  prior to host interaction effects. Different models using various implementations of baryonic physics exhibit significantly different gas density distributions and gas contents, such as between GEAR and AREPO-T, with the former containing substantially more and denser gas than the latter. During their first infall, shown in second row (b), we see how the gas contained within the satellite begins to interact with the host CGM. Interestingly, in AREPO-T we can discern clear signs of ram pressure stripping as soon as the satellite crosses the virial radius, with gas tails stripped in the direction opposite to the satellite's motion. In contrast, the initial gas distribution appears apparently unchanged in the other models. This suggests that the lower density and initial amount of gas in AREPO-T, as observed in the first row (a), makes the gas more susceptible to being stripped by ram pressure during the infall. Another detail we can observe in panel (b) for ENZO and AREPO is how, due to the outflows caused by SN feedback, we can distinguish holes in the gas distribution. This highlights how SN feedback also plays a significant role in the expulsion of gas in these low-mass halos. The fact that this is only visible in these codes in this Figure is due to the choice of snapshot. In the animation showing the full temporal evolution (see footnote \ref{footnote_movie}), these features formed in the other codes as well, although they are less prominent in GADGET-3 and GEAR. This aligns with the findings of \citetalias{Strawn_2024} and may be due to a systematic behavior of SPH codes or the different implementations of SN feedback in these models.

At the first apocenter, in the third row (c), we show the gas density remaining after the first pericenter passage where the ram pressure and tidal stripping are supposed to be maximum. The satellite in AREPO-T has undergone a complete gas removal after its first pericenter. In the cases of ART-I, ENZO and GADGET-3,  the satellite's gas has been stripped from its outskirts, retaining a dense region of gas at its center. In contrast, the satellite in GEAR retains its gas in a more efficient way, showing minimal stripping compared to the other models. It is noteworthy that in this panel, the satellite in ART-I is undergoing a high-speed encounter with another satellite, located within its virial radius along the positive Y axis. We study this event in further detail later in this section. Furthermore, the decrease in size of the virial radius for all the models indicates effective tidal stripping of less bound DM particles from the outer regions of the halo. 

In bottom row (d), just before the second pericenter, we appreciate how the satellite's gas in ENZO and GADGET-3 is close to be completely stripped, retaining only the innermost and densest regions of their ISM, similar to findings in \cite{Samuel_fire_2022}. In ART-I, the satellite has undergone complete gas removal due to a highly effective combination of stripping and harassment. In AREPO-T, the satellite does not reaccrete gas anymore so it remains quenched, while GEAR is the only model in which the satellite retains a significant amount of gas and remains largely star-forming during its second infall, even after 5 Gyr of evolution.

To analyze in detail the physical processes observed in the Figure \ref{fig:gas_density_comparison} and discussed above, we represent the evolution of key properties and parameters of the satellite during its history in Figure \ref{fig:mechanism_case_comparison}. This plot gives the essential information required to assess the contribution and significance of the quenching mechanisms under study. In the first row, we compare the trajectory of the satellite across different models. Orbits for ENZO,  GADGET-3, GEAR and AREPO-T are quite similar with some timing discrepancies: ENZO crosses the host virial radius at $\rm{z}\sim 1.3$ while the other models cross it around $\rm{z}\sim 1.1$. In contrast, ART-I shows an infall redshift of $\rm{z}\sim 0.9$ and displays a less eccentric orbit than the other models. Due to the timing discrepancies, presented in \citetalias{santi_paper_4_agora}, and differences in the final snapshots for each model, the satellite reaches two pericenters in ART-I and GADGET-3, three pericenters in ENZO and AREPO-T, and only one pericenter in GEAR. In the second and third rows, we plot the evolution of the ram pressure felt by the gas, the restoring force per unit of area exerted by the gravity of the subhalo and the ratio between them. The fourth row shows how the ram pressure and tidal radius vary during the satellite’s infall. The fifth row represents the evolution of the ratio between the energy gained from high-speed encounters and the internal energy of our satellite, we highlight $\rm{log\left(\rm{E_s/E_{int}}\right) = -1.5}$ as a threshold beyond which harassment is found to be effective \citep{Marasco_2016}. The sixth, seventh, eighth, and ninth rows display the evolution over time of the star-forming gas ($\rm{n_H > 1\,cm^{-3}}$ and $\rm{T<10^4\,K}$) mass, the gas mass fraction, the SFR, and the stellar mass, respectively. Finally, the tenth row illustrates how the inflow mass rate is affected when the subhalo is embedded within the virial-shocked host halo. 

Attending to the evolution of the satellite properties together with the evolution of the quantities describing the efficiency of the different quenching mechanisms, we analyzed the satellite quenching for each model for this case study. We highlight the features representative of the entire satellite population for each model along with the methodology used to identify the dominant quenching mechanism in the following subsections.

\paragraph{\textsc{\textbullet\hspace{0.5em}ART-I.} }
 The ram pressure felt by the satellite's gas is lower than the one experienced in the other models due to its less eccentric orbit. However, it is able to reach the restoring force during the first pericenter, achieving \( \rm{r_{ram}} \) of $\sim 10$ kpc, stripping gas from its outskirts. The gas fraction and star-forming gas mass remain constant until the first apocenter at \( \rm{z} \sim 0.5 \). At this redshift we found an abrupt drop of the restoring force, increasing the efficiency of the ram pressure stripping and leading to an rapid decrease of $\rm{r_{ram}}$ that results in the removal of the star-forming gas mass and the drop of the gas fraction, quenching the star formation.  Attending to the ratio between $\rm{E_s/E_{int}}$, we found that the drop in restoring force and star-forming gas mass coincides with a high-speed encounter with another satellite (third row (c) of Figure \ref{fig:gas_density_comparison}). This encounter causes its gas to heat up and its restoring pressure to decrease drastically, resulting in gas loss due to the combined effects of ram pressure stripping and harassment. 

The behavior of the satellite in this case study in this model is influenced by stochastic processes, such as variations in the eccentricity of its orbit compared with the other \texttt{CosmoRun} models or its high-speed encounter with another satellite. However, the satellite evolution prior to the encounter is representative of the satellite population in this model and this case allows us to exemplify a case where harassment is causing the quenching of our satellite.

\paragraph{\textsc{\textbullet\hspace{0.5em}ENZO.} }
 The ram pressure exceeds the restoring force during the first pericenter, leading to the complete stripping of its external layers. From that moment, the remaining gas is located in the inner and dense region of the subhalo (second (b) and 
 third rows (c) of Figure \ref{fig:gas_density_comparison}), as the restoring force in this inner region is higher. Although this central region of gas that the satellite still retains is relatively cold and dense, the gas is not able to compress sufficiently to reach a density of \(\rm{n_H > 1 \, \text{cm}^{-3}}\). Therefore, the satellite cannot regenerate star-forming gas and, as a result, is unable to form stars again. During the second pericenter passage, both ram pressure and tidal stripping reach their peak, as observed in $\rm{r_{ram}}$ and $\rm{r_{tidal}}$ evolution, effectively removing the remaining gas. Additionally, tidal forces during the second pericenter strip stellar particles from the ISM. 

This case study is representative of the general quenching behavior of intermediate-mass satellites in ENZO \texttt{CosmoRun} model, where the satellites experienced rapid quenching during their first infall due to ram pressure stripping. Even if they are able to retain some little gas at the inner regions, they cannot compress it enough to form stars again.

\paragraph{\textsc{\textbullet\hspace{0.5em}GADGET-3.} }
 The satellite galaxy in this \texttt{CosmoRun} model evolves similarly to the one in the ENZO model (as shown in Figure \ref{fig:gas_density_comparison}). However, in contrast to what is observed in the satellite for ENZO, the satellite in GADGET-3 is still capable of sufficiently compressing the remaining gas in the center at certain times to form stars. As a result, GADGET-3 persists in forming stars at a lower and bursty rate until it completely loses its gas during the second pericenter passage, when the gas is stripped due to the effect of ram pressure. As we can check by examining the stellar mass evolution and $\rm{r_{tidal}}$ during the second pericenter, tidal stripping is not able to strip the ISM.

As in the case of the ENZO model, this case allows us to illustrate the quenching of satellites in this mass range for this \texttt{CosmoRun} model: ram pressure stripping removes gas from the outermost regions during the first pericenter, which leads to a drop in SFR. However, while ENZO is unable to effectively compress the remaining gas in the central region to form stars, GADGET-3 successfully does so. As a result, GADGET-3 continues to form stars in a bursty behavior until the second pericenter, whereas ENZO remains quenched from the first pericenter. This difference in star formation efficiency is in line with the different stellar-halo mass relation noted in Section \ref{sec:sat_evolution} and results in ENZO displaying much shorter timescales than GADGET-3 in this mass range.

\paragraph{\textsc{\textbullet\hspace{0.5em}GEAR.} }
 The ram pressure felt by the satellite is comparable to the other models. However, the restoring force is considerably higher than in the rest of the models, due to the higher gas mass and density as we can see in the comparison between models in the 4th column of Figure \ref{fig:gas_density_comparison}. Consequently, the ram pressure is unable to exceed the restoring force, resulting in a minimum \( r_{\text{ram}} \) comparable to the minimum \( \rm{r}_{\text{tidal}} \), that are above 15 comoving kpc during the first pericenter. This indicates that stripping processes are not able to efficiently remove the gas of the ISM, and the star-forming gas mass and gas fraction remain constant even after 5 Gyr of infall, as we have observed in Figure \ref{fig:gas_density_comparison}.

This case illustrates how stripping mechanisms affect the satellites in GEAR for this mass range. Unlike the other models, despite experiencing similar ram pressure and tidal forces, the satellites of this range of mass in this \texttt{CosmoRun} model have a higher restoring pressure. This higher restoring pressure enables them to retain their gas reservoirs for a longer period since the stripping mechanisms are less efficient. Hence, quenching timescales approach the star formation depletion time, with quenching primarily driven by strangulation after the satellite has exhausted all its gas reservoirs. These findings align with those from the analysis of the CGM of the host galaxy presented in \citetalias{Strawn_2024}, which demonstrated that GEAR is the least efficient model in ejecting gas from the host halo. This lower efficiency in expelling gas from satellite galaxies accounts for the greater satellite gas mass and density observed in this model, leading to a higher restoring pressure.

\paragraph{\textsc{\textbullet\hspace{0.5em}AREPO-T.} }
 The ram pressure surpasses the restoring force during the first pericenter, similarly to ENZO and GADGET-3. However, unlike in GADGET-3 and ENZO,  the gas is completely removed just after this first pericenter, leading to the quenching of the satellite that is not able to reaccrete gas for fueling star formation. It is worth noting that the gas fraction in AREPO-T begins to decrease from \( \rm{z} \sim 2 \), when the satellite is at a distance of about \( 4\rm{R_{vir}^{host}} \) and ram pressure stripping is not yet effective. Looking at the SFR, this decrease in gas is preceded by a burst in star formation, which (as seen in the animation mentioned in footnote \ref{footnote_movie}) triggers quite effective SNe outflows. This suggests that the high efficiency of SN feedback in this model contributes to the low restoring pressure, facilitating stripping during the first pericenter.

By analyzing this case study, we can understand the general behavior of satellites in this mass range within this model. The star-forming gas (and stellar) mass in the AREPO-T model is lower than in the other models, resulting in a lower restoring pressure and facilitating gas stripping due to ram pressure. This leads to the satellites in this model experiencing faster quenching compared to the other models.

\vspace{1 em}

In addition to the processes related to gas removal from the satellite that have already been studied, we want to highlight the relevance of strangulation in satellites within a MW-Mass halo.  Attending to the last row of Figure \ref{fig:mechanism_case_comparison}, the cutoff of cold inflows for our satellite is reproduced by all our models. Thus, once the satellite's gas is stripped away, it is unable to reaccrete gas to replenish its reservoirs, leading to long-term satellite quenching.  Furthermore, in addition to the gas in our satellites potentially being stripped, it is important to note that before quenching, while star formation is still ongoing, the outflows driven by SNe will also expel gas from our satellite (with varying efficiency depending on the model), facilitating gas removal in the inner regions and contributing to quenching.

\subsubsection{Contribution of quenching mechanisms}
\label{sec:dominance_mechanisms}
\begin{figure*}
    \centering
    \includegraphics[width=0.99\linewidth]{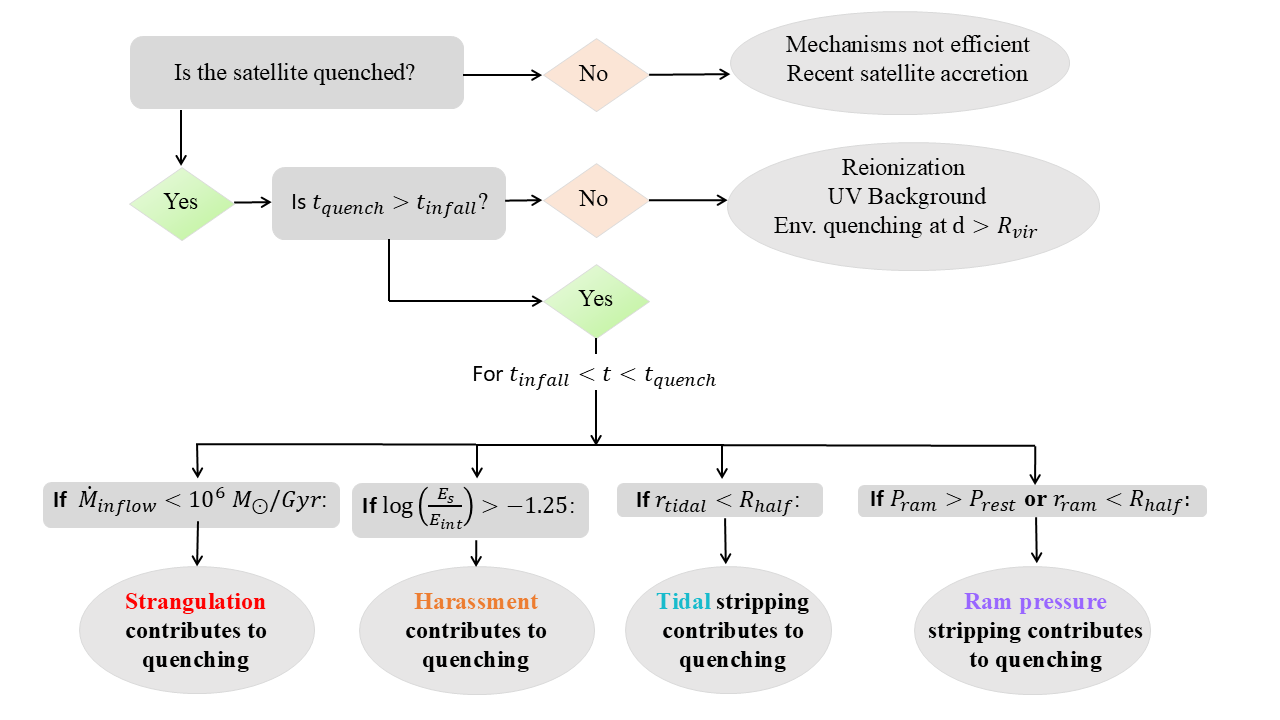}
    \caption{Flowchart illustrating the decision-making process for determining which mechanisms contribute to satellite quenching. The process is described in detail in Section \ref{sec:dominance_mechanisms}.}
    \label{fig:flowchart}
\end{figure*}

After presenting the analysis methodology in the previous section, here we examine the entire population of satellites that interacted with the host halo when $\rm{M}_{host}>5\times 10^{11}\rm{M}_\odot$  (the ones shown in Figure \ref{quench_timescales}). The evolution of each satellite was analyzed to determine which quenching mechanisms were responsible for their quenching or if they were inefficient. We followed the steps outlined in Figure \ref{fig:flowchart} to perform this analysis.

First, we determined whether our satellite galaxy is quenched at the last available snapshot for each \texttt{CosmoRun} model. If it is not quenched, this may be due to ineffective quenching mechanisms, either because the satellite was recently accreted or because the mechanisms were not able to strip the gas responsible of star formation. If the satellite is quenched, we verify whether it was quenched before or after its infall. If it was quenched after infall, the mechanisms contributing to its quenching were examined to determine if any of them dominates over the others. 

To determine when a quenching mechanism is responsible for reducing the amount of gas in our satellites, the evolution of the satellite's properties during infall is visually inspected in relation to the parameters characterizing each mechanism, as illustrated in Figure \ref{fig:flowchart}. In this way, some threshold is established for the parameters related with each mechanism, beyond which our mechanisms are typically found to influence the reduction of gas and cessation of star formation in our satellites.

The effectiveness of ram pressure stripping in removing gas is noted when \( \rm{P_{ram} > P_{rest}} \) or \( \rm{r_{ram} < r_{half,\,gas}} \). Likewise, tidal forces are significant for gas stripping when \( \rm{r_{tidal} < r_{half,\,gas}} \). It is worth mentioning that we have also tested alternative definitions of $\rm{r_{tidal}}$, such as the one used in \citealt{Marasco_2016}, yielding similar values. For harassment, the same threshold from \cite{Marasco_2016} is considered: $\rm{log\left(\frac{E_s}{E_{int}}\right)}>-1.5$, which has been found to be a reliable tracer of when high-speed satellite-satellite encounters typically lead to rapid quenching for our models. An example showing the effectiveness of these criteria was provided in Figure \ref{fig:mechanism_case_comparison}. 

For satellites quenched before infall, if quenching occurred within $4\rm{R_{host}^{vir}}$, we also assess whether any of the stripping mechanisms contributed to gas removal beyond the host's virial radius. On the other hand, for satellites quenched beyond $4\rm{R_{host}^{vir}}$, we conclude that quenching was likely due to mechanisms unrelated to the host environment (such as those discussed in Section \ref{sec:when}), as environmental quenching is expected to have little to no influence at those distances \citep{Cen_2014}. 

\begin{figure*}
    \centering
    \begin{subfigure}[b]{0.99999\linewidth}
        \includegraphics[width=\linewidth]{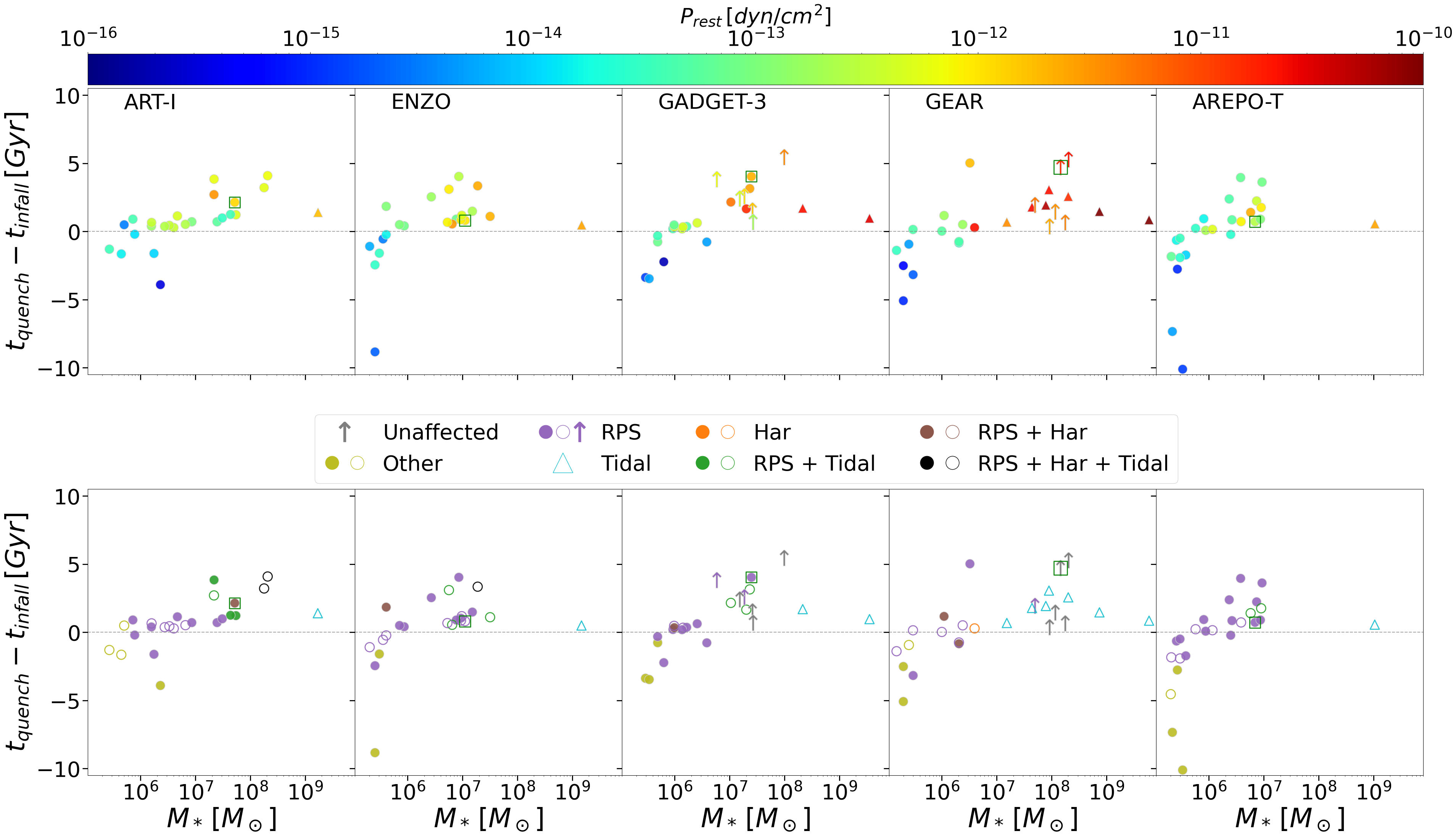}
    \end{subfigure}
    \caption{Same as Figure \ref{quench_timescales} but the color bars colored by the maximum restoring force during infall (\textbf{top}) and the quenching mechanism affecting each satellite (\textbf{bottom}). Whether a mechanism contributes to quenching is determined by following the process described in Section \ref{sec:dominance_mechanisms} and illustrated in Figure \ref{fig:flowchart}. We note that strangulation is found to be efficient for all satellites due to the virial-shocked halo; however, it is excluded here for clarity and we focus on the stripping mechanisms. If a satellite was quenched before any of the mechanisms became efficient, it is represented in olive color and labeled as quenched by other mechanisms, as the ones enumerated in Section \ref{sec:when}. Satellites not affected by any stripping mechanism and still star-forming are shown in gray and labeled as unaffected. \textcolor{black}{The open green square enclosing a single data marker in each panel indicates the same satellite across the different models and we compare its evolution in Section \ref{sec:case_study} and in Figures \ref{fig:gas_density_comparison} and \ref{fig:mechanism_case_comparison}.}}
    \label{quench_timescales_complete}
\end{figure*}

By applying this methodology to all our satellites, we determine the mechanisms contributing to satellite quenching for each model. Figure \ref{quench_timescales_complete} presents the quenching timescales of our satellites, similar to Figure \ref{quench_timescales}, but now with colorbars colored by the maximum restoring force experienced during infall and the quenching mechanisms that efficiently contribute to gas removal for each satellite.

On the top row we show the maximum restoring pressure during infall for each satellite.  For all the models, the more massive the satellite, the higher its restoring pressure, which makes stripping processes less efficient. This explains the observed trend of larger quenching timescales for more massive satellites. Comparing between models, GADGET-3 and GEAR show higher restoring pressures than the others, with GEAR exhibiting the highest. This explains the longer quenching timescales observed in these models. A more detailed analysis of the differences between restoring pressure across models and its influence in satellite quenching can be found at the end of this section.

The bottom row shows the mechanisms responsible for stripping gas from each satellite. We note that strangulation is efficient for all satellites within the host's virial radius, as it prevents the accretion of new cold gas once the satellites are embedded in the virial-shocked halo, leading to long-term quenching of star formation. However, strangulation is excluded from this plot for clarity, and we focus on stripping mechanisms. 
Some general trends can be observed across the models. The lowest-mass satellites  tend to quench before infall. According to our criteria, many of these satellites are not impacted by any of the stripping mechanisms considered. We therefore hypothesize that their quenching is driven by processes unrelated to the host, such as cosmic reionization in the early universe. A stellar mass threshold, between $5\times10^5 - 2\times 10^6 \,\rm{M}_\odot$,  depending on the model, is identified beyond which quenching occurs only after crossing the $\rm{R_{vir}^{host}}$ and due to the effect of the host. This threshold is consistent with the observed masses of satellites quenched during reionization \citep{Brown_2014_reionization, Weisz_2014_reionization, Rodriguez_Wimberly_2019_reionization}, as well as findings from similar simulations \citep{Akins_2021, Samuel_fire_2022}. Reionization process heats up the intergalactic medium
to a temperature above and around $10^4$ K. This raises the Jeans mass abruptly from about $10^6\, \rm{M_\odot}$
to about $10^{10}\, \rm{M_\odot}$ at the redshift of reionization \citep{Cen_McDonald_2002}, suppressing and hence quenching 
subsequent star formation in halos below this halo mass. For those galaxies with stellar mass of $5\times10^5 - 2\times 10^6 \,\rm{M}_\odot$, their halo masses at the time of reionization are below this threshold of the raised Jeans mass above, based on estimates of the stellar mass-halo mass relation at the relevant high redshift (e.g., \citealt{Shuntov_2024}) and are thus expected to be quenched post-reionization. Notably, some satellites in this mass range also quench before crossing $\rm{R_{vir}^{host}}$ but due to ram pressure stripping by the host's gas beyond the virial radius, as their extremely low restoring pressure allows this process to efficiently strip their gas. For satellites with stellar masses between $10^6 - 10^7 \,\rm{M}_\odot$, ram pressure stripping remains the dominant stripping mechanism across all the models. Once they cross the $\rm{R_{vir}^{host}}$ and interact with the host's warm CGM, ram pressure stripping leads to rapid quenching.

For intermediate-mass satellites with \( \rm{M_* \sim 10^7 - 10^8\,M}_\odot \), we find that the gas stripping for most of them is dominated by a combined effect of ram pressure and tidal stripping. Due to their higher restoring pressure, ram pressure alone is typically not sufficient to strip all the gas. Ram pressure stripping removes the outermost layers of gas, which are less gravitationally bound. Simultaneously, tidal forces lead to the loss of DM and some gas from the outskirts. This, along with potential morphological perturbations due to tidal forces, reduces the restoring force, making the satellite more vulnerable to ram pressure stripping after its first pericenter passage. Consequently, the quenching delay times for this mass range are closer to the crossing time of MW-mass halos $\sim 2$Gyr, as they need a full orbit to efficiently strip the gas. 

Finally, for high-mass satellites, with \( \rm{M_* > 10^8\,M}_\odot \), we find that stripping mechanisms are generally not efficient. As a result, quenching occurs when the satellite consumes its cold gas reservoirs, since it cannot replenish its gas due to strangulation. This leads to large quenching timescales, which are close to the satellite’s gas depletion time. However, for high-mass galaxies we also find that they often undergo merger without being quenched, with tidal forces dominating and causing disruption before ram pressure stripping has a chance to remove their gas. This is specially relevant in GEAR. As more massive satellites are expected to follow more eccentric orbits, characterized by lower specific angular momentum and smaller pericenters \citep{radial_orbits_2015}, tidal forces are expected to be stronger, making it more likely for the subhalo to experience disruption. 

These findings regarding the mass-dependent trends of quenching mechanisms observed in satellites within MW-mass halos align with the results from previous studies as \citealt{Fillingham_2015, Fillingham_2016} and \citealt{Rodriguez_Wimberly_2019_reionization} (see Figure 5 in said work for a schematic summary): satellites with stellar masses below \(10^6\,\rm{M}_\odot  \) are primarily quenched by mechanisms unrelated to the host, such as cosmic reionization; between \(10^6\,\rm{M}_\odot \) and \(10^8\,\rm{M}_\odot \), the dominant mechanisms are stripping and SN feedback; and above \(10^8\,\rm{M}_\odot \), the quenching timescales indicate a depletion of star-formation time, which, in conjunction with SN feedback, results in the exhaustion of the fuel for star formation, which cannot be replenished due to the cutoff of inflows caused by strangulation.

\begin{figure*}
    \centering
    \includegraphics[width=0.99\linewidth]{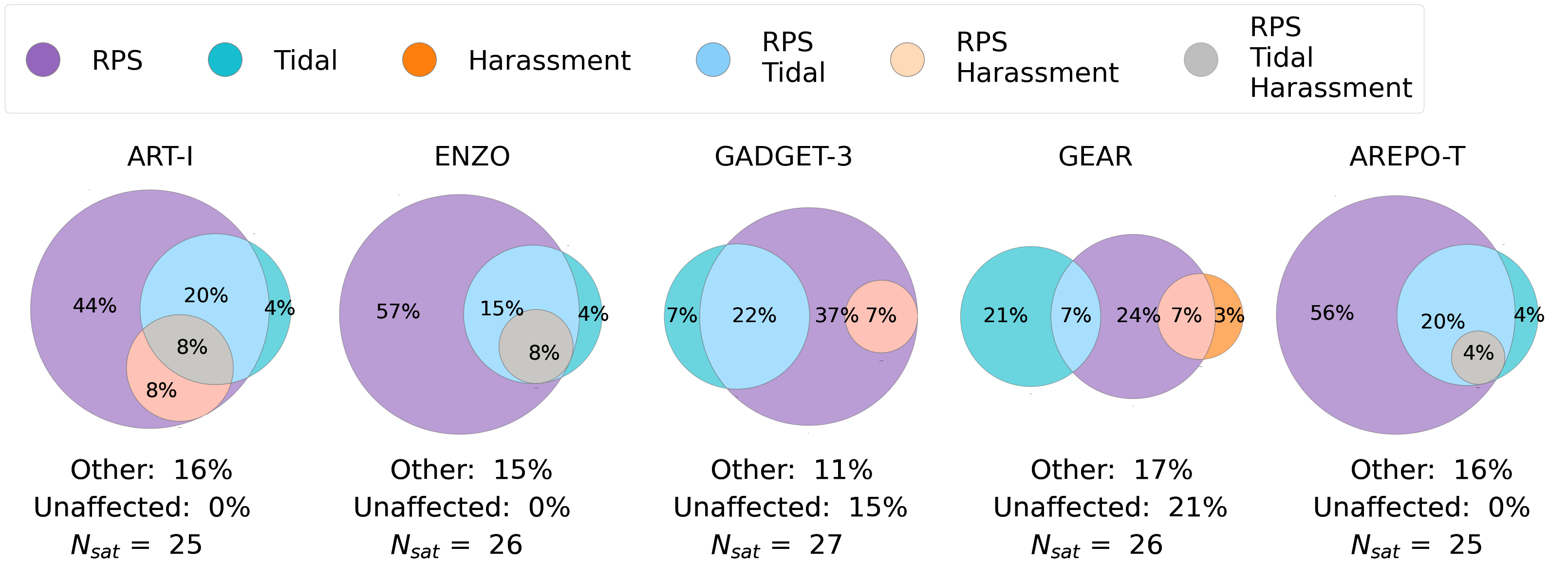}
    \caption{Venn diagram for each \texttt{CosmoRun} model analyzed in this paper showing the fraction of satellite galaxies affected by each of the mechanisms considered responsible for the stripping of the satellites' gas. All satellites shown in Figures \ref{quench_timescales} and \ref{quench_timescales_complete} are included. Please note that we found that strangulation affects all satellites by cutting off the replenishment of cold gas reservoirs; however, since it impacts 100\% of the satellites, we do not include it to enhance clarity. Whether a mechanism contributes to quenching is determined by following the process described in Section \ref{sec:dominance_mechanisms} and illustrated in Figure \ref{fig:flowchart}. Below each diagram, we present the fraction of satellites quenched by other mechanisms not considered in our analysis (e.g., the ones described described in Section \ref{sec:when}), along with the fraction of satellites that are unaffected by any mechanism and still star-forming, and the total number of satellites analyzed.}
    \label{fig:hist_contr}
\end{figure*}

In Figure \ref{fig:hist_contr}, we present Venn diagrams for each model, showing the fraction of satellites affected by each gas stripping mechanism considered. As in Figure \ref{quench_timescales_complete}, strangulation is excluded for clarity, since it impacts all satellites. Below each diagram, for completeness, we show the fraction of satellites in the lowest-mass range that are affected by other mechanisms unrelated to the host’s influence, as discussed above for Figure \ref{quench_timescales_complete}. We also include the fraction of satellites unaffected by any stripping mechanism.

\begin{figure*}
    \centering
    \includegraphics[width=0.99\linewidth]{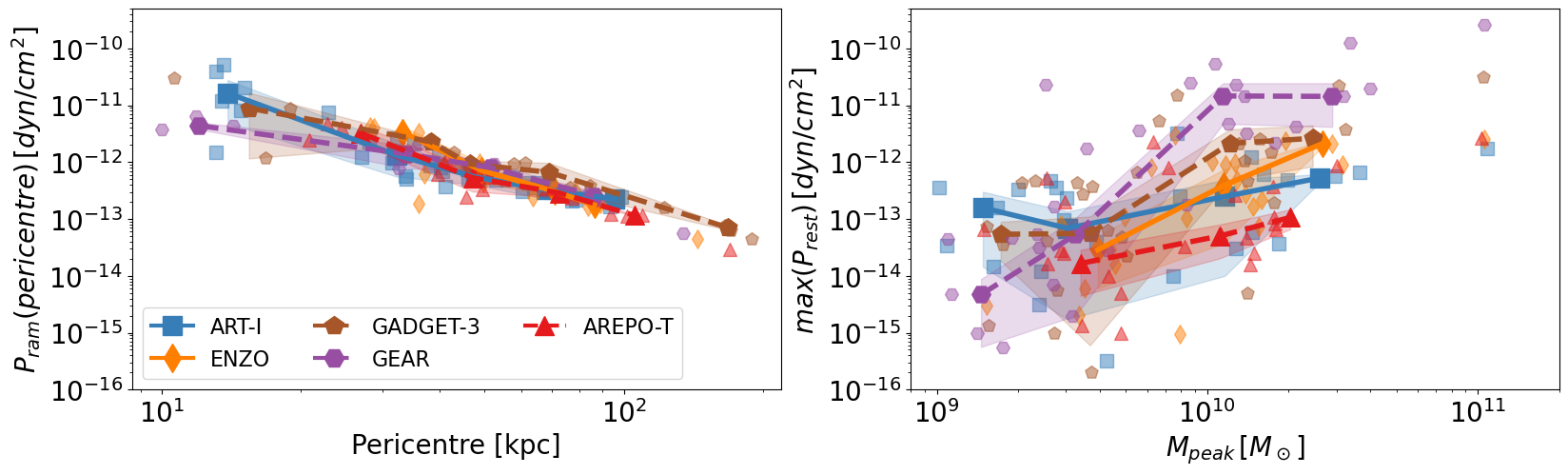}
    \caption{Comparison of intercode differences in the ram pressure experienced by satellites and their restoring pressure. \textbf{Left}: Ram pressure felt by satellites during their first pericenter as a function of pericenter distance. \textbf{Right}: Maximum restoring pressure of each satellite as a function of their peak halo mass. Each marker represents an individual satellite, while the colored lines denote the median values for each \texttt{CosmoRun} model in each bin, with the shaded regions indicating the median absolute deviation. The codes differ significantly more in the restoring pressure of each satellite than in the ram pressure exerted by the host CGM.}
    \label{fig:ramvsrest}
\end{figure*}

The most common gas stripping mechanism across all models is ram pressure stripping, either acting alone or in combination with tidal stripping, and less frequently, with harassment, similar to the findings in Figure \ref{quench_timescales_complete}. However, the efficiency of ram pressure stripping varies across codes, particularly for the most massive satellites, which explains the differences in satellite quenching observed in the previous figures. This variation in efficiency could arise from differences in the host CGM across models, leading to different ram pressures experienced by the satellite gas; or differences in the restoring pressure of each satellite, which determines their ability to resist stripping processes. To compare the influence of these effects in explaining the differences in satellite quenching between models, we present in Figure \ref{fig:ramvsrest} an intercode comparison of the ram pressure experienced by satellites and their restoring pressure. In the left panel, we plot the ram pressure experienced during the first pericenter as a function of the pericenter distance. In the right panel, we show the maximum restoring pressure of each satellite as a function of its peak halo mass. Each marker point represents an individual satellite galaxy, while the colored lines indicate the median in each bin for each model, and the shaded colored region represents the median absolute deviation. The ram pressure felt by satellites largely converges across all models, with differences of less than 1 dex in the closest pericenters ($< 20-30 \,\rm{kpc}$) between ART-I and GEAR. However, the restoring pressure of satellites differs considerably more, especially for $\rm{M_{peak}} > 3\times 10^9 \,\rm{M_\odot}$, where GEAR exceeds by 3 dex the restoring pressure of satellites in AREPO-T. This comparison allows us to conclude that the variations in satellite quenching efficiency in our models are primarily driven by differences in the restoring pressure of the satellites, rather than by differences in the host CGM across models. This highlights how different SN feedback physics and hydrodynamic methods result in satellites with significantly different capacities to retain their gas reservoirs and resist stripping processes. For example, in \citetalias{Strawn_2024}, it was shown that the feedback recipe implemented in the GEAR \texttt{CosmoRun} model is the least efficient at driving outflows. This inefficiency leads to a higher gas concentration in the central regions of satellites compared to the other models, which increases the restoring pressure and makes satellites more resistant to stripping. Consistent with this, when comparing the percentage of satellites unaffected by any stripping mechanism, GEAR shows the highest proportion at 21\%, as ram pressure stripping is inefficient for all satellites with stellar masses greater than $10^7 \rm{M}_\odot$. This is followed by GADGET-3 with 15\%, which also exhibits higher restoring pressures than the other models, as shown in top row of Figure \ref{quench_timescales_complete} and in right panel of Figure \ref{fig:ramvsrest}. Finally, for the rest of the models, whose satellites have lower restoring pressures, all satellites were affected by some stripping mechanism and were quenched on either shorter or longer timescales.

Setting aside quenching efficiency, where our models differ particularly for the more massive satellites, the results from Figures \ref{quench_timescales_complete} and \ref{fig:hist_contr} indicate that for satellite quenching influenced by MW-Mass hosts, ram pressure stripping is the most dominant mechanism, crucial for explaining the quenching of nearly all satellites across all models. Tidal stripping becomes relevant for intermediate-mass satellites that are able to retain their gas until the first pericenter passage. It is worth noting that when tidal forces dominate over ram pressure stripping, quenching does not occur because the satellite is disrupted before quenching can take place, meaning such satellites would never be observed in a quenched state. Lastly, harassment has some influence in the quenching of satellites within MW-mass halos, though it occurs less frequently than ram pressure and tidal stripping. Once the satellites have been stripped of their gas, the cessation of cold gas replenishment due to strangulation ensures long-term quenching.

This finding regarding the dominant mechanism of satellite quenching in MW-mass halos differs from the results in \cite{Marasco_2016} for clusters and galaxy groups. In \cite{Marasco_2016}, they observe that, based on a single snapshot, the most common mechanism disrupting satellite gas in clusters is ram pressure stripping, which aligns with our result. However, when they examine the evolution of all satellites instead of just a single snapshot, they find that harassment is the most common mechanism causing quenching in clusters of galaxies, closely followed by ram pressure stripping. However, this difference between MW-mass halos and clusters of galaxies may be expected. Aside from differences between the systems - where MW-mass halos have recently reached the mass needed for virial shock formation and their CGM is more inhomogeneous with less extreme temperatures compared to the intra-cluster medium (ICM) - it is expected that clusters and galaxy groups experience more high-speed satellite-satellite encounters. This is due to the statistically higher density and velocity of subhalos in these environments, which increases the likelihood of high-speed encounters between satellites.

\section{Conclusions} \label{sec:conclusion}

In this paper, we have used a subset of five codes participating in the high resolution \texttt{CosmoRun} simulations --- ART-I, ENZO, GADGET-3, GEAR, and AREPO-T --- to examine whether the results of our simulations are reproducible with respect to satellite quenching in MW-mass halos. We investigated the quenching of the satellite population and the physical processes driving it in each of our models. Each model has its own distinct feedback prescriptions and code architecture. As the halos in \texttt{CosmoRun} reach the mass predicted by theory for virial shock formation at $\rm{z}\sim 1$, studying satellite quenching at low redshift provides a suitable scenario to test how satellite galaxies are affected by the warm-hot corona that develops within the host's CGM. The main results presented in this paper are as follows:

\begin{enumerate}
    \item While the feedback strategies used by each code group result in a consistent stellar mass for the main galaxy, significant discrepancies were observed in the stellar mass formed by each individual satellite galaxy across different models. This can be interpreted as a reflection of the different feedback models employed, which have a greater impact on low-mass halos than on the main halo, where SN outflows become less efficient.

    \item The quenched fraction of satellites increases when the host halo mass approaches the expected mass for virial shock formation, as observed across all models. This suggests that the properties of satellites in Milky Way-mass halos can serve as a good tracer of the formation of the virial shock in these halos.
    
    \item Quenched fractions for all the models are compatible with the latest SAGA data within $1\sigma$ host-to-host scatter. However, there are large intercode discrepancies on the quenching of high-mass satellites. ART-I is the only model that matches the quenched fraction observed in the LG for high-mass satellites, while the other models predict lower quenched fractions for this mass bin, thus aligning more closely with the SAGA Survey results.
    
    \item Satellite quenching in ART-I, ENZO, and AREPO-T is more efficient than in GEAR and GADGET-3. The main driver of these differences between models is the higher restoring pressures of satellite galaxies in these models, particularly in GEAR. This results in lower quenched fractions and longer quenching timescales for the same satellites in GADGET-3 and GEAR. Differences in restoring pressure are primarily driven by variations in the gas masses and gas concentrations within the satellites rather than differences in their DM profiles. These findings align with the results of \citetalias{Strawn_2024} (see Figure 6), which showed that outflows in ART-I, ENZO, and AREPO-T are much faster and more efficient than in GADGET-3, and especially in GEAR.

    \item Strangulation and ram pressure stripping are the main quenching mechanisms in MW-mass halos across all models. For satellites within an MW-mass halo, strangulation is efficient at cutting off the cold inflows, whereas ram pressure stripping is the main process responsible for the gas removal. Although the efficiency of ram pressure stripping varies among the different models, leading to differences in the quenched fractions of satellites and quenching timescales, it consistently remains the dominant mechanism of gas stripping.

    \item Quenching timescales are compatible, within their broad spread, with quenching delay times estimated for the LG and the ELVES survey. 
    
    \item Relating quenching mechanisms to quenching timescales allowed us to identify some common trends across all models. For instance, low-mass satellites ($\rm{M}_* < 10^6\,\rm{M}_\odot$) tend to quench before infall by the effect of ram pressure stripping beyond the host virial radius or due to other quenching mechanisms not related to the host. Satellites in the stellar mass range $10^6-10^7\,\rm{M}_\odot$ are predominantly quenched by ram pressure stripping and strangulation in a rapid quenching timescale. Gas in  satellites in the stellar mass range $10^7-10^8\,\rm{M}_\odot$ is usually stripped by a combined effect of ram pressure and tidal stripping, showing quenching timescales close to the crossing time $2$Gyr. Finally, for the more massive satellites with \( M_* > 10^8\,\rm{M}_\odot \), we find that stripping mechanisms are generally not efficient. As a result, quenching occurs when the satellite consumes its cold gas reserves since it cannot replenish its gas due to strangulation. This leads to long quenching timescales that are close to the satellite’s depletion gas time.
    
\end{enumerate}

In general, we observed convergence across the codes we employed regarding the overall trends in quenching timescales and quenched fractions despite the use of different approaches for supernova feedback. Namely, the less massive the satellite, the faster its quenching occurs. We also found that ram pressure stripping and strangulation are the most dominant mechanisms driving satellite quenching in MW-mass halos regardless of the code or supernova feedback recipe employed (although their efficiency is highly dependent on the model). On the other hand, aside from these trends, the specific quenching timescales and the quenched fraction of satellites with stellar masses above $10^7,\rm{M}_\odot$ are highly sensitive to the different \texttt{CosmoRun} models considered. The main factor influencing satellite quenching efficiency is the restoring pressure of each satellite, which varies considerably between models and determines the satellite's ability to resist stripping mechanisms. Disentangling the influence of each feedback implementation and each code would require several feedback models for the same code (which is underway and will lead to new papers within AGORA). However, the fact that the differences are driven by variations in gas concentrations and that they align with the various outflow efficiencies for each \texttt{CosmoRun} model (see \citetalias{Strawn_2024}) allows us to identify the key role different SN feedback recipes play in dwarf satellite quenching in MW-mass halos, a mass regime for which this feedback is expected to be highly relevant. Finally, regarding the treatment of dynamic instabilities and viscous stripping, although a more in-depth analysis is needed, the convergence of our models without observing a systematic difference between hydro-methods suggests that the historical differences between SPH and AMR have been resolved in the codes used in this \texttt{CosmoRun} models or do not play a decisive role in satellite quenching in MW-mass halos.

\section{Caveats}
\label{sec:caveats}
The AGORA \texttt{CosmoRun} simulations were not specifically designed to study dwarf satellites in detail. While the resolution in this simulation suite allows us to resolve the internal dynamics of galaxies with $\rm{M}_*>10^6$ reasonably well, the lowest-mass galaxies may suffer from numerical over-quenching due to the limited resolution \citep{Hopkins_2018_overquenching}. It is worth remembering that for our lower mass galaxy limit of \( 3.39 \times 10^5 M_\odot \), galaxies will only have six stellar particles, so they will be poorly resolved. In Figure \ref{fq_evolution}, for all models, field galaxies with $\rm{M}_\mathrm{peak} < 10^{10}\,\rm{M}_\odot$ show that around 50\% are quenched at low redshift. This could be indicative of the effects of reionization and UV background heating on these low-mass galaxies, but it also suggests possible over-quenching due to resolution, leading to an overestimation of the number of quenched low-mass galaxies. Such overquenching may occur if gas is unrealistically expelled from low-mass galaxies. In \texttt{CosmoRun} simulations, for galaxies with $\rm{M}_*<10^6{\rm{M}_\odot}$, where the ISM may consist of only a few gas elements (cells or particles), a supernova event could deposit all its energy and momentum to these few elements, potentially leading to their unphysical permanent removal from the galaxy  \citep{Samuel_fire_2022}. Future high-resolution runs in AGORA are already underway, and their results will provide valuable insights into the impact of resolution on numerical simulations regarding satellite quenching.

The great strength of AGORA lies in its ability to compare how the physical predictions of models for the same halo and satellites vary or converge using different codes and supernova feedback recipes. However, since we are only simulating a single halo, we lack sufficient host statistics to generalize our results. Future runs with different initial conditions leading to various halos will provide a more representative sample, accounting for the host-to-host scatter observed both in MW-analogs and simulations. The availability of more hosts would also improve the satellite statistics in our analysis, allowing us to draw more robust conclusions about the role of different physical mechanisms in satellite quenching.

Finally, is worth to mention that \texttt{CosmoRun} simulations are not including the effects of AGN feedback, that are supposed to be relevant at low-redshift. The impact of AGN feedback in these simulations is currently being investigated within the collaboration and will lead to new AGORA projects. Nevertheless, this effect seems to be not needed to reproduce the observed quenched fraction in LG and ELVES or SAGA surveys.

\begin{acknowledgements}
    %\textcolor{red}{To be completed}
    We thank all of our colleagues participating in the AGORA Project for their collaborative spirit, which has allowed the AGORA Collaboration to remain strong as a platform to foster and launch multiple science-oriented comparison efforts. We particularly thank code leaders: S.R.F., J.H.K., K.N., J.W.P., Y.R., H.V., A.G., H.K. and A.L. for leading the effort within each code group to perform and analyze simulations. We also thank Artemi Camps Fariña and Pablo Santos Peral for their insightful comments during the work presented in this paper. We are also grateful to Volker Springel for making the GADGET-2 code public and for providing the original versions of GADGET-3 to be used in the AGORA Project. 
    This research used resources of the National Energy Research Scientific Computing Center, a DOE Office of Science User Facility supported by the Office of Science of the U.S. Department of Energy under contract No. DE-AC02-05CH11231. R.R.C., S.R.F., and J.G. acknowledge financial support from the Spanish Ministry of Science and Innovation through the research grant PID2021-123417OB-I00, funded by MCIN/AEI/10.13039/501100011033/FEDER, EU. Additionally, R.R.C. acknowledges financial support by IND2022/TIC-23643 project funded by Comunidad de Madrid. J. K.’s work was supported by the National Research Foundation of Korea (NRF) grant funded by the Korea government (MSIT) (No. 2022M3K3A1093827 and No. 2023R1A2C1003244).  His work was also supported by the National Institute of Supercomputing and Network/Korea Institute of Science and Technology Information with supercomputing resources including technical support, grants KSC-2021-CRE-0442, KSC- 2022-CRE-0355 and KSC-2024-CRE-0232. Some of the numerical computations for this study were carried out on the Cray XC50 at the Center for Computational Astrophysics, National Astronomical Observatory of Japan, and the {\sc SQUID} at the Cybermedia Center, Osaka University as part of the HPCI system Research Project (hp230089, hp240141). This work is supported in part by the MEXT/JSPS KAKENHI grant number 20H00180, 22K21349, 24H00002, and 24H00241 (K.N.). K.N. acknowledges the support from the Kavli IPMU, World Premier Research Center Initiative (WPI), UTIAS, the University of Tokyo. D.C. is supported by the Ministerio de Ciencia, Innovación y Universidades (MICIU/FEDER) under research grant PID2021-122603NB-C21. The publicly available ENZO and yt codes used in this
    work are the products of collaborative efforts by many independent scientists from numerous institutions around the world. Their commitment to open science has helped make this
    work possible.
\end{acknowledgements}



\bibliography{aa53639-24}{}

\begin{thebibliography}{114}
\expandafter\ifx\csname natexlab\endcsname\relax\def\natexlab#1{#1}\fi

\bibitem[{{Abadi} {et~al.}(1999){Abadi}, {Moore}, \& {Bower}}]{Abadi_1999}
{Abadi}, M.~G., {Moore}, B., \& {Bower}, R.~G. 1999, \mnras, 308, 947

\bibitem[{{Agertz} {et~al.}(2007){Agertz}, {Moore}, {Stadel}, {Potter}, {Miniati}, {Read}, {Mayer}, {Gawryszczak}, {Kravtsov}, {Nordlund}, {Pearce}, {Quilis}, {Rudd}, {Springel}, {Stone}, {Tasker}, {Teyssier}, {Wadsley}, \& {Walder}}]{Agertz_2007}
{Agertz}, O., {Moore}, B., {Stadel}, J., {et~al.} 2007, \mnras, 380, 963

\bibitem[{{Akins} {et~al.}(2021){Akins}, {Christensen}, {Brooks}, {Munshi}, {Applebaum}, {Engelhardt}, \& {Chamberland}}]{Akins_2021}
{Akins}, H.~B., {Christensen}, C.~R., {Brooks}, A.~M., {et~al.} 2021, \apj, 909, 139

\bibitem[{{Behroozi} {et~al.}(2015){Behroozi}, {Knebe}, {Pearce}, {Elahi}, {Han}, {Lux}, {Mao}, {Muldrew}, {Potter}, \& {Srisawat}}]{Behroozi_2015_majormergerscomparison}
{Behroozi}, P., {Knebe}, A., {Pearce}, F.~R., {et~al.} 2015, \mnras, 454, 3020

\bibitem[{Behroozi {et~al.}(2012{\natexlab{a}})Behroozi, Wechsler, \& Wu}]{Rockstar_behroozi_2012}
Behroozi, P.~S., Wechsler, R.~H., \& Wu, H.-Y. 2012{\natexlab{a}}, ApJ, 762, 109

\bibitem[{Behroozi {et~al.}(2012{\natexlab{b}})Behroozi, Wechsler, Wu, Busha, Klypin, \& Primack}]{Consistent_behroozi_2012}
Behroozi, P.~S., Wechsler, R.~H., Wu, H.-Y., {et~al.} 2012{\natexlab{b}}, ApJ, 763, 18

\bibitem[{Benítez-Llambay {et~al.}(2015)Benítez-Llambay, Navarro, Abadi, Gottlöber, Yepes, Hoffman, \& Steinmetz}]{Benitez-llambay_2015}
Benítez-Llambay, A., Navarro, J.~F., Abadi, M.~G., {et~al.} 2015, MNRAS, 450, 4207

\bibitem[{{Binney} \& {Tremaine}(2008)}]{BinneyTremaine2008}
{Binney}, J. \& {Tremaine}, S. 2008, {Galactic Dynamics: Second Edition}

\bibitem[{{Birnboim} \& {Dekel}(2003)}]{Birnboim_dekel_2003}
{Birnboim}, Y. \& {Dekel}, A. 2003, \mnras, 345, 349

\bibitem[{{Brown} {et~al.}(2014){Brown}, {Tumlinson}, {Geha}, {Simon}, {Vargas}, {VandenBerg}, {Kirby}, {Kalirai}, {Avila}, {Gennaro}, {Ferguson}, {Mu{\~n}oz}, {Guhathakurta}, \& {Renzini}}]{Brown_2014_reionization}
{Brown}, T.~M., {Tumlinson}, J., {Geha}, M., {et~al.} 2014, \apj, 796, 91

\bibitem[{Brown {et~al.}(2014)Brown, Tumlinson, Geha, Simon, Vargas, VandenBerg, Kirby, Kalirai, Avila, Gennaro, Ferguson, Muñoz, Guhathakurta, \& Renzini}]{Brown_2014}
Brown, T.~M., Tumlinson, J., Geha, M., {et~al.} 2014, ApJ, 796, 91

\bibitem[{{Brummel-Smith} {et~al.}(2019){Brummel-Smith}, {Bryan}, {Butsky}, {Corlies}, {Emerick}, {Forbes}, {Fujimoto}, {Goldbaum}, {Grete}, {Hummels}, {Kim}, {Koh}, {Li}, {Li}, {Li}, {OShea}, {Peeples}, {Regan}, {Salem}, {Schmidt}, {Simpson}, {Smith}, {Tumlinson}, {Turk}, {Wise}, {Abel}, {Bordner}, {Cen}, {Collins}, {Crosby}, {Edelmann}, {Hahn}, {Harkness}, {Harper-Clark}, {Kong}, {Kritsuk}, {Kuhlen}, {Larrue}, {Lee}, {Meece}, {Norman}, {Oishi}, {Paschos}, {Peruta}, {Razoumov}, {Reynolds}, {Silvia}, {Skillman}, {Skory}, {So}, {Tasker}, {Wagner}, {Wang}, {Xu}, \& {Zhao}}]{Brummel_smith_2019_enzo}
{Brummel-Smith}, C., {Bryan}, G., {Butsky}, I., {et~al.} 2019, JOSS, 4, 1636

\bibitem[{{Bryan} \& {Norman}(1998)}]{Byan_Norman_1998}
{Bryan}, G.~L. \& {Norman}, M.~L. 1998, \apj, 495, 80

\bibitem[{{Bryan} {et~al.}(2014){Bryan}, {Norman}, {O'Shea}, {Abel}, {Wise}, {Turk}, {Reynolds}, {Collins}, {Wang}, {Skillman}, {Smith}, {Harkness}, {Bordner}, {Kim}, {Kuhlen}, {Xu}, {Goldbaum}, {Hummels}, {Kritsuk}, {Tasker}, {Skory}, {Simpson}, {Hahn}, {Oishi}, {So}, {Zhao}, {Cen}, {Li}, \& {Enzo Collaboration}}]{Bryan_2014_enzo}
{Bryan}, G.~L., {Norman}, M.~L., {O'Shea}, B.~W., {et~al.} 2014, \apjs, 211, 19

\bibitem[{{Bullock} {et~al.}(2000){Bullock}, {Kravtsov}, \& {Weinberg}}]{Bullock_2000_Reionization}
{Bullock}, J.~S., {Kravtsov}, A.~V., \& {Weinberg}, D.~H. 2000, \apj, 539, 517

\bibitem[{Carlsten {et~al.}(2020)Carlsten, Greco, Beaton, \& Greene}]{Carlsten_2020}
Carlsten, S.~G., Greco, J.~P., Beaton, R.~L., \& Greene, J.~E. 2020, ApJ, 891, 144

\bibitem[{Carlsten {et~al.}(2022)Carlsten, Greene, Beaton, Danieli, \& Greco}]{Carlsten_2022}
Carlsten, S.~G., Greene, J.~E., Beaton, R.~L., Danieli, S., \& Greco, J.~P. 2022, ApJ, 933, 47

\bibitem[{{Cen}(2014)}]{Cen_2014}
{Cen}, R. 2014, \apj, 781, 38

\bibitem[{{Cen} \& {McDonald}(2002)}]{Cen_McDonald_2002}
{Cen}, R. \& {McDonald}, P. 2002, \apj, 570, 457

\bibitem[{Cramer {et~al.}(2024)Cramer, Noble, Rudnick, Pigarelli, Wilson, Bahé, Cooper, Demarco, Matharu, Miller, Muzzin, Nantais, Sportsman, van Kampen, Webb, \& Yee}]{Cramer_2024}
Cramer, W.~J., Noble, A.~G., Rudnick, G., {et~al.} 2024, ApJ, 975, 144

\bibitem[{{Dekel} \& {Birnboim}(2006)}]{DB06}
{Dekel}, A. \& {Birnboim}, Y. 2006, \mnras, 368, 2

\bibitem[{{Dekel} {et~al.}(2019){Dekel}, {Lapiner}, \& {Dubois}}]{dekel2019_goldenmass}
{Dekel}, A., {Lapiner}, S., \& {Dubois}, Y. 2019, arXiv e-prints, arXiv:1904.08431

\bibitem[{{Dekel} \& {Silk}(1986)}]{Dekel_1986_SN}
{Dekel}, A. \& {Silk}, J. 1986, \apj, 303, 39

\bibitem[{{Diemer}(2021)}]{Diemer_2021_splashback}
{Diemer}, B. 2021, \apj, 909, 112

\bibitem[{{Diemer} {et~al.}(2024){Diemer}, {Behroozi}, \& {Mansfield}}]{Diemer_sparta_2023}
{Diemer}, B., {Behroozi}, P., \& {Mansfield}, P. 2024, \mnras, 533, 3811

\bibitem[{{Donnari} {et~al.}(2021){Donnari}, {Pillepich}, {Nelson}, {Marinacci}, {Vogelsberger}, \& {Hernquist}}]{Donnari_2021_TNG}
{Donnari}, M., {Pillepich}, A., {Nelson}, D., {et~al.} 2021, \mnras, 506, 4760

\bibitem[{Fillingham {et~al.}(2016)Fillingham, Cooper, Pace, Boylan-Kolchin, Bullock, Garrison-Kimmel, \& Wheeler}]{Fillingham_2016}
Fillingham, S.~P., Cooper, M.~C., Pace, A.~B., {et~al.} 2016, MNRAS, 463, 1916

\bibitem[{Fillingham {et~al.}(2015)Fillingham, Cooper, Wheeler, Garrison-Kimmel, Boylan-Kolchin, \& Bullock}]{Fillingham_2015}
Fillingham, S.~P., Cooper, M.~C., Wheeler, C., {et~al.} 2015, MNRAS, 454, 2039

\bibitem[{{Font} {et~al.}(2022){Font}, {McCarthy}, {Belokurov}, {Brown}, \& {Stafford}}]{font_2022}
{Font}, A.~S., {McCarthy}, I.~G., {Belokurov}, V., {Brown}, S.~T., \& {Stafford}, S.~G. 2022, \mnras, 511, 1544

\bibitem[{{Gabor} \& {Dav{\'e}}(2015)}]{Gabor_2015_strang}
{Gabor}, J.~M. \& {Dav{\'e}}, R. 2015, \mnras, 447, 374

\bibitem[{{Geha} {et~al.}(2024){Geha}, {Mao}, {Wechsler}, {Asali}, {Kado-Fong}, {Kallivayalil}, {Nadler}, {Tollerud}, {Weiner}, {de los Reyes}, {Wang}, \& {Wu}}]{Geha_2024}
{Geha}, M., {Mao}, Y.-Y., {Wechsler}, R.~H., {et~al.} 2024, \apj, 976, 118

\bibitem[{Geha {et~al.}(2017)Geha, Wechsler, Mao, Tollerud, Weiner, Bernstein, Hoyle, Marchi, Marshall, Muñoz, \& Lu}]{Geha_2017}
Geha, M., Wechsler, R.~H., Mao, Y.-Y., {et~al.} 2017, ApJ, 847, 4

\bibitem[{{Girelli} {et~al.}(2020){Girelli}, {Pozzetti}, {Bolzonella}, {Giocoli}, {Marulli}, \& {Baldi}}]{Girelli_2020}
{Girelli}, G., {Pozzetti}, L., {Bolzonella}, M., {et~al.} 2020, \aap, 634, A135

\bibitem[{{Grand} {et~al.}(2021){Grand}, {Marinacci}, {Pakmor}, {Simpson}, {Kelly}, {G{\'o}mez}, {Jenkins}, {Springel}, {Frenk}, \& {White}}]{Grand_2021}
{Grand}, R. J.~J., {Marinacci}, F., {Pakmor}, R., {et~al.} 2021, \mnras, 507, 4953

\bibitem[{{Grcevich} \& {Putman}(2009{\natexlab{a}})}]{Grcevich_Putman_HI_2009}
{Grcevich}, J. \& {Putman}, M.~E. 2009{\natexlab{a}}, \apj, 696, 385

\bibitem[{{Grcevich} \& {Putman}(2009{\natexlab{b}})}]{Grcevich_2009_hi}
{Grcevich}, J. \& {Putman}, M.~E. 2009{\natexlab{b}}, \apj, 696, 385

\bibitem[{Green {et~al.}(2021)Green, van~den Bosch, \& Jiang}]{Green_2021}
Green, S.~B., van~den Bosch, F.~C., \& Jiang, F. 2021, MNRAS, 509, 2624

\bibitem[{{Greene} {et~al.}(2023){Greene}, {Danieli}, {Carlsten}, {Beaton}, {Jiang}, \& {Li}}]{Greene_2023_elves_quenchingtimes}
{Greene}, J.~E., {Danieli}, S., {Carlsten}, S., {et~al.} 2023, \apj, 949, 94

\bibitem[{Greene {et~al.}(2023)Greene, Danieli, Carlsten, Beaton, Jiang, \& Li}]{Greene_2023}
Greene, J.~E., Danieli, S., Carlsten, S., {et~al.} 2023, ApJ, 949, 94

\bibitem[{{Gunn} \& {Gott}(1972)}]{Gunn_Gott_1972}
{Gunn}, J.~E. \& {Gott}, J.~Richard, I. 1972, \apj, 176, 1

\bibitem[{Haardt \& Madau(2012)}]{Haardt_2012}
Haardt, F. \& Madau, P. 2012, ApJ, 746, 125

\bibitem[{Han {et~al.}(2017)Han, Cole, Frenk, Benitez-Llambay, \& Helly}]{Han_2018}
Han, J., Cole, S., Frenk, C.~S., Benitez-Llambay, A., \& Helly, J. 2017, MNRAS, 474, 604

\bibitem[{Han {et~al.}(2012)Han, Jing, Wang, \& Wang}]{Han_HBT_2012}
Han, J., Jing, Y.~P., Wang, H., \& Wang, W. 2012, MNRAS, 427, 2437

\bibitem[{{Hausammann} {et~al.}(2019){Hausammann}, {Revaz}, \& {Jablonka}}]{Haussaman_2019}
{Hausammann}, L., {Revaz}, Y., \& {Jablonka}, P. 2019, \aap, 624, A11

\bibitem[{Henriques \& Thomas(2010)}]{henriques_tidal_2010}
Henriques, B. M.~B. \& Thomas, P.~A. 2010, MNRAS, 403, 768

\bibitem[{Hopkins {et~al.}(2018)Hopkins, Wetzel, Kereš, Faucher-Giguère, Quataert, Boylan-Kolchin, Murray, Hayward, Garrison-Kimmel, Hummels, Feldmann, Torrey, Ma, Anglés-Alcázar, Su, Orr, Schmitz, Escala, Sanderson, Grudić, Hafen, Kim, Fitts, Bullock, Wheeler, Chan, Elbert, \& Narayanan}]{Hopkins_2018_overquenching}
Hopkins, P.~F., Wetzel, A., Kereš, D., {et~al.} 2018, MNRAS, 480, 800

\bibitem[{Jiang {et~al.}(2015)Jiang, Cole, Sawala, \& Frenk}]{radial_orbits_2015}
Jiang, L., Cole, S., Sawala, T., \& Frenk, C.~S. 2015, MNRAS, 448, 1674

\bibitem[{{Jung} {et~al.}(2024){Jung}, {Roca-F{\`a}brega}, {Kim}, {Genina}, {Hausammann}, {Kim}, {Lupi}, {Nagamine}, {Powell}, {Revaz}, {Shimizu}, {Vel{\'a}zquez}, {Ceverino}, {Primack}, {Quinn}, {Strawn}, {Abel}, {Dekel}, {Dong}, {Oh}, {Teyssier}, \& {AGORA Collaboration}}]{minyong_paper5}
{Jung}, M., {Roca-F{\`a}brega}, S., {Kim}, J.-H., {et~al.} 2024, \apj, 964, 123

\bibitem[{Kallivayalil {et~al.}(2013)Kallivayalil, van~der Marel, Besla, Anderson, \& Alcock}]{Kallivayalil_2013}
Kallivayalil, N., van~der Marel, R.~P., Besla, G., Anderson, J., \& Alcock, C. 2013, ApJ, 764, 161

\bibitem[{{Karunakaran} {et~al.}(2021){Karunakaran}, {Spekkens}, {Oman}, {Simpson}, {Fattahi}, {Sand}, {Bennet}, {Crnojevi{\'c}}, {Frenk}, {G{\'o}mez}, {Grand}, {Jones}, {Marinacci}, {Mutlu-Pakdil}, {Navarro}, \& {Zaritsky}}]{karunakaran_2021}
{Karunakaran}, A., {Spekkens}, K., {Oman}, K.~A., {et~al.} 2021, \apjl, 916, L19

\bibitem[{Katz {et~al.}(2020)Katz, Ramsoy, Rosdahl, Kimm, Blaizot, Haehnelt, Michel-Dansac, Garel, Laigle, Devriendt, \& Slyz}]{katz_2020}
Katz, H., Ramsoy, M., Rosdahl, J., {et~al.} 2020, MNRAS, 494, 2200

\bibitem[{{Kauffmann} {et~al.}(2003){Kauffmann}, {Heckman}, {White}, {Charlot}, {Tremonti}, {Brinchmann}, {Bruzual}, {Peng}, {Seibert}, {Bernardi}, {Blanton}, {Brinkmann}, {Castander}, {Cs{\'a}bai}, {Fukugita}, {Ivezic}, {Munn}, {Nichol}, {Padmanabhan}, {Thakar}, {Weinberg}, \& {York}}]{Kauffman_2003b}
{Kauffmann}, G., {Heckman}, T.~M., {White}, S. D.~M., {et~al.} 2003, \mnras, 341, 33

\bibitem[{Kereš {et~al.}(2005)Kereš, Katz, Weinberg, \& Davé}]{keres_2005}
Kereš, D., Katz, N., Weinberg, D.~H., \& Davé, R. 2005, MNRAS, 363, 2

\bibitem[{{Kim} {et~al.}(2014){Kim}, {Abel}, {Agertz}, {Bryan}, {Ceverino}, {Christensen}, {Conroy}, {Dekel}, {Gnedin}, {Goldbaum}, {Guedes}, {Hahn}, {Hobbs}, {Hopkins}, {Hummels}, {Iannuzzi}, {Keres}, {Klypin}, {Kravtsov}, {Krumholz}, {Kuhlen}, {Leitner}, {Madau}, {Mayer}, {Moody}, {Nagamine}, {Norman}, {Onorbe}, {O'Shea}, {Pillepich}, {Primack}, {Quinn}, {Read}, {Robertson}, {Rocha}, {Rudd}, {Shen}, {Smith}, {Szalay}, {Teyssier}, {Thompson}, {Todoroki}, {Turk}, {Wadsley}, {Wise}, {Zolotov}, \& {AGORA Collaboration29}}]{Kim_2014_paper1}
{Kim}, J.-h., {Abel}, T., {Agertz}, O., {et~al.} 2014, \apjs, 210, 14

\bibitem[{{Kim} {et~al.}(2016){Kim}, {Agertz}, {Teyssier}, {Butler}, {Ceverino}, {Choi}, {Feldmann}, {Keller}, {Lupi}, {Quinn}, {Revaz}, {Wallace}, {Gnedin}, {Leitner}, {Shen}, {Smith}, {Thompson}, {Turk}, {Abel}, {Arraki}, {Benincasa}, {Chakrabarti}, {DeGraf}, {Dekel}, {Goldbaum}, {Hopkins}, {Hummels}, {Klypin}, {Li}, {Madau}, {Mandelker}, {Mayer}, {Nagamine}, {Nickerson}, {O'Shea}, {Primack}, {Roca-F{\`a}brega}, {Semenov}, {Shimizu}, {Simpson}, {Todoroki}, {Wadsley}, {Wise}, \& {AGORA Collaboration}}]{Kim_2016_paper2}
{Kim}, J.-h., {Agertz}, O., {Teyssier}, R., {et~al.} 2016, \apj, 833, 202

\bibitem[{{Kirby} {et~al.}(2013){Kirby}, {Cohen}, {Guhathakurta}, {Cheng}, {Bullock}, \& {Gallazzi}}]{kirby_2013}
{Kirby}, E.~N., {Cohen}, J.~G., {Guhathakurta}, P., {et~al.} 2013, \apj, 779, 102

\bibitem[{{Knebe} {et~al.}(2011){Knebe}, {Knollmann}, {Muldrew}, {Pearce}, {Aragon-Calvo}, {Ascasibar}, {Behroozi}, {Ceverino}, {Colombi}, {Diemand}, {Dolag}, {Falck}, {Fasel}, {Gardner}, {Gottl{\"o}ber}, {Hsu}, {Iannuzzi}, {Klypin}, {Luki{\'c}}, {Maciejewski}, {McBride}, {Neyrinck}, {Planelles}, {Potter}, {Quilis}, {Rasera}, {Read}, {Ricker}, {Roy}, {Springel}, {Stadel}, {Stinson}, {Sutter}, {Turchaninov}, {Tweed}, {Yepes}, \& {Zemp}}]{Knebe_2011}
{Knebe}, A., {Knollmann}, S.~R., {Muldrew}, S.~I., {et~al.} 2011, \mnras, 415, 2293

\bibitem[{Knebe {et~al.}(2012)Knebe, Libeskind, Pearce, Behroozi, Casado, Dolag, Dominguez-Tenreiro, Elahi, Lux, Muldrew, \& Onions}]{Knebe_2012_galaxycomparison}
Knebe, A., Libeskind, N.~I., Pearce, F., {et~al.} 2012, MNRAS, 428, 2039

\bibitem[{{Kravtsov} {et~al.}(1997){Kravtsov}, {Klypin}, \& {Khokhlov}}]{Kravtsov_97_art}
{Kravtsov}, A.~V., {Klypin}, A.~A., \& {Khokhlov}, A.~M. 1997, \apjs, 111, 73

\bibitem[{{Legrand} {et~al.}(2019){Legrand}, {McCracken}, {Davidzon}, {Ilbert}, {Coupon}, {Aghanim}, {Douspis}, {Capak}, {Le F{\`e}vre}, \& {Milvang-Jensen}}]{Legrand_2019}
{Legrand}, L., {McCracken}, H.~J., {Davidzon}, I., {et~al.} 2019, \mnras, 486, 5468

\bibitem[{{Leroy} {et~al.}(2008){Leroy}, {Walter}, {Brinks}, {Bigiel}, {de Blok}, {Madore}, \& {Thornley}}]{Leroy_2008}
{Leroy}, A.~K., {Walter}, F., {Brinks}, E., {et~al.} 2008, \apj, 136, 2782

\bibitem[{Lovell {et~al.}(2021)Lovell, Wilkins, Thomas, Schaller, Baugh, Fabbian, \& Bahé}]{lovell2021}
Lovell, C.~C., Wilkins, S.~M., Thomas, P.~A., {et~al.} 2021, MNRAS, 509, 5046

\bibitem[{{Mansfield} {et~al.}(2024){Mansfield}, {Darragh-Ford}, {Wang}, {Nadler}, {Diemer}, \& {Wechsler}}]{Symfind_Mansfield_2023}
{Mansfield}, P., {Darragh-Ford}, E., {Wang}, Y., {et~al.} 2024, \apj, 970, 178

\bibitem[{{Mao} {et~al.}(2024){Mao}, {Geha}, {Wechsler}, {Asali}, {Wang}, {Kado-Fong}, {Kallivayalil}, {Nadler}, {Tollerud}, {Weiner}, {de los Reyes}, \& {Wu}}]{mao2024}
{Mao}, Y.-Y., {Geha}, M., {Wechsler}, R.~H., {et~al.} 2024, \apj, 976, 117

\bibitem[{Mao {et~al.}(2021)Mao, Geha, Wechsler, Weiner, Tollerud, Nadler, \& Kallivayalil}]{Mao_2021}
Mao, Y.-Y., Geha, M., Wechsler, R.~H., {et~al.} 2021, ApJ, 907, 85

\bibitem[{Marasco {et~al.}(2016)Marasco, Crain, Schaye, Bahé, van der Hulst, Theuns, \& Bower}]{Marasco_2016}
Marasco, A., Crain, R.~A., Schaye, J., {et~al.} 2016, MNRAS, 461, 2630–2649

\bibitem[{{Mayer} {et~al.}(2001){Mayer}, {Governato}, {Colpi}, {Moore}, {Quinn}, {Wadsley}, {Stadel}, \& {Lake}}]{Mayer_2001_tidal_morph}
{Mayer}, L., {Governato}, F., {Colpi}, M., {et~al.} 2001, \apj, 559, 754

\bibitem[{{Mayer} {et~al.}(2006){Mayer}, {Mastropietro}, {Wadsley}, {Stadel}, \& {Moore}}]{Mayer_2006}
{Mayer}, L., {Mastropietro}, C., {Wadsley}, J., {Stadel}, J., \& {Moore}, B. 2006, \mnras, 369, 1021

\bibitem[{McCarthy {et~al.}(2007)McCarthy, Frenk, Font, Lacey, Bower, Mitchell, Balogh, \& Theuns}]{mccarthy_2018}
McCarthy, I.~G., Frenk, C.~S., Font, A.~S., {et~al.} 2007, MNRAS, 383, 593

\bibitem[{{McConnachie}(2012)}]{Mcconnachie_2012}
{McConnachie}, A.~W. 2012, \apj, 144, 4

\bibitem[{Merluzzi {et~al.}(2016)Merluzzi, Busarello, Dopita, Haines, Steinhauser, Bourdin, \& Mazzotta}]{MERLUZZI_2016}
Merluzzi, P., Busarello, G., Dopita, M.~A., {et~al.} 2016, MNRAS, 460, 3345

\bibitem[{{Moore} {et~al.}(1996){Moore}, {Katz}, {Lake}, {Dressler}, \& {Oemler}}]{Moore_96_harassment}
{Moore}, B., {Katz}, N., {Lake}, G., {Dressler}, A., \& {Oemler}, A. 1996, \nat, 379, 613

\bibitem[{Muldrew {et~al.}(2010)Muldrew, Pearce, \& Power}]{Muldrew_2010_subhalodetection}
Muldrew, S.~I., Pearce, F.~R., \& Power, C. 2010, MNRAS, 410, 2617

\bibitem[{{Munshi} {et~al.}(2021){Munshi}, {Brooks}, {Applebaum}, {Christensen}, {Quinn}, \& {Sligh}}]{Munshi_2021}
{Munshi}, F., {Brooks}, A.~M., {Applebaum}, E., {et~al.} 2021, \apj, 923, 35

\bibitem[{{Nelson} {et~al.}(2018){Nelson}, {Pillepich}, {Springel}, {Weinberger}, {Hernquist}, {Pakmor}, {Genel}, {Torrey}, {Vogelsberger}, {Kauffmann}, {Marinacci}, \& {Naiman}}]{Nelso_2018_bimodality_TNG}
{Nelson}, D., {Pillepich}, A., {Springel}, V., {et~al.} 2018, \mnras, 475, 624

\bibitem[{{Nulsen}(1982)}]{Nulsen_1982_viscous}
{Nulsen}, P.~E.~J. 1982, \mnras, 198, 1007

\bibitem[{{Onions} {et~al.}(2012){Onions}, {Knebe}, {Pearce}, {Muldrew}, {Lux}, {Knollmann}, {Ascasibar}, {Behroozi}, {Elahi}, {Han}, {Maciejewski}, {Merch{\'a}n}, {Neyrinck}, {Ruiz}, {Sgr{\'o}}, {Springel}, \& {Tweed}}]{Onions_2012_subhalofindercomparison}
{Onions}, J., {Knebe}, A., {Pearce}, F.~R., {et~al.} 2012, \mnras, 423, 1200

\bibitem[{{Peng} {et~al.}(2015){Peng}, {Maiolino}, \& {Cochrane}}]{Peng_2015_strang}
{Peng}, Y., {Maiolino}, R., \& {Cochrane}, R. 2015, \nat, 521, 192

\bibitem[{{Peng} {et~al.}(2010){Peng}, {Lilly}, {Kova{\v{c}}}, {Bolzonella}, {Pozzetti}, {Renzini}, {Zamorani}, {Ilbert}, {Knobel}, {Iovino}, {Maier}, {Cucciati}, {Tasca}, {Carollo}, {Silverman}, {Kampczyk}, {de Ravel}, {Sanders}, {Scoville}, {Contini}, {Mainieri}, {Scodeggio}, {Kneib}, {Le F{\`e}vre}, {Bardelli}, {Bongiorno}, {Caputi}, {Coppa}, {de la Torre}, {Franzetti}, {Garilli}, {Lamareille}, {Le Borgne}, {Le Brun}, {Mignoli}, {Perez Montero}, {Pello}, {Ricciardelli}, {Tanaka}, {Tresse}, {Vergani}, {Welikala}, {Zucca}, {Oesch}, {Abbas}, {Barnes}, {Bordoloi}, {Bottini}, {Cappi}, {Cassata}, {Cimatti}, {Fumana}, {Hasinger}, {Koekemoer}, {Leauthaud}, {Maccagni}, {Marinoni}, {McCracken}, {Memeo}, {Meneux}, {Nair}, {Porciani}, {Presotto}, \& {Scaramella}}]{Peng_2010_quenching}
{Peng}, Y.-j., {Lilly}, S.~J., {Kova{\v{c}}}, K., {et~al.} 2010, \apj, 721, 193

\bibitem[{{Price}(2008)}]{Price_2008_SPH}
{Price}, D.~J. 2008, JCP, 227, 10040

\bibitem[{{Putman} {et~al.}(2021){Putman}, {Zheng}, {Price-Whelan}, {Grcevich}, {Johnson}, {Tollerud}, \& {Peek}}]{Putman_2021}
{Putman}, M.~E., {Zheng}, Y., {Price-Whelan}, A.~M., {et~al.} 2021, \apj, 913, 53

\bibitem[{{Quilis} {et~al.}(2000){Quilis}, {Moore}, \& {Bower}}]{Quilis_2000_viscous}
{Quilis}, V., {Moore}, B., \& {Bower}, R. 2000, Science, 288, 1617

\bibitem[{{Read} {et~al.}(2010){Read}, {Hayfield}, \& {Agertz}}]{Read_2010_sph}
{Read}, J.~I., {Hayfield}, T., \& {Agertz}, O. 2010, \mnras, 405, 1513

\bibitem[{{Read} {et~al.}(2006){Read}, {Wilkinson}, {Evans}, {Gilmore}, \& {Kleyna}}]{Read_2006_tidal}
{Read}, J.~I., {Wilkinson}, M.~I., {Evans}, N.~W., {Gilmore}, G., \& {Kleyna}, J.~T. 2006, \mnras, 366, 429

\bibitem[{{Revaz} \& {Jablonka}(2012)}]{Revaz_Jablonka_2012_gear}
{Revaz}, Y. \& {Jablonka}, P. 2012, \aap, 538, A82

\bibitem[{{Roca-F{\`a}brega} {et~al.}(2021){Roca-F{\`a}brega}, {Kim}, {Hausammann}, {Nagamine}, {Lupi}, {Powell}, {Shimizu}, {Ceverino}, {Primack}, {Quinn}, {Revaz}, {Vel{\'a}zquez}, {Abel}, {Buehlmann}, {Dekel}, {Dong}, {Hahn}, {Hummels}, {Kim}, {Smith}, {Strawn}, {Teyssier}, {Turk}, \& {AGORA Collaboration}}]{santi_paper3_agora}
{Roca-F{\`a}brega}, S., {Kim}, J.-H., {Hausammann}, L., {et~al.} 2021, \apj, 917, 64

\bibitem[{{Roca-F{\`a}brega} {et~al.}(2024){Roca-F{\`a}brega}, {Kim}, {Primack}, {Jung}, {Genina}, {Hausammann}, {Kim}, {Lupi}, {Nagamine}, {Powell}, {Revaz}, {Shimizu}, {Strawn}, {Vel{\'a}zquez}, {Abel}, {Ceverino}, {Dong}, {Quinn}, {Shin}, {Segovia-Otero}, {Agertz}, {Barrow}, {Cadiou}, {Dekel}, {Hummels}, {Oh}, {Teyssier}, \& {AGORA Collaboration}}]{santi_paper_4_agora}
{Roca-F{\`a}brega}, S., {Kim}, J.-H., {Primack}, J.~R., {et~al.} 2024, \apj, 968, 125

\bibitem[{{Rodriguez Wimberly} {et~al.}(2019){Rodriguez Wimberly}, {Cooper}, {Fillingham}, {Boylan-Kolchin}, {Bullock}, \& {Garrison-Kimmel}}]{Rodriguez_Wimberly_2019_reionization}
{Rodriguez Wimberly}, M.~K., {Cooper}, M.~C., {Fillingham}, S.~P., {et~al.} 2019, \mnras, 483, 4031

\bibitem[{{Roediger} \& {Hensler}(2005)}]{Roediger_2005}
{Roediger}, E. \& {Hensler}, G. 2005, \aap, 433, 875

\bibitem[{{Samuel} {et~al.}(2023){Samuel}, {Pardasani}, {Wetzel}, {Santistevan}, {Boylan-Kolchin}, {Moreno}, \& {Faucher-Gigu{\`e}re}}]{Samuel_2023}
{Samuel}, J., {Pardasani}, B., {Wetzel}, A., {et~al.} 2023, \mnras, 525, 3849

\bibitem[{Samuel {et~al.}(2022)Samuel, Wetzel, Santistevan, Tollerud, Moreno, Boylan-Kolchin, Bailin, \& Pardasani}]{Samuel_fire_2022}
Samuel, J., Wetzel, A., Santistevan, I., {et~al.} 2022, MNRAS, 514, 5276–5295

\bibitem[{Schaller {et~al.}(2015)Schaller, Frenk, Bower, Theuns, Jenkins, Schaye, Crain, Furlong, Dalla~Vecchia, \& McCarthy}]{schaller2015}
Schaller, M., Frenk, C.~S., Bower, R.~G., {et~al.} 2015, MNRAS, 451, 1247

\bibitem[{{Schaye} {et~al.}(2010){Schaye}, {Dalla Vecchia}, {Booth}, {Wiersma}, {Theuns}, {Haas}, {Bertone}, {Duffy}, {McCarthy}, \& {van de Voort}}]{Schaye_2010_physics_quench}
{Schaye}, J., {Dalla Vecchia}, C., {Booth}, C.~M., {et~al.} 2010, \mnras, 402, 1536

\bibitem[{{Schulz} \& {Struck}(2001)}]{Schulz_2001_viscous}
{Schulz}, S. \& {Struck}, C. 2001, \mnras, 328, 185

\bibitem[{{Shuntov} {et~al.}(2024){Shuntov}, {Ilbert}, {Toft}, {Arango-Toro}, {Akins}, {Casey}, {Franco}, {Harish}, {Kartaltepe}, {Koekemoer}, {McCracken}, {Paquereau}, {Laigle}, {Bethermin}, {Dubois}, {Drakos}, {Faisst}, {Gozaliasl}, {Gillman}, {Hayward}, {Hirschmann}, {Huertas-Company}, {Jespersen}, {Jin}, {Kokorev}, {Lambrides}, {Le Borgne}, {Liu}, {Magdis}, {Massey}, {McPartland}, {Mercier}, {McCleary}, {McKinney}, {Oesch}, {Rhodes}, {Rich}, {Robertson}, {Sanders}, {Trebitsch}, {Tresse}, {Valentino}, {Vijayan}, {Weaver}, {Weibel}, \& {Wilkins}}]{Shuntov_2024}
{Shuntov}, M., {Ilbert}, O., {Toft}, S., {et~al.} 2024, arXiv e-prints, arXiv:2410.08290

\bibitem[{{Simpson} {et~al.}(2018){Simpson}, {Grand}, {G{\'o}mez}, {Marinacci}, {Pakmor}, {Springel}, {Campbell}, \& {Frenk}}]{Simpson_2018_ram_pressure_MW}
{Simpson}, C.~M., {Grand}, R. J.~J., {G{\'o}mez}, F.~A., {et~al.} 2018, \mnras, 478, 548

\bibitem[{Smith {et~al.}(2016)Smith, Bryan, Glover, Goldbaum, Turk, Regan, Wise, Schive, Abel, Emerick, O'Shea, Anninos, Hummels, \& Khochfar}]{smith_2017}
Smith, B.~D., Bryan, G.~L., Glover, S. C.~O., {et~al.} 2016, MNRAS, 466, 2217

\bibitem[{{Spekkens} {et~al.}(2014){Spekkens}, {Urbancic}, {Mason}, {Willman}, \& {Aguirre}}]{Spekkens_2014}
{Spekkens}, K., {Urbancic}, N., {Mason}, B.~S., {Willman}, B., \& {Aguirre}, J.~E. 2014, \apjl, 795, L5

\bibitem[{Springel(2005)}]{Springel_2005_gadget2}
Springel, V. 2005, MNRAS, 364, 1105

\bibitem[{Springel(2010)}]{Springel_2010}
Springel, V. 2010, MNRAS, 401, 791

\bibitem[{{Springel} {et~al.}(2001){Springel}, {White}, {Tormen}, \& {Kauffmann}}]{Subfind_springel_2001}
{Springel}, V., {White}, S. D.~M., {Tormen}, G., \& {Kauffmann}, G. 2001, \mnras, 328, 726

\bibitem[{{Strawn} {et~al.}(2024){Strawn}, {Roca-F{\`a}brega}, {Primack}, {Kim}, {Genina}, {Hausammann}, {Kim}, {Lupi}, {Nagamine}, {Powell}, {Revaz}, {Shimizu}, {Vel{\'a}zquez}, {Abel}, {Ceverino}, {Dong}, {Jung}, {Quinn}, {Shin}, {Barrow}, {Dekel}, {Oh}, {Mandelker}, {Teyssier}, {Hummels}, {Maji}, {Man}, {Mayerhofer}, \& {The Agora Collaboration}}]{Strawn_2024}
{Strawn}, C., {Roca-F{\`a}brega}, S., {Primack}, J.~R., {et~al.} 2024, \apj, 962, 29

\bibitem[{{Wadsley} {et~al.}(2008){Wadsley}, {Veeravalli}, \& {Couchman}}]{Wadsley_2008_SPH}
{Wadsley}, J.~W., {Veeravalli}, G., \& {Couchman}, H.~M.~P. 2008, \mnras, 387, 427

\bibitem[{{Wang} {et~al.}(2024){Wang}, {Nadler}, {Mao}, {Wechsler}, {Abel}, {Behroozi}, {Geha}, {Asali}, {de los Reyes}, {Kado-Fong}, {Kallivayalil}, {Tollerud}, {Weiner}, \& {Wu}}]{Wang_saga_5_2024}
{Wang}, Y., {Nadler}, E.~O., {Mao}, Y.-Y., {et~al.} 2024, \apj, 976, 119

\bibitem[{{Weisz} {et~al.}(2014){Weisz}, {Dolphin}, {Skillman}, {Holtzman}, {Gilbert}, {Dalcanton}, \& {Williams}}]{Weisz_2014_reionization}
{Weisz}, D.~R., {Dolphin}, A.~E., {Skillman}, E.~D., {et~al.} 2014, \apj, 789, 148

\bibitem[{Wetzel {et~al.}(2015{\natexlab{a}})Wetzel, Deason, \& Garrison-Kimmel}]{Wetzel_2015a}
Wetzel, A.~R., Deason, A.~J., \& Garrison-Kimmel, S. 2015{\natexlab{a}}, ApJ, 807, 49

\bibitem[{{Wetzel} {et~al.}(2013){Wetzel}, {Tinker}, {Conroy}, \& {van den Bosch}}]{Wetzel_2013}
{Wetzel}, A.~R., {Tinker}, J.~L., {Conroy}, C., \& {van den Bosch}, F.~C. 2013, \mnras, 432, 336

\bibitem[{Wetzel {et~al.}(2015{\natexlab{b}})Wetzel, Tollerud, \& Weisz}]{Wetzel_2015b}
Wetzel, A.~R., Tollerud, E.~J., \& Weisz, D.~R. 2015{\natexlab{b}}, ApJ Letters, 808, L27

\bibitem[{Wheeler {et~al.}(2014)Wheeler, Phillips, Cooper, Boylan-Kolchin, \& Bullock}]{Wheeler_2014}
Wheeler, C., Phillips, J.~I., Cooper, M.~C., Boylan-Kolchin, M., \& Bullock, J.~S. 2014, MNRAS, 442, 1396

\bibitem[{{White} \& {Frenk}(1991)}]{White_frenk_91}
{White}, S. D.~M. \& {Frenk}, C.~S. 1991, \apj, 379, 52

\bibitem[{{Wright} {et~al.}(2022){Wright}, {Lagos}, {Power}, {Stevens}, {Cortese}, \& {Poulton}}]{Wright_2022}
{Wright}, R.~J., {Lagos}, C. d.~P., {Power}, C., {et~al.} 2022, \mnras, 516, 2891

\bibitem[{{Zaritsky} {et~al.}(1993){Zaritsky}, {Smith}, {Frenk}, \& {White}}]{zaritsky_93}
{Zaritsky}, D., {Smith}, R., {Frenk}, C., \& {White}, S. D.~M. 1993, \apj, 405, 464

\bibitem[{Zaritsky {et~al.}(1997)Zaritsky, Smith, Frenk, \& White}]{Zaritsky_1997}
Zaritsky, D., Smith, R., Frenk, C., \& White, S. D.~M. 1997, ApJ, 478, 39

\bibitem[{{Zhu} {et~al.}(2024){Zhu}, {Tonnesen}, {Bryan}, \& {Putman}}]{zhu2024itsbreezecircumgalacticmedium}
{Zhu}, J., {Tonnesen}, S., {Bryan}, G.~L., \& {Putman}, M.~E. 2024, \apj, 974, 142

\end{thebibliography}
\bibliographystyle{aa}

\newpage
\appendix
\section{Determining gas content and inflows: Comparing grid-based versus particle-based codes}
\label{appendix:inflows}

When comparing particle-based and grid-based codes, it is crucial to ensure that the methods used don't introduce any bias. Regarding this, the simulation analysis code yt provides a solid framework in AGORA to ensure a consistent approach across all codes. As mentioned in the paper, when it comes to determining the gas mass that is effectively bound to a subhalo, particle-based codes are particularly useful since they allow us to determine the number of particles that are gravitationally bound to a subhalo. However, as explained in Section \ref{gas_assigment}, this is not feasible in grid-based codes. Therefore, we employed a method that corrects the total gas mass within the subhalo’s virial radius by subtracting the gas associated with the host CGM during its infall. This process is outlined in Section \ref{gas_assigment}. By applying this method to particle-based codes, we can verify whether it converges with the mass of gas that is gravitationally bound, helping us assess whether the subhalo retains gas during its interaction. This comparison can be found in the second row of Figure \ref{fig:appendix}, where we show the evolution of the total gas mass contained within $\rm{R_{vir}^{sub}}$ and of the gas that is effectively bound, calculated using the gravitationally bound gas particles (particle-based method) and the gas obtained by subtracting the host CGM gas (grid-based method). As we can observe in this example, our approach for determining the bound gas in grid-based codes, by subtracting the CGM gas, largely converges with the results obtained by calculating the number of gravitationally bound particles in particle-based codes. This ensures that we accurately capture the evolution in the gas content of our subhalos during the interaction, independently of the code.

\begin{figure*}
    \centering
    \includegraphics[width=0.99\linewidth]{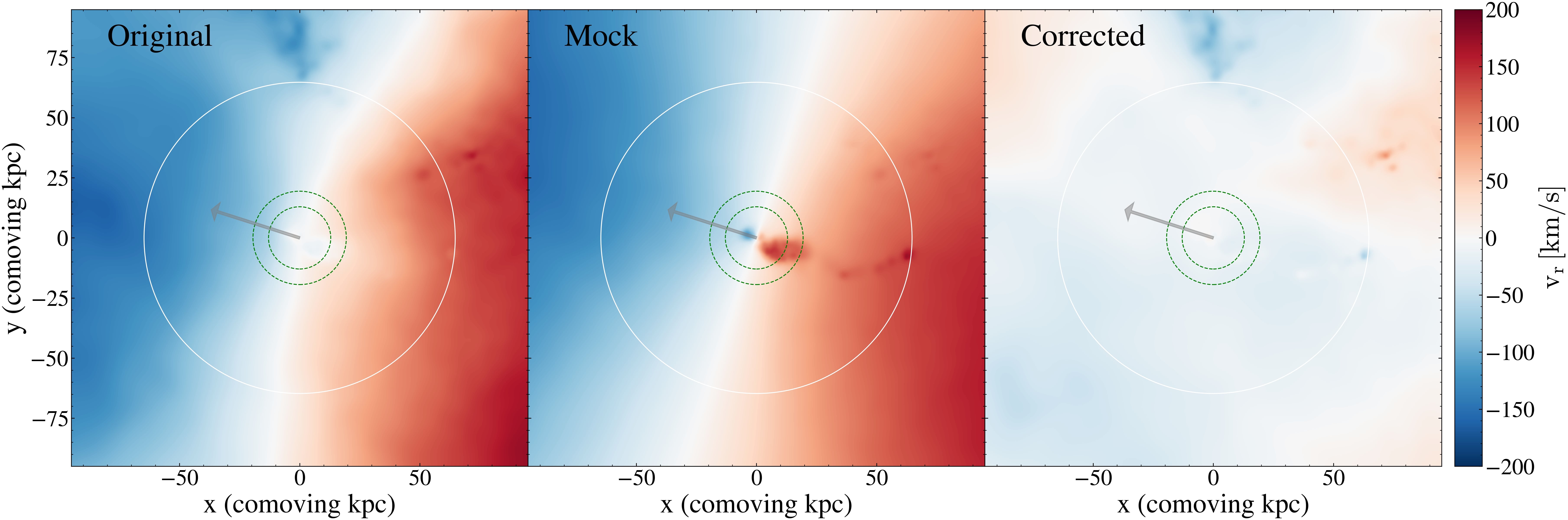}
    \caption{Radial velocity maps for a single snapshot of a subhalo during its infall into the host halo for one of the codes. The white solid circle represents \(\rm{R_{vir}^{sub}}\), while the dashed green circles indicate \(0.2\rm{R_{vir}^{sub}}\) and \(0.3\rm{R_{vir}^{sub}}\). We calculate the rate of inflowing gas in the \(0.1\rm{R_{vir}^{sub}}\) sized shell at \(0.2\rm{R_{vir}^{sub}}\) radii from the center of the subhalo. \textbf{Left:} Original radial velocity map before corrections. As our subhalo moves through the host CGM, CGM gas in the direction of motion will be observed as infalling gas with a negative radial velocity, while the gas it leaves behind will be seen as an outflow with a positive radial velocity. \textbf{Center:} Mock map showing how the radial velocity map would appear exclusively due to the effect of the subhalo's motion relative to the host CGM. \textbf{Right:} Radial velocity map after subtracting the radial velocity values from the mock map for all gas cells except those associated with the subhalo (i.e., cells with velocities relative to the subhalo \(< 2\sigma_{\text{sub}}\) and within \(\rm{R_{vir}^{sub}}\)). After this correction, we can accurately identify the radial velocity of the gas relative to the satellite without introducing the effects of the CGM gas motion during its infall, allowing us to determine the subhalo's inflow mass rate during infall.}

    \label{fig:rv_correction}
\end{figure*}

\begin{figure*}
    \centering
    \includegraphics[width=0.99\linewidth]{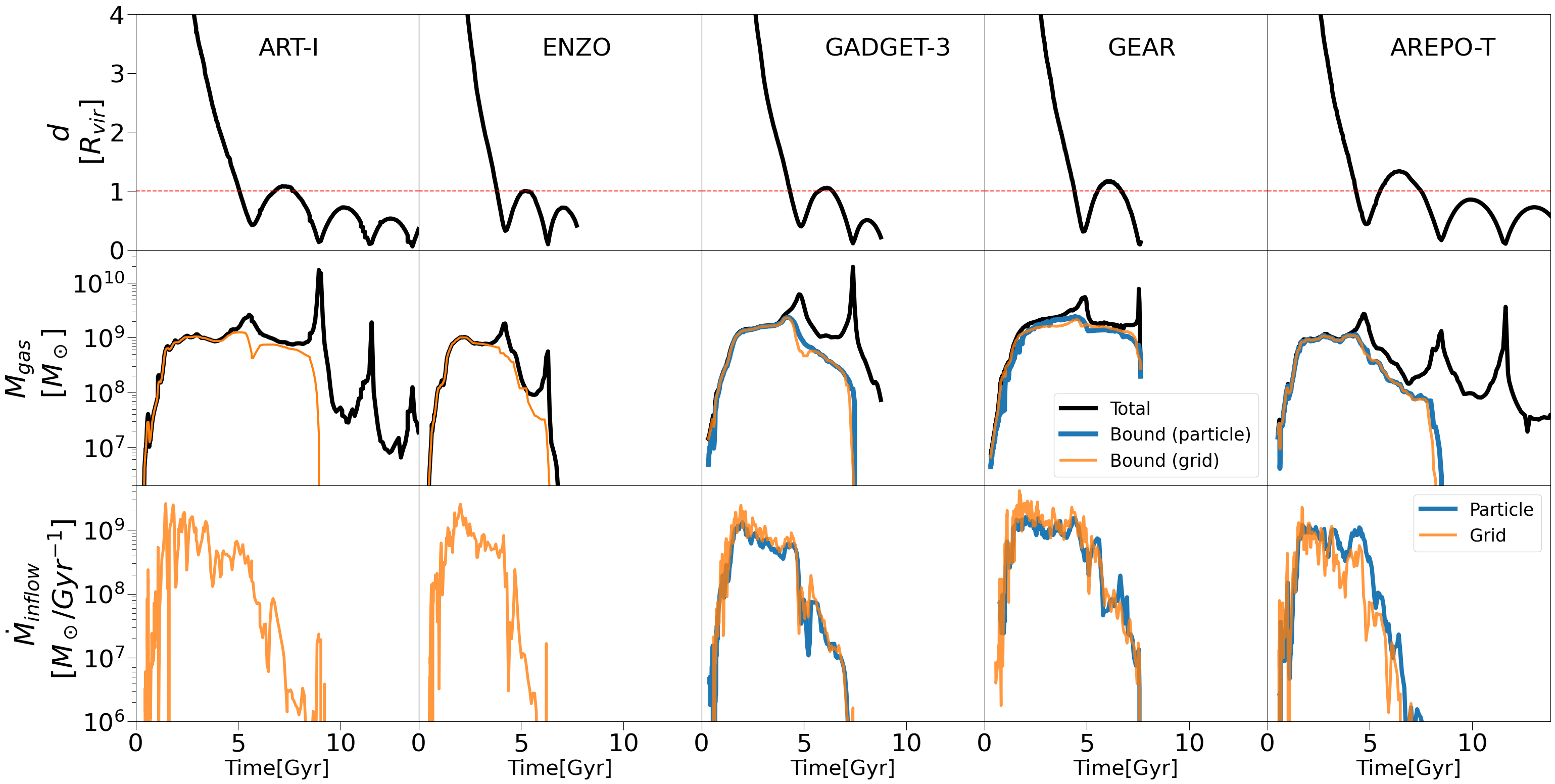}
    \caption{Comparison between determination of satellite (bound) gas content and satellite inflows mass rate using both particle and grid approaches, plotted as blue and orange lines, respectively. Both approaches are described in Section \ref{gas_assigment} and Appendix \ref{appendix:inflows}. The evolution of the properties for the same satellite identified across the different \texttt{CosmoRun} models is shown. \textbf{Top:} Trajectory in $\rm{R_{vir}^{host}}$ units. \textbf{Center:} Evolution of total and bound gas mass for the satellite, comparing bound gas content determined using both particle and grid approaches. \textbf{Bottom:} Evolution of the satellite's inflow mass rate, determined using both particle and grid approaches.}
    \label{fig:appendix}
\end{figure*}

When it comes to determining the amount of gas that is accreted through inflows, we have a similar situation. As we explained, for particle-based codes, we have the IDs of the gas particles. This allows us to compare the IDs of the particles contained (and bound) within \(0.2\rm{R_{vir}^{sub}}\) with temperature below \(T < 10^{4.5}K\) and density $\rm{n_H > 10^{-2.5}\,cm^{-3}}$ in consecutive snapshots to determine the accreted inflow ratio for our subhalo. However, as we discussed in Section \ref{sec:how?}, in the case of grid-based codes, we do not have gas particles to track. The alternative method explained in Section \ref{sec:how?} is based on using the gas mass contained in the cells, so we will consider as inflows all cells that meet the following conditions: i) they reside between \(0.2 \, \rm{R_{vir}^{sub}}\) and \(0.3 \, \rm{R_{vir}^{sub}}\), ii) they move with a sufficiently negative radial velocity to cross the shell between \(0.2 \, \rm{R_{vir}^{sub}}\) and \(0.3 \, \rm{R_{vir}^{sub}}\) within 100 Myr, and iii) they have a temperature lower than \(10^{4.5}\) K and density $\rm{n_H > 10^{-2.5}\,cm^{-3}}$. However, because our satellite is falling with a certain velocity toward the host's gravitational well, all CGM gas in the direction of motion will be observed as infalling gas with a negative radial velocity, while the gas it leaves behind will be seen as an outflow with a positive radial velocity (both with a magnitude close to the satellite's velocity relative to the CGM). To avoid attributing CGM gas that is approaching as inflow, we correct for the effect of the subhalo's velocity relative to the CGM by creating a mock map of how the radial velocity map would appear solely due to this effect. Subsequently, we correct the radial velocity of the cells by subtracting the value that the mock map has for each cell. To prevent correcting gas that is truly associated with the subhalo and assigning it incorrect radial velocities, we excluded from this correction the cells that have a velocity relative to the subhalo of less than \(2 \sigma_{\text{sub}}\) and that are within \(\rm{R_{vir}^{sub}}\). Once this correction is applied, we use the above-mentioned conditions to determine the inflows. An example of this correction for a snapshot during a subhalo's infall for one of our codes can be seen in Figure \ref{fig:rv_correction}. In this figure, we can see how this correction allows us to accurately identify the radial velocity of the gas relative to the satellite without introducing the motion relative to the CGM gas during its infall. By comparing the mass inflow rate determined for a satellite using both particle and grid approaches, we can verify that our method successfully captures the behavior described by the tracked gas particles.  The convergence between the two approaches, both in terms of gas content and inflow mass rate, ensures that we are not introducing any bias in our analysis between particle-based and grid-based codes.

\end{document}